%% file: m33ppak_accepted.tex
\documentclass[useAMS,usenatbib,usegraphicx,referee]{mn2e}
\usepackage[usenames,dvipsnames]{xcolor}
\usepackage{threeparttable}
\usepackage{lscape}
\usepackage{amsmath}
\usepackage{comment}
\usepackage{times}

\newsavebox{\fmbox}

\newcommand{\degree}{\ensuremath{^\circ}}
\newcommand{\arcseconds}{\ensuremath{^{\prime\prime}}}

\newcommand{\Ha}{\ensuremath{{\rm H\alpha}}}
\newcommand{\Hb}{\ensuremath{{\rm H\beta}}}

\newcommand{\Ne}{\ensuremath{n_{\rm e}}}
\newcommand{\Te}{\ensuremath{T_{\rm e}}}

\newcommand{\HII}{H{\textsc{ii}}}

\newcommand{\OIII}{O{\textsc{iii}}}
\newcommand{\OII}{O{\textsc{ii}}}

\newcommand{\SIII}{S{\textsc{iii}}}
\newcommand{\SII}{S{\textsc{ii}}}
\newcommand{\NII}{N{\textsc{ii}}}

\newcommand{\HeII}{He{\textsc{ii}}}

\newcommand{\ArIII}{Ar{\textsc{iii}}}

\def\apj{{ApJ}}
\def\apjs{{ApJS}}
\def\aj{{AJ}}
\def\aap{{A\&A}}
\def\mnras{{MNRAS}}
\def\araa{{ARA\&A}}
\def\pasp{{PASP}}
\def\physscr{{At. Data Nucl. Data Tables}}
\def\aaps{{A\&AS}}
\def\apjl{{ApJL}}
\def\nar{{NewAR}}
\def\pasj{{PASJ}}               

\title[IFS of \HII\ regions in M 33]
{Integral field spectroscopy of H{\normalsize II} regions in M33}
\author[Jes\'{u}s L\'{o}pez-Hern\'{a}ndez et al.] 
{Jes\'{u}s~L\'{o}pez-Hern\'{a}ndez$^1$,\thanks{jlopez@inaoep.mx}
Elena Terlevich$^{1}$, Roberto Terlevich$^{1,2}$, 
\newauthor
Daniel Rosa-Gonz\'{a}lez$^{1}$, \'{A}ngeles D\'{i}az$^{3}$,
Rub\'{e}n Garc\'{i}a-Benito$^{4}$, 
\newauthor
Jos\'{e} V\'{i}lchez$^{4}$, Guillermo H\"agele$^{3,5,6}$ \\
$^{1}$Instituto Nacional de Astrof\'{i}sica, \'{O}ptica y Electr\'{o}nica, 
Luis Enrique Erro 1. Tonantzintla, Puebla, C.P. 72840, M\'{e}xico.  \\
$^{2}$Institute of Astronomy, Madingley Rd., Cambridge, CB3 0HA, UK.\\
$^{3}$Departamento de F\'{i}sica Te\'{o}rica, C-XI,
Universidad Aut\'{o}noma de Madrid, 28049 Madrid, Spain.\\
$^{4}$Instituto de Astrof\'{i}sica de Andaluc\'{i}a (CSIC), 
Glorieta de la Astronom\'{i}a s. n., 18008 Granada, Spain. \\
$^{5}$ Consejo Nacional de Investigaciones Cient\'{i}ficas y T\'{ecnicas}
(CONICET), Argentina. \\
$^{6}$ Facultad de Ciencias Astron\'{o}micas y Geof\'{i}sicas,
Universidad Nacional de La Plata,
Paseo del Bosque s/n, 1900 La Plata, Argentina. }

\begin{document}

\date{Accepted yyyy month dd. Received yyyy month dd; in original form yyyy Month ss}

\pagerange{\pageref{firstpage}--\pageref{lastpage}} \pubyear{2012}

\maketitle
\newpage
\label{firstpage}

\begin{abstract}

Integral field spectroscopy (IFS) is presented for star 
forming regions in M 33. A central area of 300 x 500 
pc$^2$ and the external \HII\ region IC 132, at a 
galactocentric distance $\sim {\rm 19}^{\prime}$ 
(4.69 kpc) were observed with the Potsdam Multi 
Aperture Spectrophotometer (PMAS) instrument at 
the 3.5 m telescope of the Calar Alto Hispano-Alem\'{a}n 
observatory (CAHA). The spectral coverage goes 
from 3600 \AA\ to 1$\mu$m to include from [\OII]$\lambda$3727 \AA\ to the 
near infrared lines required for deriving  sulphur electron temperature and 
abundance diagnostics.

Local conditions within individual \HII\ regions  are 
presented in the form of emission line fluxes and 
physical conditions for each spatial 
resolution element (spaxel) and for segments with
similar \Ha\ surface brightness.

A clear dichotomy is observed when comparing the central 
to outer disc \HII\ regions. While the external \HII\ region has
higher electron temperature plus larger \Hb\ equivalent width,
size and excitation, the central region has higher extinction
and metal content.

The dichotomy extends to the BPT diagnostic diagrams 
that show two orthogonal broad distributions of points. 
By comparing with pseudo-3D photoionization models 
we conclude that the bulk observed differences are probably 
related to a different ionization parameter and metallicity.

Wolf-Rayet features are detected in  IC~132, and resolved 
into two concentrations whose integrated spectra were 
used to estimate the characteristic number of WR stars. 
No WR features were detected in the central \HII\ regions
despite their higher metallicity.

\end{abstract}
\begin{keywords}
ISM: abundances \HII\ regions - Galaxies: individual: M 33
\end{keywords}

\section{Introduction}
Extragalactic \HII\ regions provide an excellent laboratory to 
study star formation processes, evolution of massive stars and 
the properties of the surrounding interstellar medium (ISM).
A wealth of information can be obtained from the spectral 
analysis of the bright emission lines and the stellar continuum.

The spectral information from extragalactic \HII\ regions has 
been traditionally obtained from single aperture or long slit 
observations of the brightest part of the region. The data  is  
then used to derive the physical conditions of the gas 
(temperatures and densities) and to estimate abundances and 
ionization conditions, as well as characteristics of the 
ionizing star clusters [masses, ages, effective temperatures (T$_*$)].  These results 
constitute the main body of our knowledge regarding the 
evolution of disc galaxies \citep[see e.g.][]
{aller42,searle71,smith75,pageletal79, diazetal87,
vilchezetal88,dinerstein90,skillman98,kennicutt98,bresolin04}.

Underpinning this proven methodology to study the emission 
line spectra lies the tacit assumption that the observations 
and the derived measurements are representative of the whole
\HII\ region, basically internal variations within the nebula 
are assumed to be minimal or non existent. The limitation of 
this assumption has been long recognized 
\citep{diazetal87,rubin89,castanedavilcop92}, 
however in the majority of the cases any solution has been 
precluded by limitations in the observation technology and
optimization of the assigned observing time. Nevertheless some
options have been available to combine spectral information
with spatial resolution.  A simple solution is  to sweep the \HII\ 
region following a direction perpendicular to the slit elongation 
\citep{kosugi95,maizperhes04} in steps separated by one slit width.
A different approach involves the use of a  set of narrow band filters 
to obtain monochromatic images centred on the emission lines that, 
when combined with neighbouring continuum bands, allows the 
extraction of  pure emission images. Tunable filters with 
Fabry-Perot interferometers offer a more flexible scheme by 
allowing the use of various pass bands, centred at different 
wavelengths. 

Of particular interest for this work is the technique of integral 
field spectroscopy \cite[][IFS]{asmith06}, also known as 3D or 
area spectroscopy, in which by the use of integral field unit 
instruments (IFUs) it is possible to obtain simultaneous spectral 
and spatial information over the observed object. The main 
advantage of the IFS approach over other techniques is that the 
observation is simultaneous in space and wavelength, producing 
an homogeneous set of data. The drawback of IFS is the small 
field of view (FOV), thus the selection of the optimal method 
(tuneable filters, long slit sweeping, IFS) for spatially resolved 
observations depends on the objective of each specific project. 

IFS has proved its utility in observations of luminous and 
ultraluminous infrared galaxies \citep{aherreroetal09,
gmarinetal09,aherreroetal10}, the nuclei of active galaxies 
\citep{barbosaetal09}, circumnuclear star forming regions 
\citep{dorsetal08} and blue compact galaxies 
\citep{cairosetal09a,cairoestal09b,lagosetal09,jamesetal10,pmonteroEa11}.
Some \HII\ regions have been studied with IFS; observations 
of \HII\ regions in the outer disc  of NGC 6946 \citep[with the same
instrument and setup as in this work]{gbenitoetal10} 
among them. IFS of \HII\ regions in M 33 has been already obtained 
\citep[for NGC 588 and NGC 595 respectively]{miberoetal11,
relanoetl10}, in both cases the PMAS (Potsdam MultiAperture 
Spectrophotometer)  instrument at the 3.5m telescope of the 
Observatorio Astron\'omico Hispano-Alem\'an in Calar Alto 
(CAHA) was used.  They compared the distribution maps for 
different parameters against the reported values in the 
literature, obtaining a good agreement when the IFS data 
is integrated in a single spectrum and noting  that 
for some quantities (e.g.~extinction) to assume a single 
value for the whole region is not strictly correct. 
For some of the empirical abundance estimators a spatial
dispersion is present, however the variation is within the 
estimated observational error and in such case a uniform 
distribution may be assumed.

The general objectives of this work are, using IFS data for 
the centre and an external region of M 33, to characterize the 
extent of the internal variations in \HII\ regions for measured 
and derived values; to test whether the description of the whole 
\HII\ region with a single value for different parameters is 
valid; to map the massive star content using characteristic 
spectral features (i. e. Wolf Rayet stars) and to compare internal 
variations in high metallicity (inner) {\it vs.\ } low metallicity 
(outer) \HII\ regions. Also, given that the central region remains 
an ill defined zone with scarce spectroscopic observations, IFS 
may contribute to its better characterization.

The organization of the paper is as follows. The observations,
data reduction and integrated emission line maps are described in Section 2.  Section 3  
describes the physical conditions of the gas. Section 4
concerns the chemical abundance determinations. Section 5 is 
devoted to the discussion of  diagnostic diagrams. The detection 
of Wolf-Rayet features is presented in section 6. 
Analysis in projected shells is shown in Section 7. Finally the 
conclusions  are presented in Section 8.

\section{Observations and data reduction}
\subsection{Object selection}

M 33 is the third brightest member of the Local Group 
\citep{vanden00}. Given its proximity, large angular size and 
low inclination, it is an ideal candidate for IFS observations, 
having a rich content of \HII\ regions in the central region and 
across the disc at different galactocentric distances allowing 
the exploration of recent star formation activity in a wide range 
of physical conditions. Hundreds of nebulae have been charted 
in M 33 \citep{HoSkAs02} and included in abundance and 
gradient studies 
\citep{vilchezetal88,crockettetal06,rosolowskysimon08,bresolinea10}.
Table \ref{tbl1m33stats} summarizes the main properties 
of  the galaxy. Values are obtained from NED\footnote{This 
research has made use of the NASA/IPAC Extragalactic 
Database (NED) which is operated by the Jet Propulsion 
Laboratory, California Institute of Technology, under 
contract with the National Aeronautics and Space 
Administration (NASA).}  wherever no references are
given.

\begin{table}
\centering
\caption{M 33 properties.  }
\label{tbl1m33stats}
\begin{tabular}{ll}
\hline
Designations & Messier 033, Triangulum Galaxy, \\
\ & NGC 0598,UGC 01117, PGC 005818 \\
Classification & SA(s)cd        \HII  \\
Major Diameter & 70.8 Arcmin \\
Minor diameter &  41.7 arcmin\\
Position Angle & 23 $\deg$ \\
Distance &  840 kpc \citep{freedmanetal01}\\
Redshift & -0.000597 $\pm$ 0.000010 \\
Inclination angle & 53.52 $\deg$  \citep{corbellisalucci00}\\
PPak scale  & 4.07 pc/arcsec (10.91 pc/fibre ) \\

\hline
\multicolumn{2}{l}{{\footnotesize Source is NED
 unless otherwise specified.}}\\
\end{tabular}
\end{table}

As a first step of a larger project aimed to trace the chemical 
abundances, gradient and dispersion of physical conditions with
high spatial resolution across M 33, we obtained  IFS data of the 
centre and the external \HII\ region IC~132, located to the NW of 
the centre of M 33 at a galactocentric distance of $\sim$ 19$^{\prime}$  
or 4.69 kpc (see Fig.~\ref{f33regsidgal}).

The observed locations represent two scenarios of star 
formation in spiral galaxies with extreme different environments: 
outer spiral disc and central regions. Notorious differences between the 
two (disc vs central) exist regarding the star formation rate 
(20 vs 1000 M$_\odot$ year$^{-1}$),  star formation timescales 
(1Myr vs 1 Gyr), electron density (10 vs 10$^4$ cm$^{-3}$) 
among others \citep{kennicuttetal89,kennicutt98}.

\subsubsection{The central region of M 33}

The central region of M 33 has been the subject of various 
studies given its notorious star formation activity  
\citep{keel83}. From the fact that the dominant stellar 
population in the centre of M 33 is young ($<$ 1 Gyr) 
and metal rich (up to solar metallicities), \cite{vandenbergh91} 
suggested that the stars are the product of intense 
star forming periods originated by inflow of gas.

Star formation in the centre of spiral galaxies has been 
associated with the existence of bulges and bars. However 
the presence of a bulge in M 33 is controversial. The excess 
in surface brightness in the central region of M 33 
\citep{kent87} is not necessarily due to the existence of a 
bulge but may be explained by a large scale diffuse halo 
\citep[based on 12 $\mu$m photometry]{bothun92}. By 
using H-band photometry of individual stars \cite{minnitietal93} 
concluded that a bulge exists and that it has experienced 
a period of star formation less than 1 Gyr ago. The existence 
of a bar in M 33 is also an open issue, \cite{reganvogel94} 
propose that if a bar exists it has a length of 3$^{\prime}$ 
with a position angle of 90$^{\circ}$. They propose also that 
the bar-like emission could be due to the continuation of the spiral 
arms. However they claim that there is a lack of  evidence 
in order to decide in favour of one or the other explanation. 
In any case, in M 33 the bar-like structure contributes  only 
4\% to the total light out to 1$^{\prime}$.5, in contrast with 
a 40\% contribution in strong barred galaxies \citep{blackman83}.
Also the existence of a massive central black hole has been 
discarded \citep{kormmclure93,masseyetal96} and no 
AGN-like emission line ratios have been reported. However 
the centre of M 33 harbours M 33 X-8, the brightest 
ultraluminous X-ray source (ULX) in the local group. The 
observational evidence indicates that M 33 X-8 emission  
originates in a stellar-mass black hole 
(M$_{BH} \sim$ 10M$\odot$) with super-Eddington accretion 
rate \citep{foschiniea04,wengEa09,middletonEa11}. M 33 
X-8 position which coincides with the photometric centre 
of the galaxy, is indicated with a cross symbol in the left 
panel of Fig.~\ref{f33regsid}.

The central region is an excellent scenario to study the
star formation process in relatively high metallicity and 
density and low excitation environments, offering a 
challenging scenario deviating from the conditions of the 
majority of \HII\ region studies. Such conditions represent 
a test to the different abundance calibrators, generally 
based on low metallicity \HII\ region samples 
\citep{kinkelrosa94,bresolin04} and extrapolated to 
the high metallicity regime.

\subsubsection{The outer disc region IC 132}
IC~132 is one of the most external \HII\ regions in M 33, 
located 4.69 kpc ($\sim {\rm 19}^{\prime}$) away from the 
centre. It ranks among the brightest in the galaxy and has 
been included in chemical composition and gradient 
studies \citep{aller42,searle71,smith75,kwitteraller81,
magrinietal07}.  Wolf-Rayet features were detected from 
narrow band interference filter surveys 
\citep{boksenbergetal77,dodorbenv83}. Metallicity has
been reported ranging from 12+log(O/H)=7.85 to 
12+log(O/H)=8.08 with typical errors $\sim$ 0.05 
\citep{magrinietal07b,maginietal10}.

IC~132 shows the high excitation, low density and
low abundance environment typical of disc regions, 
for which most of the \HII\ studies, physical conditions
and abundance estimators have been developed.  

\subsection{Observations}

Observations were obtained with the CAHA 3.5m telescope, 
using the PMAS instrument in fibre Package (PPak) mode 
\citep{rothetal05}. With this configuration the IFU features 
331 fibres  for the science object packed in an hexagonal array 
with a  FOV of 74$^{\prime\prime}\times$64$^{\prime\prime}$, 
surrounded by 6 bundles of 6 fibres  each at a distance of 
72$^{\prime\prime}$ from the centre to sample the sky. 15 extra 
fibres, not on the focal plane, can be illuminated separately for 
calibration purposes. Although the unit contains more fibres  to 
provide protection against mechanical stress, the total number 
of active fibres  adds to 382. Each fibre has a projected diameter 
of 2$^{\prime\prime}$.68 in the sky or 10.91 pc at the assumed 
distance of M 33. The filling factor of the science packet is 60\%. 
The detector used was a thinned, blue enhanced, 2Kx4K CCD 
(SITe ST002A) with 15$\mu$m pixels \citep{kelzetal06}.

\begin{figure}
\centering
\includegraphics[width=12cm]{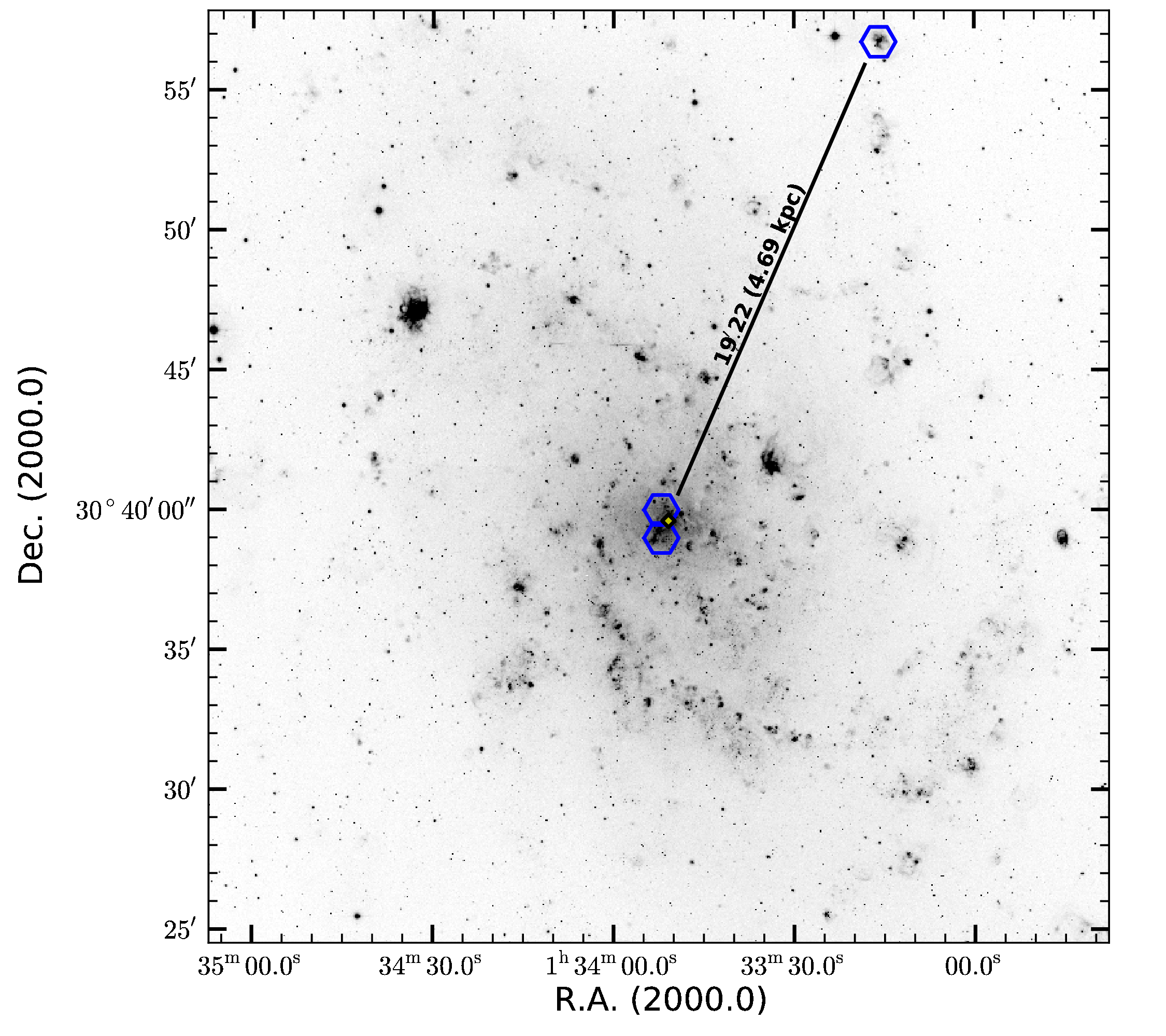}
\caption{Position of the observed fields over M 33. Hexagons 
are 74$^{\prime\prime}$ in diameter. The diamond marks the 
centre of the galaxy with coordinates 
RA=01$^h$33$^m$50$^s$.89, 
DEC=+30$^o$39$^\prime$36.$^{\prime\prime}$80 (J2000). 
The outer region IC 132 has coordinates 
RA=01$^h$33$^m$15$^s$.90, 
DEC=+30$^o$56$^\prime$44$^{\prime\prime}$.00 (J2000). 
North is  up and east is left.}
\label{f33regsidgal}
\end{figure}

\begin{figure}
\begin{tabular}{cc}
\includegraphics[width=8cm]{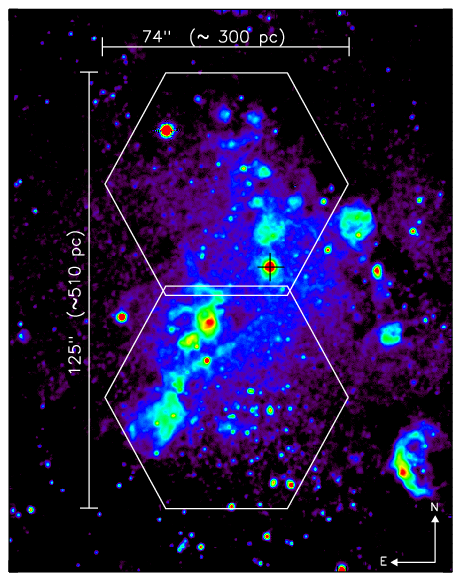} &
\includegraphics[width=8cm]{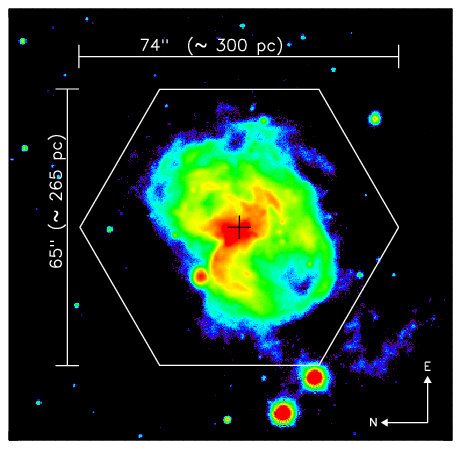} \\
\end{tabular}
\caption{Identification of the observed fields. Images are 
narrow filters centred in H$\alpha$ from the NOAO Science 
Archive \protect\citep{masseyetal06}. Overlaid is the 
PMAS-PPak hexagonal FOV. For the centre of M 33 (left) two 
consecutive fields are overlapped by one fibre line. The cross in the 
central regions indicates the centre of the galaxy coincident 
with the ultraluminous X-ray source M 33 X-8; for IC 132 (right) 
the cross marks the centre of the region. The scale in parsecs 
corresponds to the assumed  distance to M 33 of 840 kpc.}
\label{f33regsid}
\end{figure}

The centre of the galaxy and the external region IC~132 were 
observed between September 7th and 9th  2007.  The position 
of the fields over an H$\alpha$ image \citep{chengetal97} are 
in Fig.~\ref{f33regsidgal}. Detailed identification is in 
Fig.~\ref{f33regsid}, the instrument FOV is overplotted on 
NOAO Science Archive images \citep{masseyetal06} taken 
with a 50 \AA~FWHM filter centred on H$\alpha$. 
The fields for the centre were overlaid  by one fibre width.

Observations were performed at two different grating angles 
(GROT) with the 300 grooves mm$^{-1}$ (V300) grating. The 
value GROT=-75 was selected to cover from 3591 to 6996 \AA~
(optical) with a dispersion of 3.4 \AA~pixel$^{-1}$, and  
GROT=-72 to cover from 6873 to 10186 \AA~(near infrared) 
and a dispersion of 3.2 \AA~ pixel$^{-1}$.  In addition high 
spectral resolution spectra were obtained using the V1200 
grating at GROT=-55.5, covering from 6100 to 6650  
\AA\ with a dispersion of 0.64 \AA~ pixel$^{-1}$.  In all cases 
the CCD was sampled in a 2x2 binning mode and the  values quoted above
are the effective ones after the resampling. A journal of 
observations is shown in table \ref{t33osblog}. The  stars 
BD+17 4708 and  BD+28 4211 were observed for flux 
calibration in the optical and NIR ranges respectively. 
Various continuum illuminated and HeHgCs+ThAr lamp 
frames were acquired, interleaved and as close as possible 
to the science observations to account for flexures in the 
telescope. Sky flats were acquired at the beginning of the night 
and biases at the end. For each object 3 dithered pointings were 
taken in order to fully cover the instrument FOV. The positions 
are D1=(0,0), D2=(1.56,0.78) and D3=(1.56,-0.78), where the 
displacements are in arc seconds relative to the initial position 
D1.  The average seeing was 1.0$^{\prime\prime}$; higher values as
well as the presence of clouds are indicated in the last column of 
table  \ref{t33osblog}. Half of the third night was lost due to 
weather which limited severely the observing time and neither 
bias nor sky frames were taken. In this case the bias was 
determined from the overscan region of each frame.

\begin{table}
\centering
\caption{Journal of observations. }
\label{t33osblog}
\begin{tabular}{cccccc}
\hline
Object & Spectral range & Dispersion              & Exposure  Time   &   Median  &   \  \\
\            & [\AA]                    &  [\AA~pix$^{-1}$] &  [s]                         &  Airmass & Notes\\

\hline
centre P1 D1 &  3591-6996 & 3.4 & 500 $\times$ 3 &  \         & seeing $\simeq$1\arcseconds.46 \\
centre P1 D2 &  3591-6996 & 3.4 & 500 $\times$ 3 &  1.49  & seeing $\simeq$1\arcseconds.43 \\
centre P1 D3 &  3591-6996 & 3.4 & 500 $\times$ 3 &  \         &  \  \\
centre P2 D1 &  3591-6996 & 3.4 & 500 $\times$ 3 &  \         &  \   \\
centre P2 D2 &  3591-6996 & 3.4 & 500 $\times$ 3 &  1.01  &  \   \\
centre P2 D3 &  3591-6996 & 3.4 & 500 $\times$ 3 &  \         &  \    \\
IC~132 D1     &  3591-6996 & 3.4 & 500 $\times$ 3 &  \         &  \    \\
IC~132 D2     &  3591-6996 & 3.4 & 500 $\times$ 3 &  1.02  &  \    \\
IC~132 D3     &  3591-6996 & 3.4 & 500 $\times$ 3 &  \         &  \    \\
\hline
centre P1 D1 &  6873-10186 & 3.2 & 500 + 700 $\times$ 2 &       &  seeing $\simeq$1\arcseconds.33\\
centre P1 D2 &  6873-10186 & 3.2 & 700 $\times$ 3             &   1.49    & \ \\
centre P1 D3 &  6873-10186 & 3.2 & 700 $\times$ 3             &   \    & \ \\
centre P2 D1 &  6873-10186 & 3.2 & 700 $\times$ 3             &    \   & \ \\
centre P2 D2 &  6873-10186 & 3.2 & 700 $\times$ 3             &   1.04    & clouded, seeing $\simeq$1\arcseconds.48 \\
centre P2 D3 &  6873-10186 & 3.2 & 700 $\times$ 3             &    \   &  \ \\
IC~132 D1 &  6873-10186 & 3.2 & 500 $\times$ 3                 &   \    & clouded \\
IC~132 D2 &  6873-10186 & 3.2 & 500 $\times$ 3                 &   1.07    & clouded \\
IC~132 D3 &  6873-10186 & 3.2 & 500 $\times$ 3                 &    \   & clouded \\
\hline
centre P1 D1 &  6100-6650  & 0.64 & 500 $\times$ 3            &    \  & \ \\
centre P1 D2 &  6100-6650  & 0.64 & 500 $\times$ 3            &   1.19   & \ \\
centre P1 D3 &  6100-6650  & 0.64 & 500 $\times$ 3            &    \  &  guiding stopped momentarily   \\

centre P2 D1 &  6100-6650  & 0.64 & 500 $\times$ 3            &   \   & \ \\
centre P2 D2 &  6100-6650  & 0.64 & 500 $\times$ 3            &   1.01    & \   \\
centre P2 D3 &  6100-6650  & 0.64 & 500 $\times$ 3            &    \   & \  \\

IC~132 D1 &  6100-6650  & 0.64 & 500 $\times$ 3                &    \    &  clouded, seeing 1\arcseconds.50 \\
IC~132 D2 &  6100-6650  & 0.64 & 500 $\times$ 3                &   1.04     & clouded,  seeing 1\arcseconds.40 \\
IC~132 D3 &  6100-6650  & 0.64 & 500                                      &    \     & clouded, seeing 1\arcseconds.60 \\
\hline
\multicolumn{6}{l}{{The median seeing was 1\arcseconds.0, those pointings with
seeing above 1\arcseconds.3 are indicated as well as other non photometric 
conditions. }}\\
\multicolumn{6}{l}{{The median airmass is given for each 3 dithering positions. 
For the third night no bias or sky flat frames were taken.}} \\
\end{tabular}
\end{table}

\newpage
\
\newpage
\subsection{Data reduction}

The data reduction  is based  on the set of routines 
and tools  grouped in the  E3D \citep{sanchez04,sanchezetal04}  
and R3D \citep{sscard05} packages. These were complemented 
with visualization and analysis tools developed by us, together 
with standard IRAF\footnote{IRAF is distributed by the National 
Optical Astronomy Observatory, which is operated by the 
Association of Universities for Research in Astronomy (AURA) 
under cooperative agreement with the National Science Foundation.} 
utilities. 

The pipeline includes spectra tracing, extraction, fiber flat correction, 
flux calibration, sky subtraction and datacube building. 
Once the 3D datacube was assembled, it was additionally corrected
for differential atmospheric refraction, atmospheric correction for 
the [\SIII] near infrared lines and extinction. Details of the reduction 
can be found in \cite{gbenitoetal10} and in the Complementary 
Online Material.

\begin{figure}
\centering
\includegraphics[width=16.0cm]{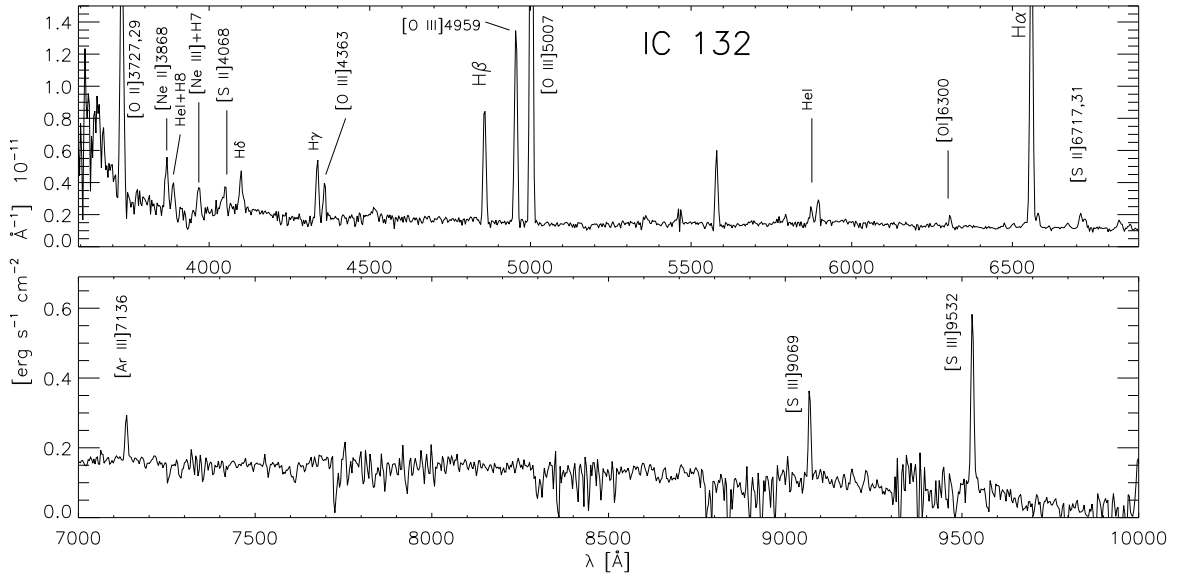}
\caption{Integrated visual and near-IR spectra for IC 132. 
Relevant emission lines are identified. See text regarding 
special considerations for [\OIII]$\lambda$4363\AA\ and 
Garc\'\i a-Benito et al. 2010 for the  treatment of the near 
IR [\SIII] lines.}
 \label{fkspsic132}
\end{figure}

The final integrated spectra for the IC 132 region is in figure 
\ref{fkspsic132} with relevant lines identified in the optical 
and NIR spectral ranges.

\subsection{Emission Line Maps}
\label{sec_emission_line_maps}
Emission lines were fitted with a gaussian,  and continuum 
bands at both sides of the emission were measured additionally 
to estimate the error in the continuum placement.
Given the low resolution of the data (FWHM effective resolution 
R$_{FWHM}$ $\simeq$ 700 at $\lambda$ = 5500 \AA\ and 
R$_{FWHM}$ $\simeq$ 1100 at $\lambda$ = 8500 \AA) 
for the first and second nights, some of the lines had to be 
deblended by fitting multiple gaussians. This was the 
case for the triple blend of H$\alpha$ and the two [\NII] 
lines at $\lambda\lambda$ 6548,6584 \AA\ and the  [\SII] 
$\lambda\lambda$ 6717,6731 \AA\ doublet. In each case 
a single gaussian is assigned to each line and the best 
solution is determined using a non-linear least-squares 
fit. For the third night, the higher resolution (R$_{FWHM}$ 
$\simeq$ 3100 at $\lambda$ = 6500 \AA) allows the separation of the 
H$\alpha$-[\NII]$\lambda\lambda$ 6548,6584~\AA\ system
as well as the measurement of the faint [\SIII]$\lambda$6312
line, at least for the brightest spaxels of IC~132. 

\begin{landscape}
\begin{figure}
\begin{tabular}{ccc}
\includegraphics[width=6.0cm]{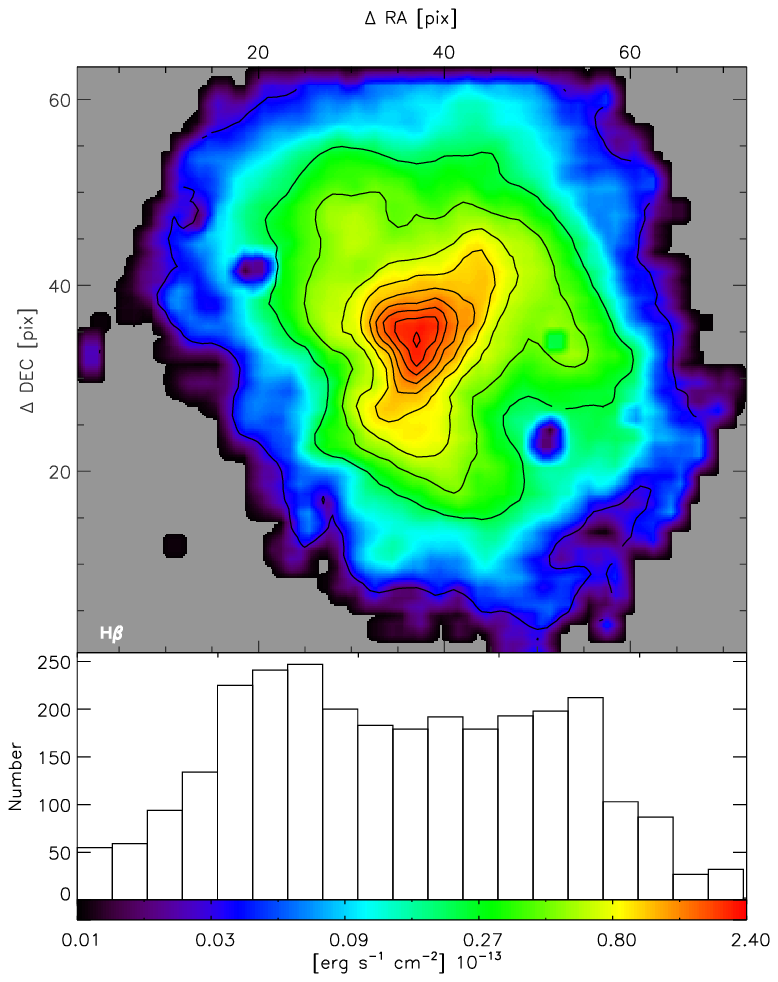} &
\includegraphics[width=6.0cm]{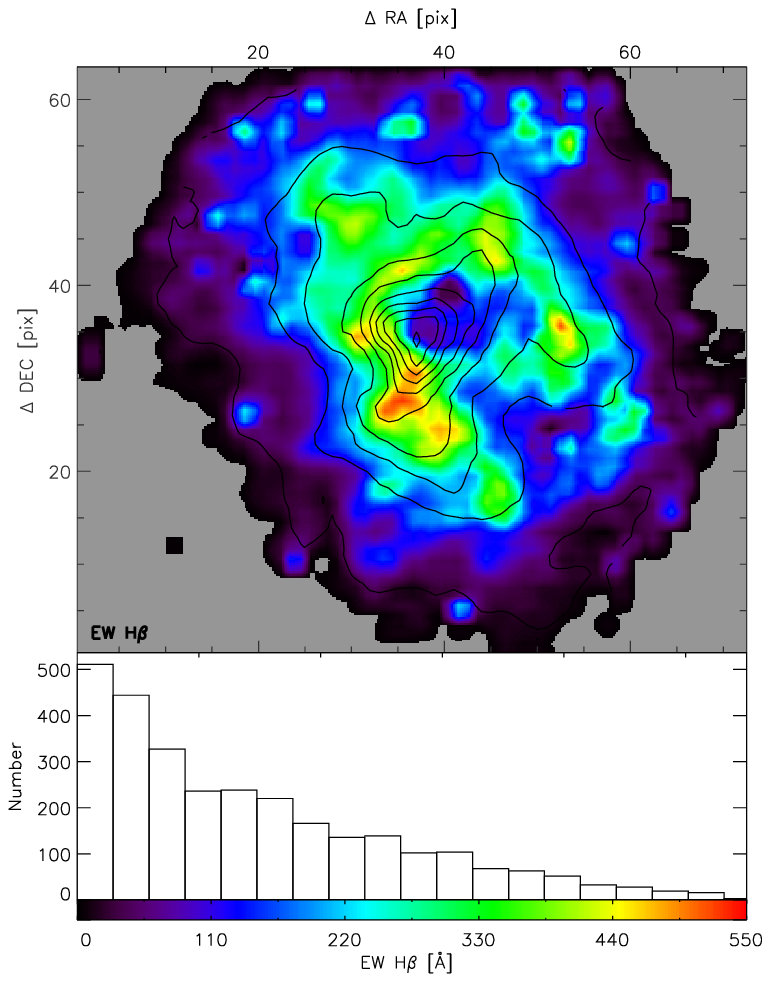}
\includegraphics[width=6.0cm]{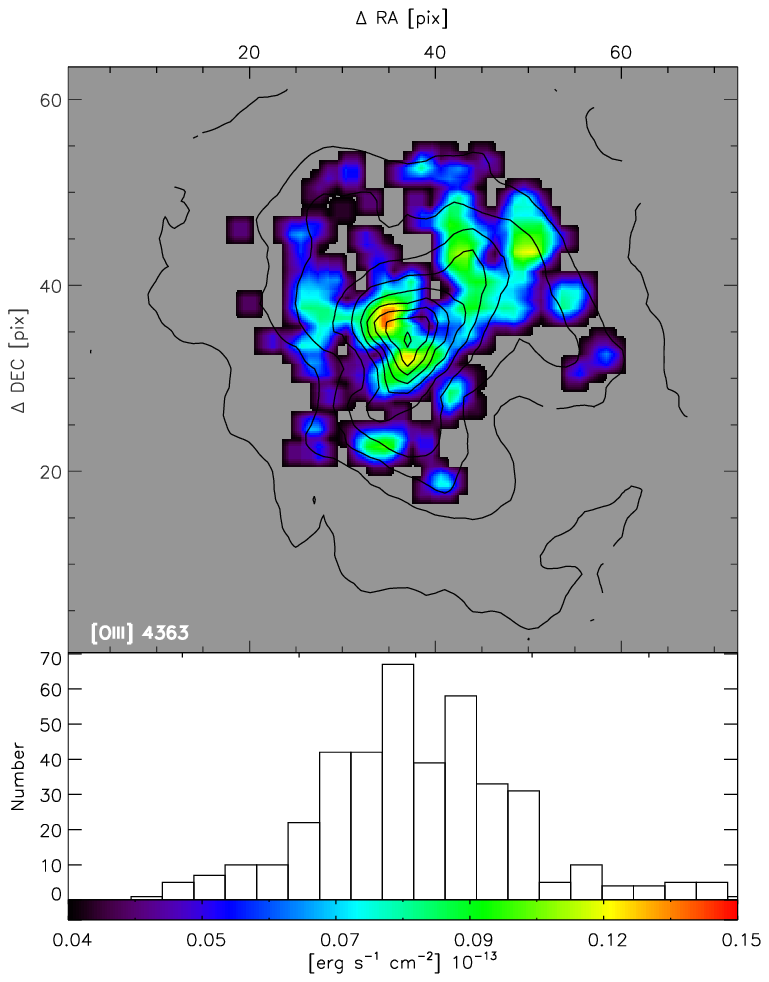} \\
\end{tabular}
\caption{Example of maps for IC 132: \Hb, \Hb\ equivalent width [EW(\Hb)] and [\OIII]
4363 with S/N $>$ 2. The colour table has a logarithmic scale to increase the dynamical
range. In this and all the other maps, the isocontours are from the \Ha\ map.}
\label{ficel}
\end{figure}
\end{landscape}

\begin{landscape}
\begin{figure}
\begin{tabular}{ccc}
\includegraphics[width=6.0cm]{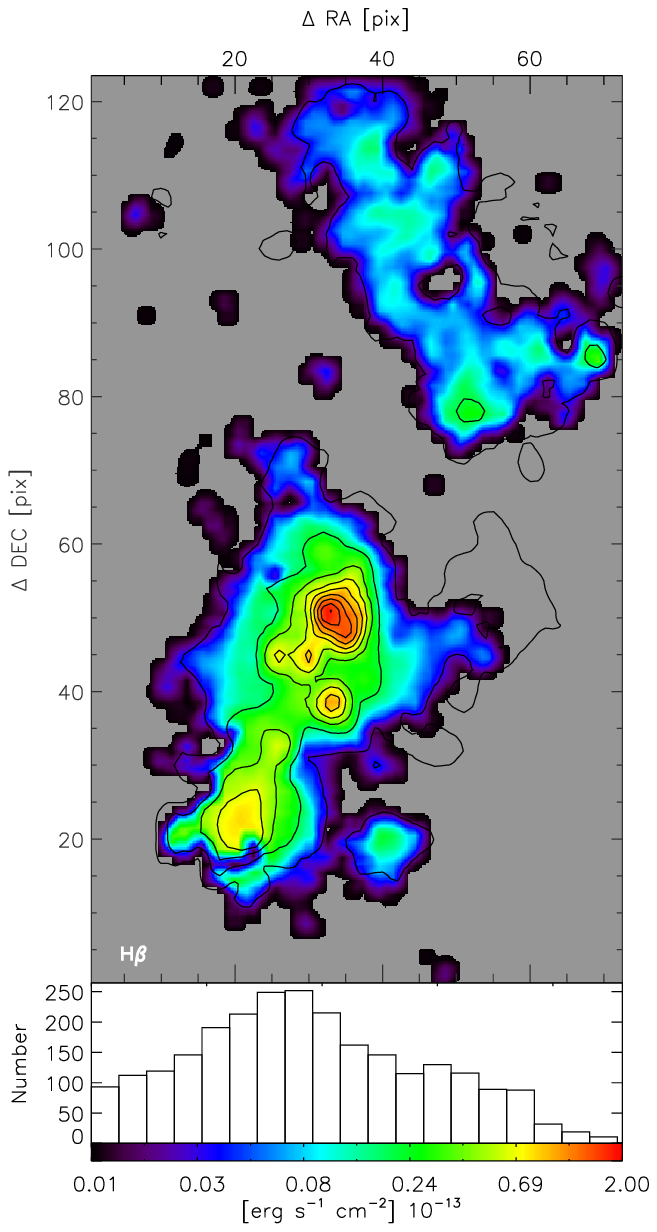} &
\includegraphics[width=6.0cm]{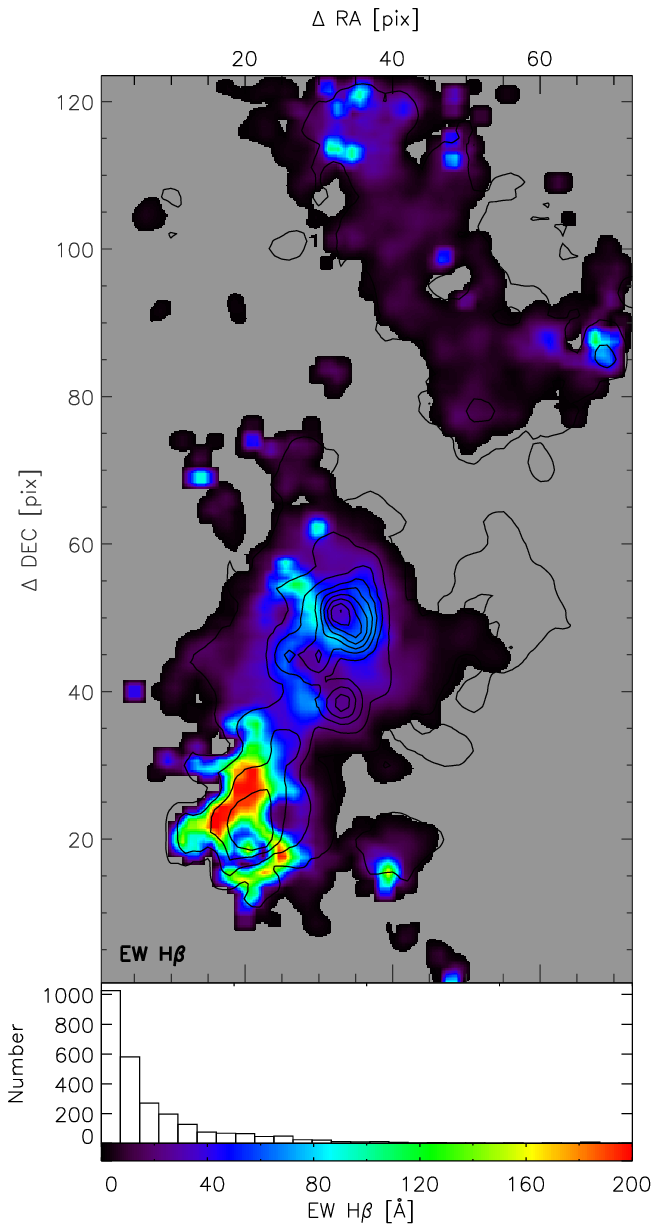} &
\includegraphics[width=6.0cm]{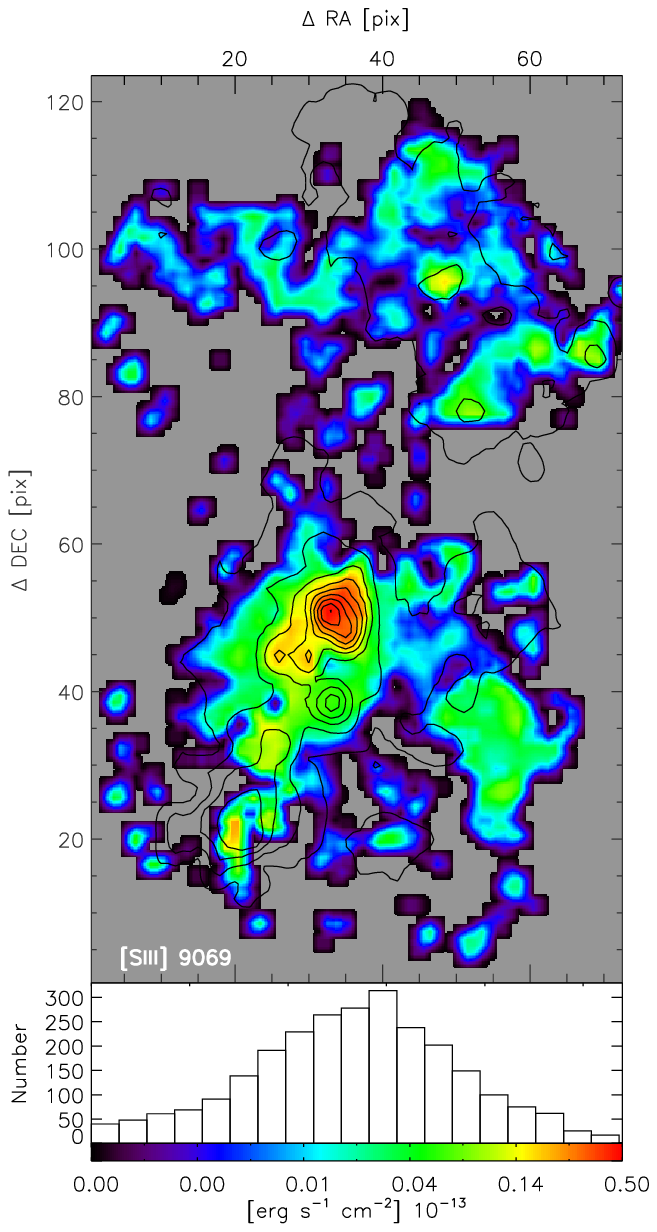} \\
\end{tabular}
\caption{Example of maps for the central zone: \Hb, EW(\Hb) and 
[\SIII] $\lambda$9069 with S/N $>$ 2. The colour table has a logarithmic scale 
to increase the dynamical range. }
\label{fkerel}
\end{figure}
\end{landscape}

Examples of 2D maps are shown in figure \ref{ficel} for IC 132: 
I(\Hb),  \Hb\ equivalent width [EW(\Hb)]  and 
I([\OIII]$\lambda$ 4363), and in figure \ref{fkerel} for the central 
zone: I(\Hb), EW(\Hb) and I([\SIII]9069). The line maps have S/N $>$ 2.0.


\section{Physical conditions of the gas}
Physical parameters describing the ionized gas were derived 
using the line fluxes measured in   \ref{sec_emission_line_maps}. 
To obtain these parameters we used the TEMDEN 
routine included in the STSDAS nebular package for IRAF. 
This task solves the statistical equilibrium equation for 
the  5-level atom approximation \citep{shawdufour95}. The 
transition probabilities have been updated with recent values 
computed for the [\OII], [\OIII], [\NII], [\SII] and [\SIII] ions 
by \cite{fischer04,fischer06}. The collision strength values 
were taken form \cite{tayal07} for [OII], \cite{aggkeen99} 
for [OIII], \cite{hudsonbell05} for [NII], \cite{ramsbottom96} 
for [\SII] and \cite{tayalgupta99} for [\SIII]. 

Computations were made for each individual spaxel where 
the required lines are detected with enough S/N and also 
for the shells analysis described in section 
\S \ref{sec_integrated_properties}.  The S/N $>$ 5 requirement
was met by the intense lines ([\OIII]$\lambda$5007, \Ha, \Hb) 
for all the spaxels, while for the  auroral lines [\OIII]$\lambda$4363  and 
[\SIII]$\lambda$6312, S/N $>$ 2 was reached only in the central 
part of IC 132.

The electron density (\Ne) was estimated from the 
[\SII]$\lambda$6717/[\SII]$\lambda$6731 line 
ratio, and for the electron temperature (\Te) we used
the faint auroral [\OIII]$\lambda$4363 and [\SIII]$\lambda$6312
lines, allowing the use of the 
[\OIII]$\lambda\lambda$4959,5007/[\OIII]$\lambda$4363 
and [\SIII]$\lambda\lambda$9069,9532/[\SIII]$\lambda$6312 
ratios to compute \Te([\OIII]) and \Te(\SIII) respectively. Due o strong atmospheric absorptions around 9300~\AA , we defined the [SIII] 9532 line by its theoretical ratio to the 9069~\AA\ line.

\subsection{Electron densities}
\label{electrodensity}
\begin{figure}
\centering
\begin{tabular}{cc}
\includegraphics[width=6cm]{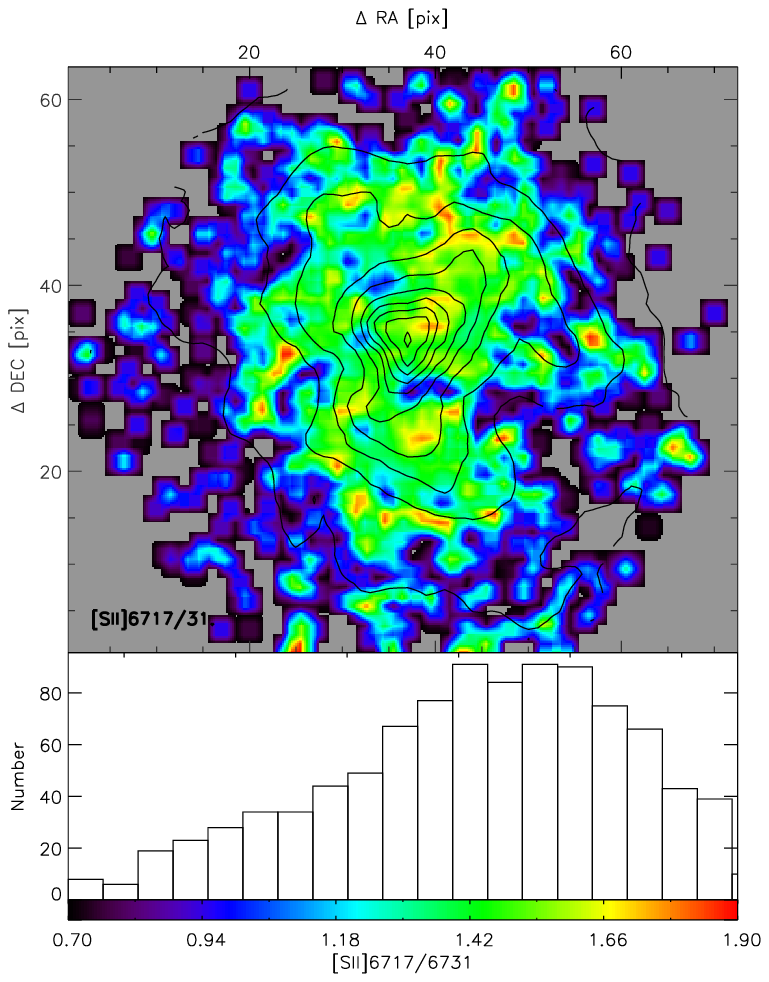} &
\includegraphics[width=6cm]{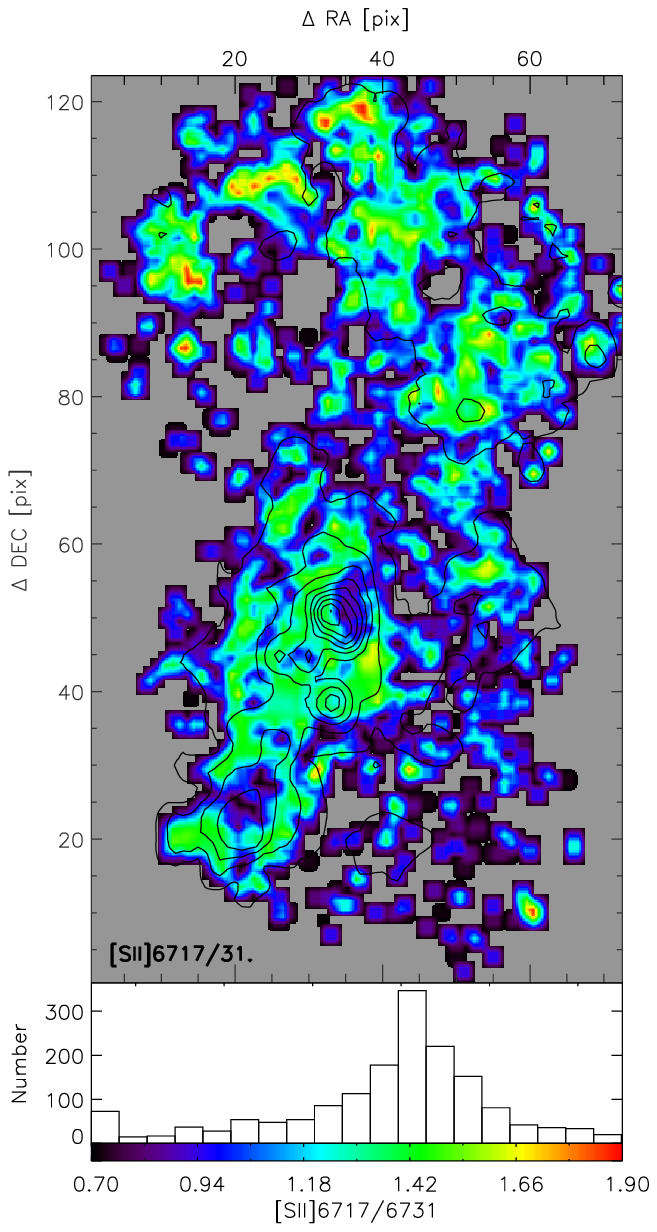} \\
\end{tabular}
\caption{[SII]$\lambda$6717/[SII]$\lambda$6731 maps for  IC~132 
(left) and for the central zone of M33  (right). 
Notice the different vertical scale in the histograms.}
\label{f33siiratio}
\end{figure}

To estimate \Ne\ we used the ratio  RS2 $\equiv$ [SII]$\lambda$6717/[SII]$\lambda$ 6731
and the TEMDEN routine in IRAF \citep{shawdufour95}.
Fig.~\ref{f33siiratio} shows the RS2 
ratio for  the external region IC 132 and for the centre of M33.
For IC 132 the RS2 distribution is quite uniform with a mean 
value around 1.45 and although the RS2 ratio shows variations 
in the brightest parts, the conversion to \Ne\ would translate 
into variations within the low density regime,  which is of no 
significant impact in further parameter estimations.

The central region of M33 shows a broad distribution with 
the brightest  H$\alpha$ region having RS2 values around 
1.4 indicating \Ne\ values also in the low density regime, while 
the northern much fainter  H$\alpha$ extension shows 
some spaxels with unrealistically large RS2 ratios. 
From the RS2 ratio it seems that a structure exists and in
such case the hypothesis of a uniform value across the nebula 
would be  too simplistic. 

From the \Ne\ maps it is obvious that many pixels show an
RS2 ratio above the theoretical value for low density limit and that it decreases down
to the lower ratio limit, indicating that the region spans the whole
range of  \Ne\ from very high to very low densities.  There is positively some 
uncertainty from deblending the [\SII] lines, and this is not 
exclusive of our data. Previous RS2 determinations from IFS 
data also show ratios with significant excursions above the 
theoretical limit, from data obtained with the same instrument
\citep{relanoetl10,miberoetal11} or other integral field units 
(IFU) in a different telescope  \citep{lagosetal09}, even with 
the [\SII] lines resolved.

However the problem does not seem to be exclusive of IFS data.  From 
a literature compilation and their own observations, all obtained
with long slit spectroscopy, Kennicutt et al.~ (1989, figure 5)
reports 
an RS2 ratio peaked at 1.3 and with a significant number of 
\HII\ regions, mainly in spiral disks, above the 1.4 theoretical 
limit. Also from long slit spectra, Zaritsky et al.~(1994, figure 4)
report a similar behaviour for RS2.

When RS2 $\geq$ 1.4 is obtained, it may be assumed that \Ne\ is 
below 10 cm$^{-3}$ however, the assignment of a true density is 
very uncertain (it begins to be uncertain as RS2 reaches the asymptotic
limit even before 1.4). Although the safe way to proceed in such 
cases is to assume \Ne\ =100 cm$^{-3}$ - in any case for temperature
and abundance determinations the density plays a second order 
role - it seems that the theoretical ground for the determination 
of \Ne\ may need an adjustment to the atomic data and further 
investigation in the topic is required which is beyond the scope of this work. 

This problem, that seems to be affecting both ends of the theoretical 
calibration, could be traced to uncertainties in the determination of the 
atomic data used, specifically the collision strengths for the low-end and the 
transition probabilities for the high-end, not only for [\SII] but for density 
indicators based on different line ratios 
 \citep[Gary Ferland, private communication]{mclaughlin98,coppetiwritzl02,wangetal04}.
We concur with Wang and collaborators, that a full independent study with ad-hoc 
data to validate previous results will be crucial.

\subsection{Electron temperatures}
We have determined the \Te([\OIII]) and \Te([\SIII]) electron 
temperatures using the direct method based on the line ratios  
[\OIII]$\lambda\lambda$4959,5007/$\lambda$4363 and 
[\SIII]$\lambda\lambda$9069,9532/$\lambda$6312 
respectively.

Only in the brightest parts of the core of IC~132 it was possible 
to derive reliable \Te([\OIII]) and \Te([\SIII]). Unfortunately the 
sky at Calar Alto suffers from light pollution produced by street 
lamps, and among the several emissions in the visible range is 
the Hg I 4358.3 line, which is strong enough \citep{sanchezetal07} 
to limit the use of the [\OIII]$\lambda$4363.2 line in the 
temperature determination by the direct method in low redshift
systems. This is  
particularly bad for M33 because of its  blue-shifted spectrum 
(z=-0.000597)  that brings the [\OIII] line to $\lambda$=4360.61, 
close enough for it to blend with the HgI line. 

\begin{landscape}
\begin{figure}
\begin{tabular}{ccc}
\includegraphics[width=6.0cm]{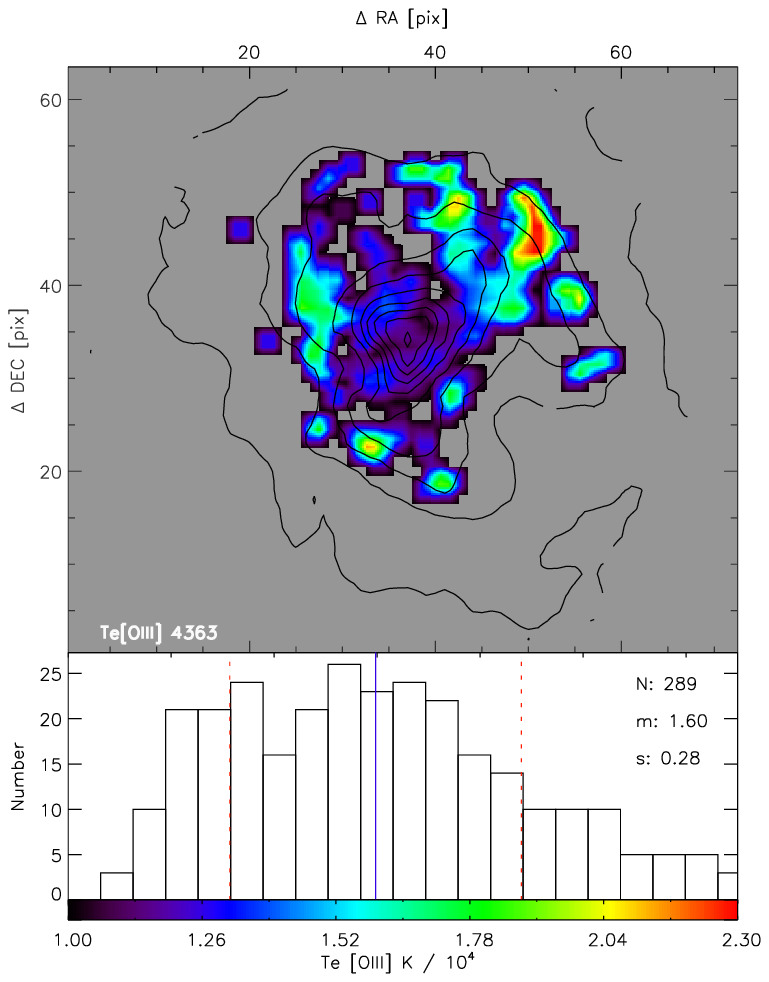}&
\includegraphics[width=6.0cm]{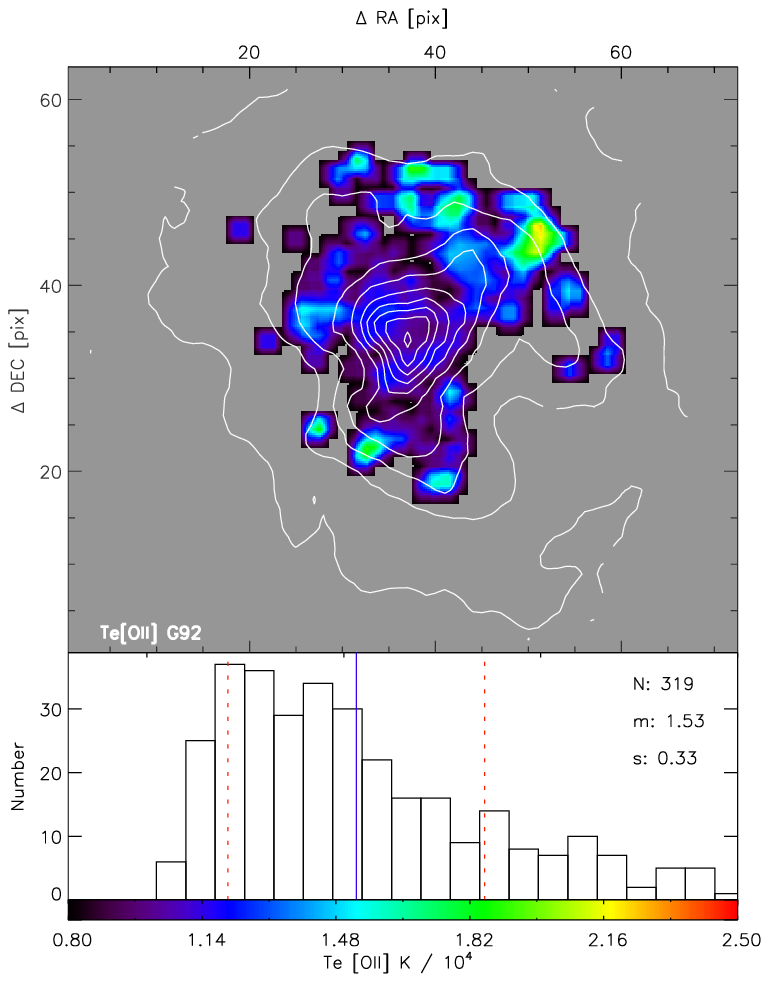} &
\includegraphics[width=6.0cm]{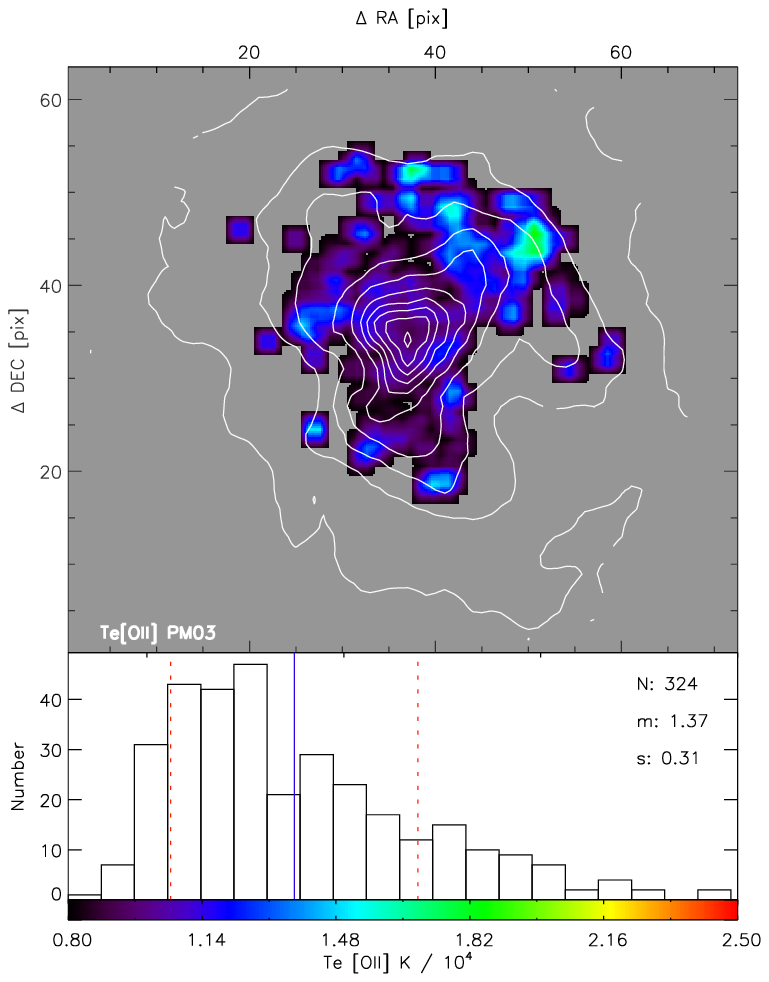} \\

\includegraphics[width=6.0cm]{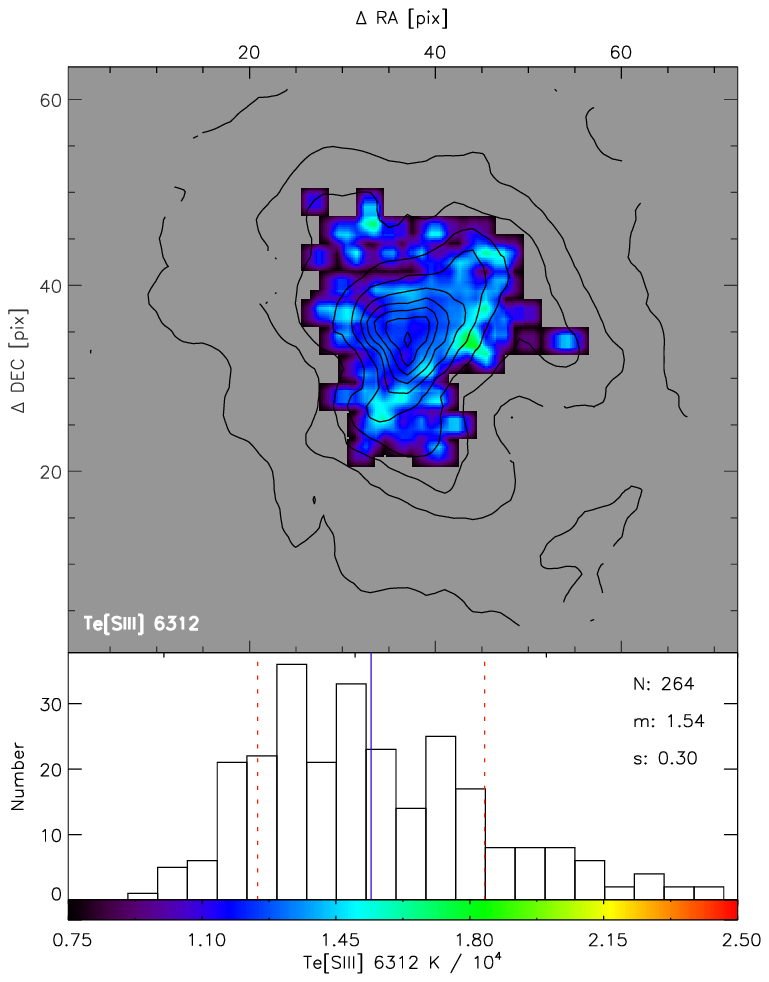} &
\includegraphics[width=6.0cm]{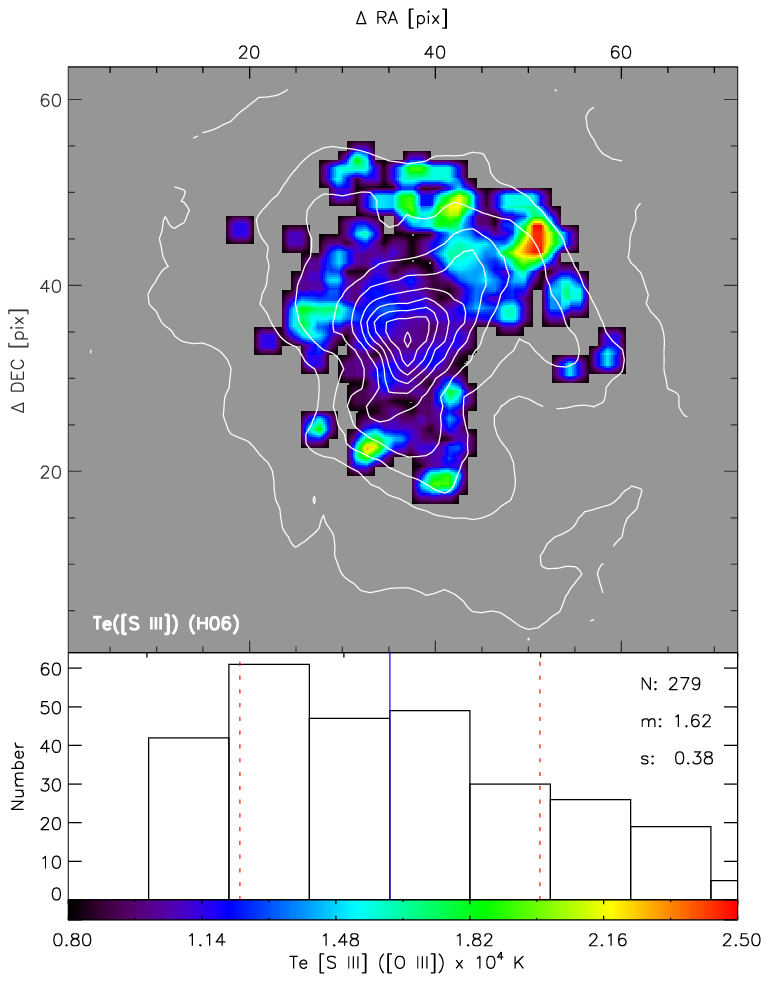} &
\includegraphics[width=6.0cm]{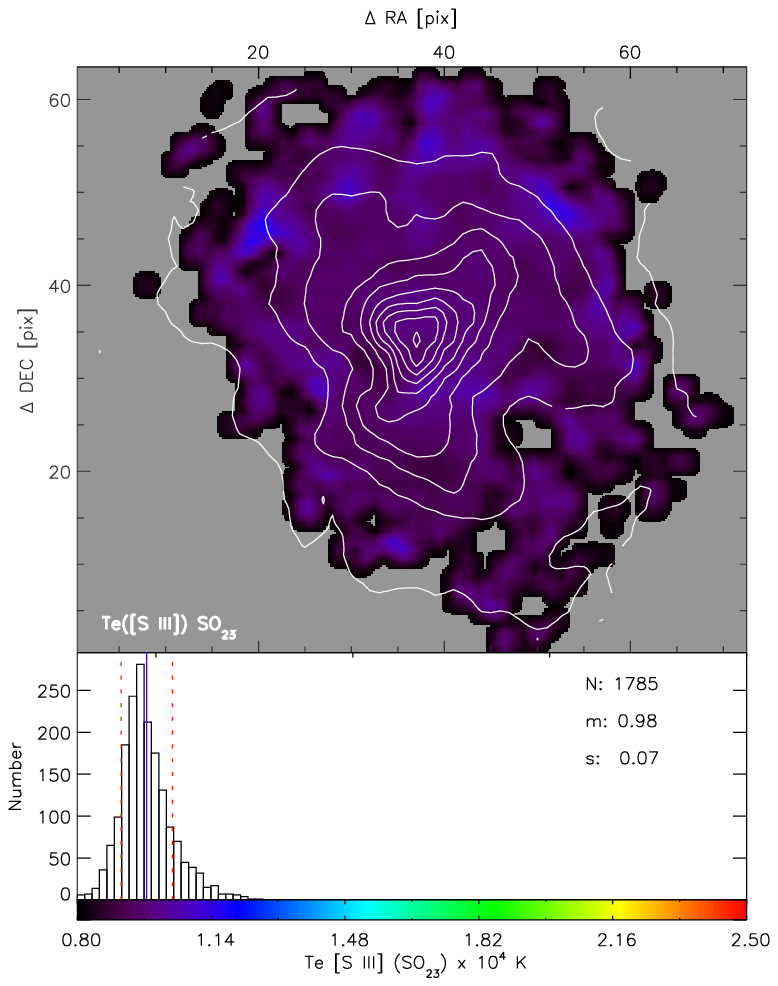} \\
\end{tabular}
\caption{Top left to right: \Te ([O III]) upper limit for IC~132.
\Te([\OII]) estimated from the relations provided by \protect\cite{garnett92} 
and \protect\cite{permodiaz03}.  
Bottom left to right: \Te([\SIII]) obtained from the direct method, using 
\protect\cite{hageleetal06} and \protect\cite{diazetal07}.
Here and in following figures, the histogram of the distribution also 
indicates the total number of pixels used (N), the mean (m and solid 
blue line) and the standard deviation (s and red dashed lines). }
\label{ficTOTS}
\end{figure}
\end{landscape}

Nevertheless, by performing a careful sky subtraction, we were able to 
obtain upper limits for the [\OIII]4363 line. This upper limit was used
to compute an upper limit of \Te([\OIII]) and to estimate the
Ionization Correction Factor (ICF) for S.  \Te([\OIII]) was 
computed using the ratio R$_{o3}$ = ([\OIII]4959+[\OIII]5007)/[\OIII]4363
and the result is shown in figure \ref{ficTOTS} top left panel.

For  \Te ([\OII]) we face the problem that while the [\OII]$\lambda\lambda$ 3727,3729\AA\
doublet was detected,  the [\OII]$\lambda$7320 and 
$\lambda$7330 lines were not. Thus we have to relay on results 
from fits to photoionization models and their comparison with 
data with direct \Te ([\OII]) and \Te ([\OIII]) 
determinations.
\cite{diazetal07} assume T([\OII]) $\simeq$ T([\SIII]), while
Izotov et al. 2006, using \cite{stasinska90} models, suggest 
taking \Te([\OII]) = \Te([\OIII]) if no other \Te([\OII]) determination 
is possible, however it is pointed out that in such case, the 
error based in photoionization models with Z=Z$_\odot$ is 
equivalent to underestimating [OII] 
$\lambda$$\lambda$3727,29 by about 40-50\%. In some cases 
\Te ([\OII]) = \Te ([\NII]), however to estimate \Te ([\NII]), the
weak [\NII] $\lambda$5755 line is required and in
this case also an underestimation of [\OII] is obtained 
\citep{hageleetal08}. In any case, [\NII] $\lambda$5755 is not detected
in our observations.

It seems that no strong correlation exists between the two 
temperatures in \HII\ regions or in \HII\ galaxies and that 
the effect of metallicity, density and maybe even other 
factors should be taken into account.
\cite{kenbregar03} find only ``a hint of correlation", using 
their observations and data from the literature. Nevertheless, 
they use the \cite{garnett92} expression to estimate T([\OII]). 
\cite{hageleetal06,hageleetal08} also show that the correlation 
between  \Te([\OII]) and \Te([\OIII]) is very weak and favour the 
use of the density dependent \cite{permodiaz03} relation. 

Here we apply the widely used relation from \cite{garnett92} and 
the more recent fits from \cite{permodiaz03} to estimate \Te~([\OII]) 
from the upper limit of \Te ([\OIII]). Fig.~\ref{ficTOTS} 
shows the results from both estimates in the top middle and right panels. 
For \Te~([\OII])$_{PM03}$ \Ne=100 was used for the whole region.

The electron temperature determined using the \cite{garnett92} method 
gives  a mean \Te([\OII]) about 1600 K higher than \cite{permodiaz03}.
This difference may be traced to the fact that for \Ne=100 and \Te([\OIII])
$>$ 15000, the \cite{permodiaz03} model is systematically below the 
\cite{garnett92} relation.  
If \Ne=10 is assumed, then the \Te([\OII]) would increase at 
least 2000K, bringing the two methods to better agreement.

The sulphur electron temperature \Te([\SIII]) was computed for 
spaxels where [\SIII]$\lambda$6312 was detected with 
S/N  $>$ 2, using the ratio R$_{S23}$=([\SIII]9069+9532)/[\SIII]6312.
The intensity of [\SIII]9532 was obtained from the 
theoretical relation [\SIII]9532 = 2.44$\times$ [\SIII]9069, as the 
$\lambda$9532 \AA\ line is more affected in these spectra by telluric 
emissions and absorptions.

\Te ([\SIII]) was also estimated using the fit obtained by 
\cite[H06]{hageleetal06} from a large set of observations that 
include \HII\ galaxies, Giant Extragalactic \HII\ regions and 
diffuse \HII\ regions in the Galaxy and the Magellanic clouds, 
given as  \Te([\SIII]) = 1.19\Te([\OIII]) - 0.32

The set includes objects with \Te([\SIII]) up to 24,000 K. This 
extends the relation to higher temperatures but also increases 
the uncertainty in the fit given that at the high temperature 
regime fewer objects exist and they exhibit large errors. In 
this case we remark that \Te([\OIII]) is an upper limit and 
caution should be taken for its interpretation.

We also used the SO$_{23}$ parameter \citep{diazpm00,diazetal07},
defined as
\begin{equation}
SO_{23} = \frac{S_{23}}{O_{23}} = 
\frac{[\SII] \lambda\lambda 6717,6731 + [\SIII] \lambda\lambda 9069,9532}
{[\OII] \lambda\lambda3727,3729 + [\OIII] \lambda\lambda4959,5007}
\end{equation}

\noindent 
which has weak dependence on the ionization parameter. \Te\ is
then obtained as  \Te([\SIII]) = 0.596 - 0.283(log~SO$_{23}$) + 
0.199(log~$SO_{23}$)$^2$.

Although this calibration was obtained for high metallicity objects and 
its use is justified for the central region, we also applied it for IC~132
given that the SO$_{23}$ values obtained are within the parameter validity
range of the 
original fit. 

Fig.~\ref{ficTOTS} maps the distribution of  \Te([\SIII]) computed
using H06 and SO$_{23}$   methods for IC 132. In the first case 
the dependency on \Te([\OIII]) and consequently on 
[\OIII]$\lambda$ 4363 is reflected in the facts that 
only the area where  [\OIII] is detected with S/N $>$ 2 has 
\Te ([\SIII]) and that these values are higher than the ones obtained
with the other methods, being \Te([\OIII]) an upper limit.

Besides the fact that SO$_{23}$ has low sensitivity to the 
ionization parameter, the difference in scatter may also be 
traced to the difference in errors associated with the lines 
used. For  the H06 method the large error of the 
[\OIII]$\lambda$ 4363 line is propagated to give a standard 
deviation of about 3800 K in \Te([\SIII]), while for SO$_{23}$ 
all the involved lines having S/N $>$ 10, give a standard 
deviation of just 700 K in \Te([\SIII]). The low mean  
\Te([\SIII])=9800 may have its root in the fact that the 
SO$_{23}$ calibration was obtained mainly with high metallicity
objects.

\begin{figure}
\centering
\includegraphics[width=6cm]{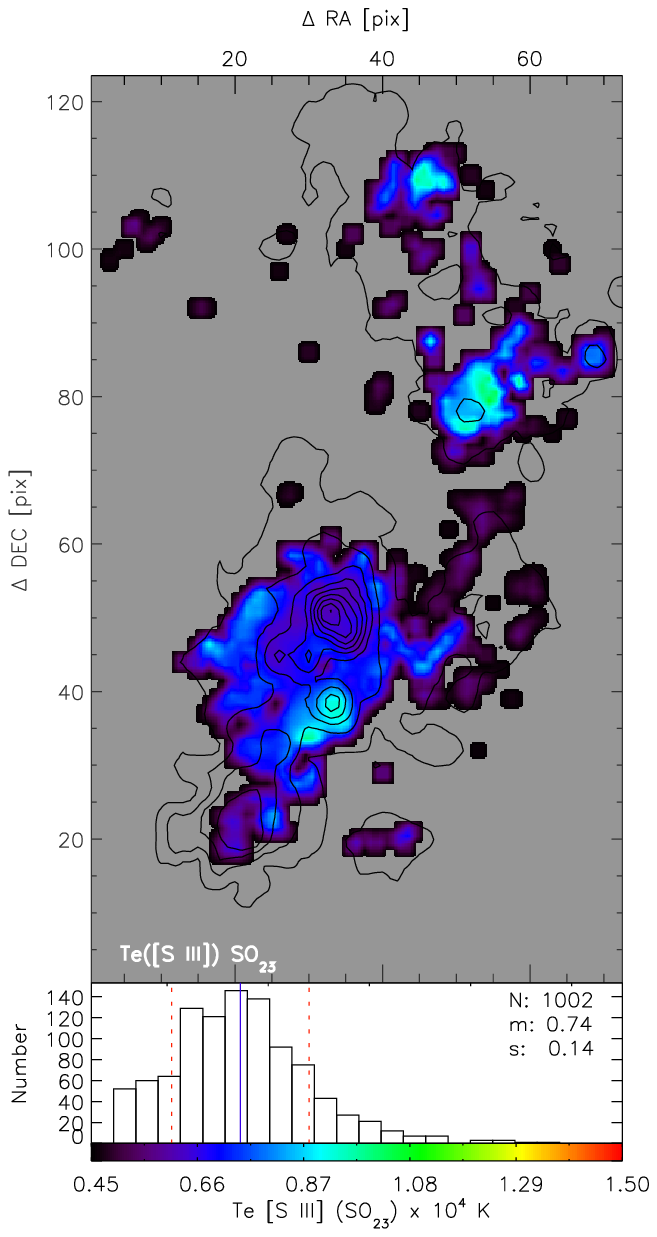} 
\caption{\Te([\SIII]) obtained as in \protect\cite{diazetal07}  for the centre of M33.}
\label{f3kerTSIII}
\end{figure}

Fig.~\ref{f3kerTSIII} shows the temperature distribution for 
the central region of M33 using  the SO$_{23}$ method. It gives 
also a very uniform temperature for all the measured spaxels in 
the field. 

If the scatter in temperature can be mainly associated to the 
ionization structure and the SO$_{23}$ effectively cancels it
then it would be worth to investigate the inclusion of more low 
metallicity objects in the \Te([\SIII]) SO$_{23}$ fit. This may 
modify the relation to produce higher temperatures in the low
metallicity range [see figure 9 of \cite{diazetal07}].

\section{Chemical abundances}
\subsection{Direct method}
\begin{landscape}
\begin{figure}
\centering
\begin{tabular}{ccc}

\includegraphics[width=6.0cm]{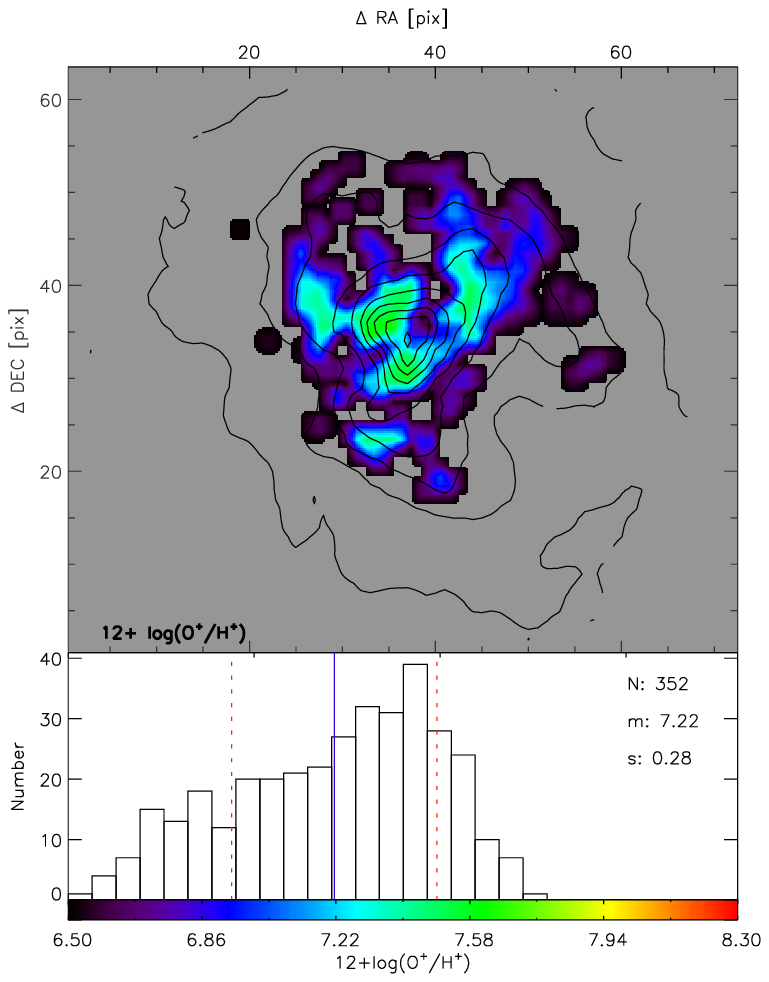} &
\includegraphics[width=6.0cm]{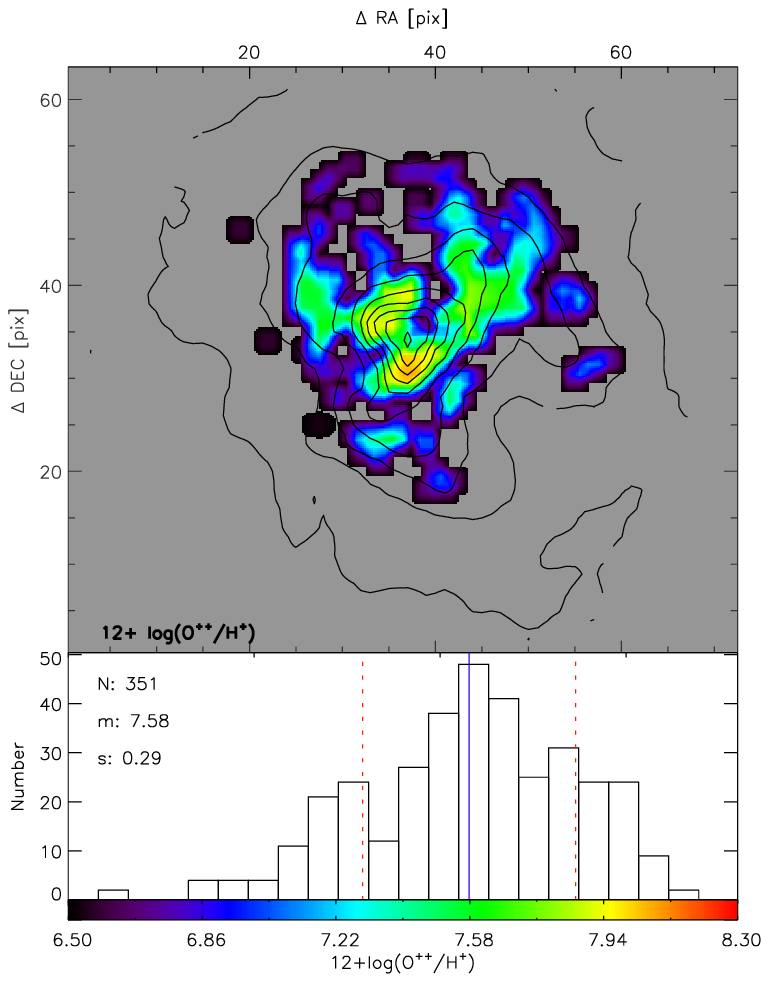} &
\includegraphics[width=6.0cm]{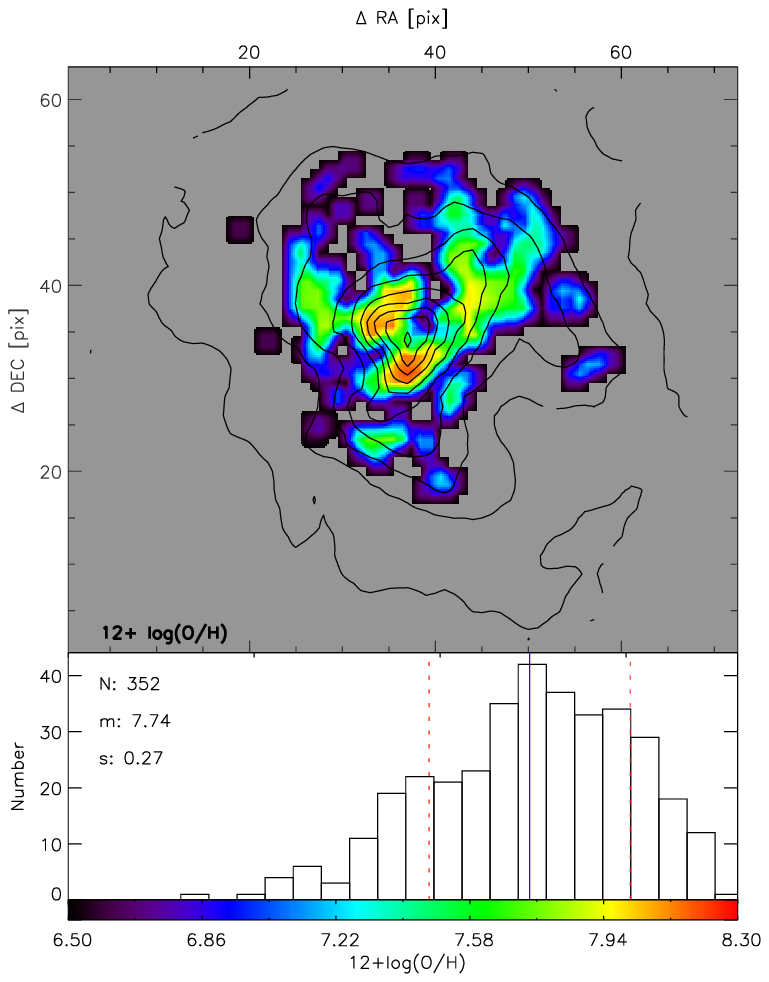} \\

\includegraphics[width=6.0cm]{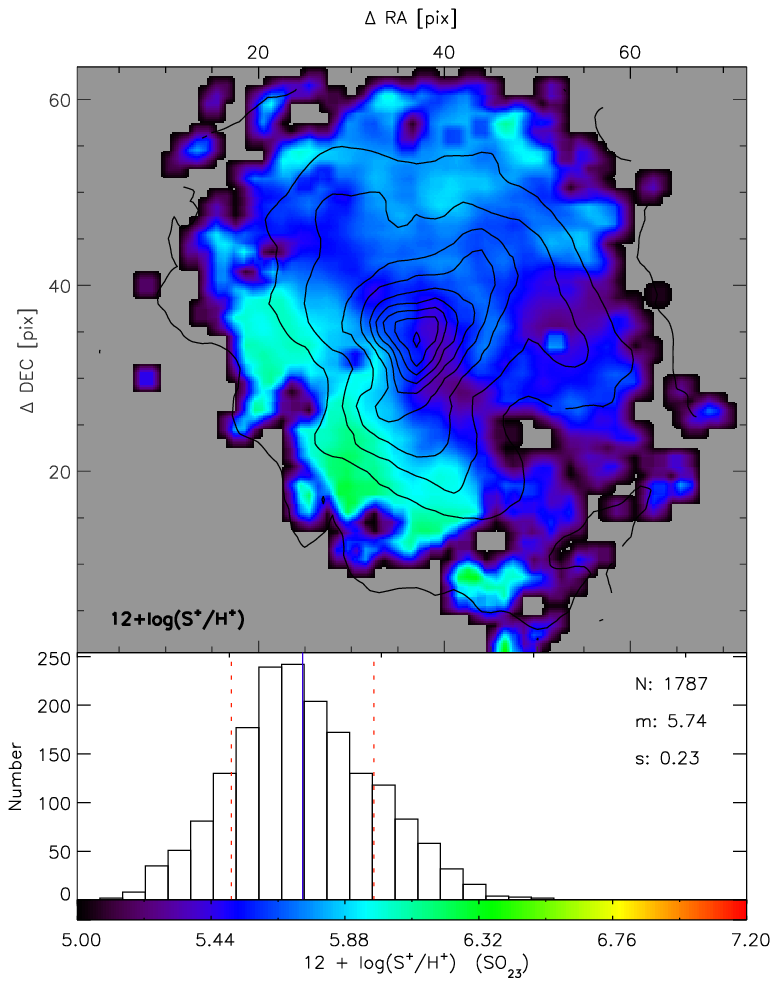}&
\includegraphics[width=6.0cm]{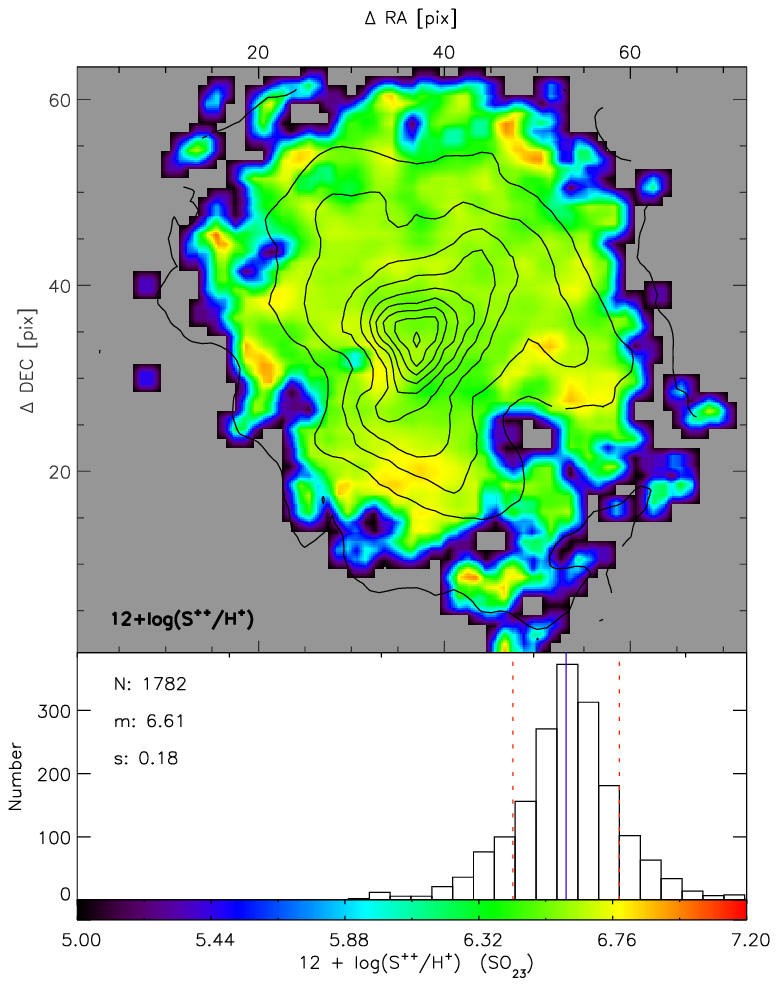}&
\includegraphics[width=6.0cm]{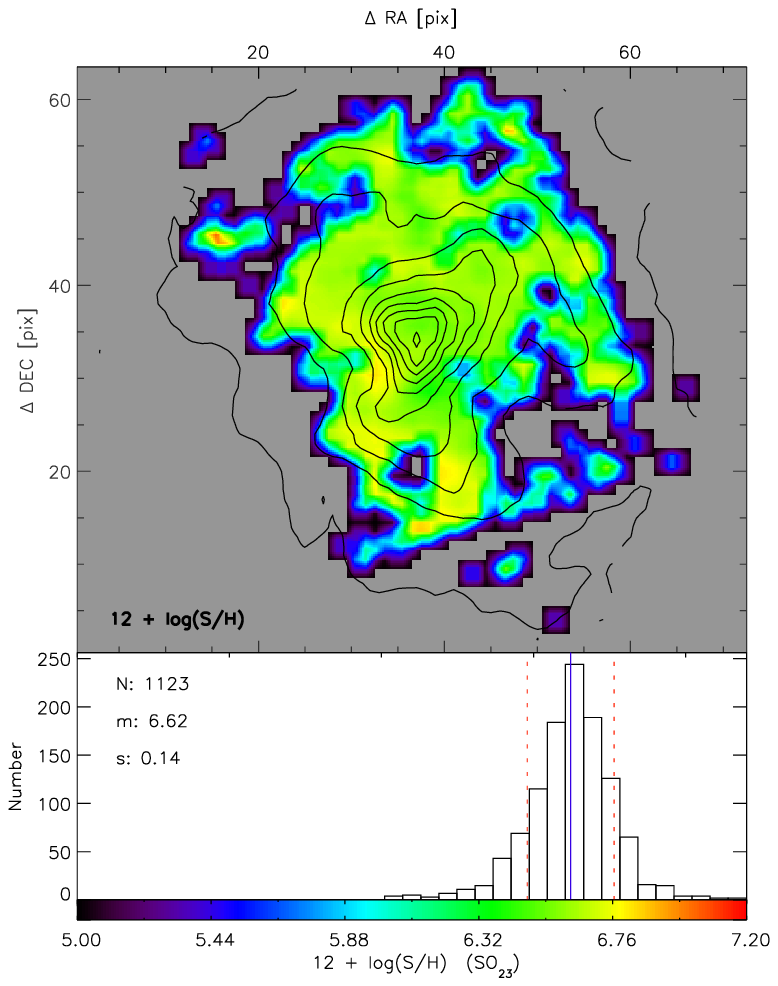}\\

\end{tabular}
\caption{IC 132 ionic and total lower limit abundances for oxygen (top) and 
sulphur (bottom)}
\label{f33icsulphabund}
\end{figure}
\end{landscape}

\noindent{\bf Oxygen}\\
Taking the upper limit temperatures obtained for the central zone 
of IC 132, we computed the lower limit ionic abundances using 
the IONIC task from  the NEBULAR package for IRAF, also based on the 
5-level atom statistical equilibrium approximation \citep{temden87}. 
This task requires additionally the electron density \Ne\   and the 
emission line flux relative to H$\beta$ of an atom in specific 
ionization stages. For O$^{++}$/H$^+$  we used the upper limit 
\Te ([\OIII]) while for O$^+$/H$^+$  we adopted the \Te([\OII]) 
obtained from the relations given by \cite{permodiaz03}.  The total 
oxygen abundance was obtained as 

\begin{equation}
\frac{O}{H}=\frac{O^+}{H^+} +  \frac{O^{++}}{H^+}
\end{equation}
\noindent and is displayed, together with the ionic values, at the 
top of figure~\ref{f33icsulphabund}.

\noindent{\bf Sulphur}\\
Sulphur ionic abundances were obtained  using the expressions 
derived by \cite{diazetal07} from fits to results from the IONIC
task. The unobserved S$^{3+}$ can represent an important contribution
in high ionization zones, thus in order to estimate the total sulphur
abundance an ionization correction factor (ICF) must be included.
We used the ICF from \cite{barker80}. Although $O^+$ and  $O^{++}$ 
are lower limits, their inclusion in the sulphur ICF yields a better 
S estimation than just ignoring the correction factor.  

\begin{landscape}
\begin{figure}
\centering
\begin{tabular}{ccc}
\includegraphics[width=6cm]{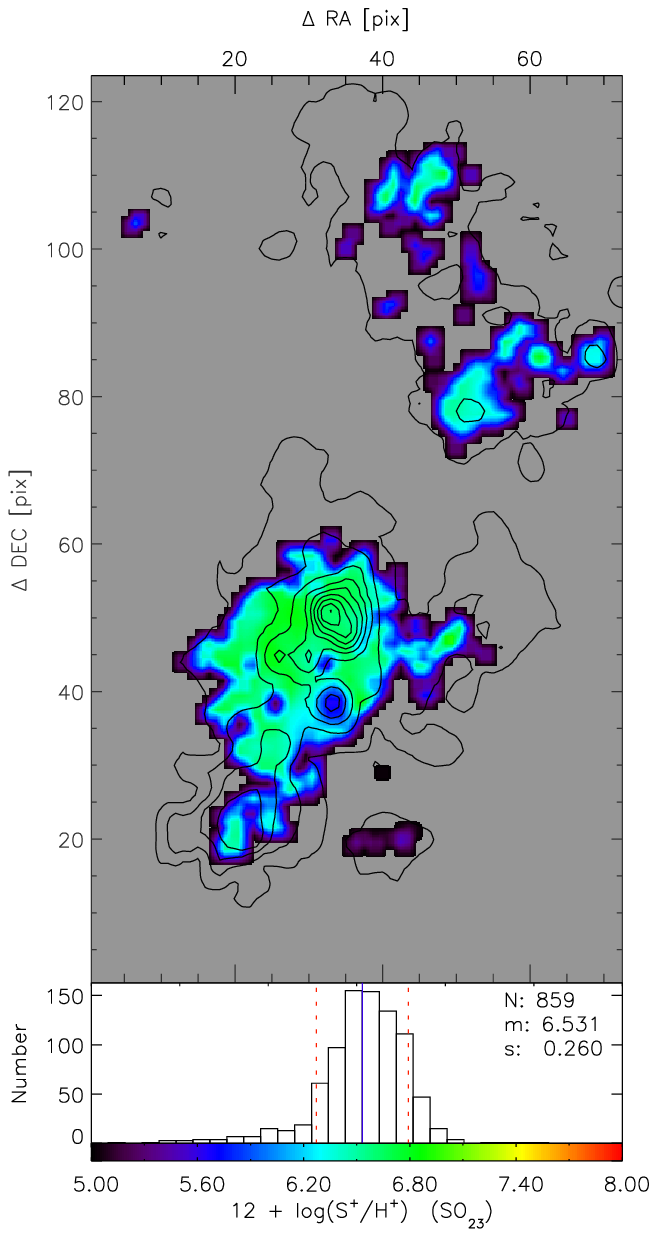}
\includegraphics[width=6cm]{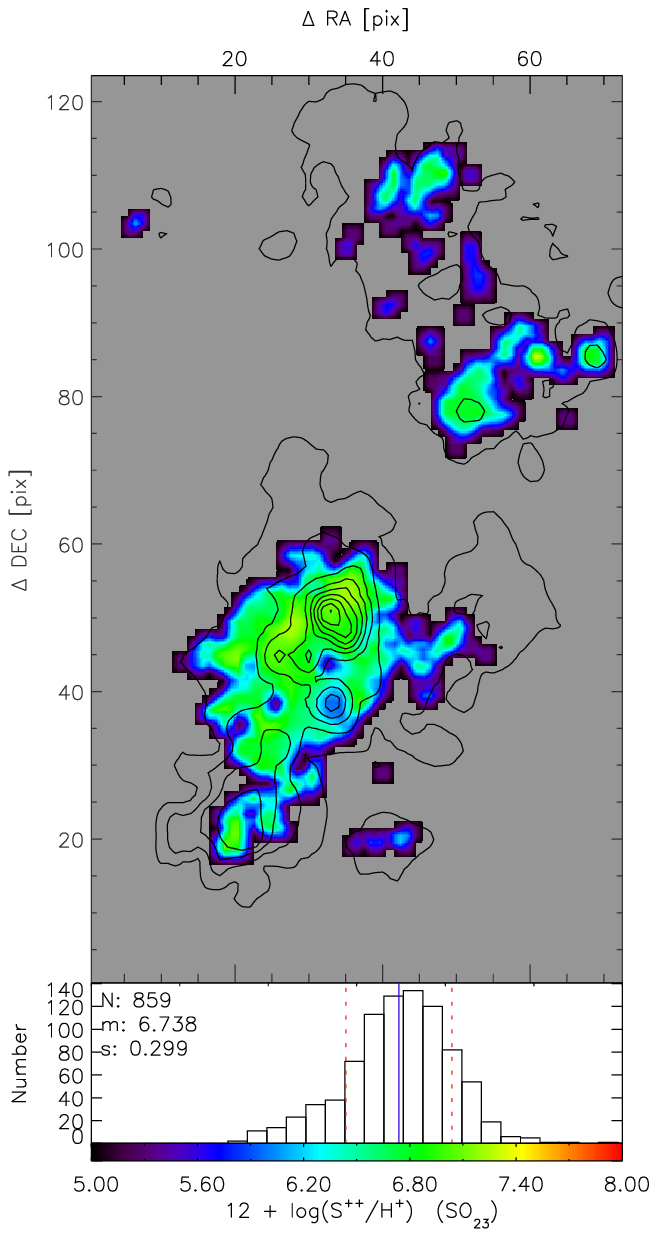}
\includegraphics[width=6cm]{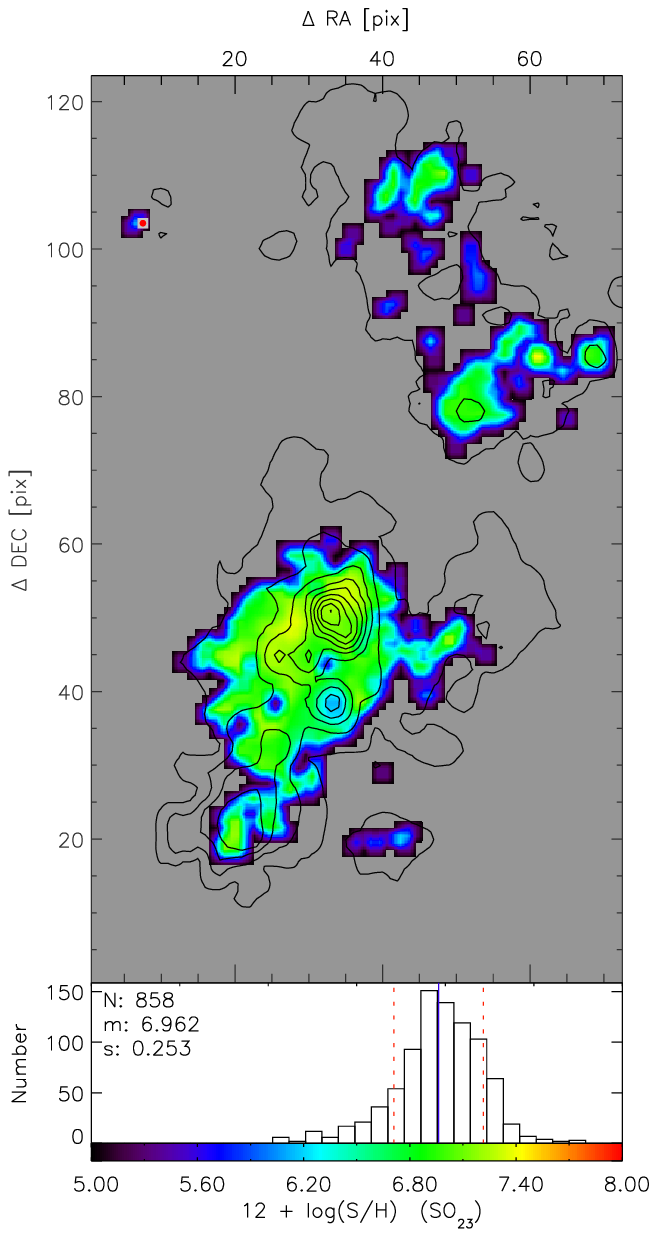}
\end{tabular}
\caption{Ionic (S$^+$/H$^+$) and (S$^{++}$/H$^+$)
and total sulphur abundances for the centre of M33.}
\label{f33kersulphabund}
\end{figure}
\end{landscape}

The bottom of figure~\ref{f33icsulphabund} shows the ionic and total 
sulphur abundances for IC~132. For the central region the corresponding
maps are in figure  \ref{f33kersulphabund}.  The spatial distibution in 
IC 132 seems to have a wider dynamical range than the central zone in 
S$^{+}$ and S$^{++}$. This variation is primarily produced by a diagonal
zone in the lower left corner of IC 132, which we call a ``wall" that could 
probably be attributed to a zone of diffuse radiation (see figure \ref{f33io
nizationratiosic132}).

\clearpage
\subsection{Empirical methods}

The direct method is useful in a restricted range of abundances. For systems with 
oxygen abundance larger than $\sim$ 0.2(O/H)$_\odot$,  as the 
metallicity increases, so does the cooling by IR fine 
structure lines and the [\OIII]$\lambda$4363\AA\ 
line becomes extremely weak, requiring high S/N 
observations to be detected. [\cite{kenbregar03} 
estimate oxygen abundances $\sim$ 0.5(O/H)$_\odot$ 
as the practical upper limit to detect [\OIII]$\lambda$4363\AA ].

The difficulty of direct abundance estimation has led to the 
development of {\it empirical} or {\it strong line} 
methods amply used in the literature. 

\subsubsection{The R$_{23}$ parameter}

The empirical parameter R$_{23}$ =([\OII]3727,29+[\OIII]4959, 5007)/\Hb , 
also known as $O_{23}$, was 
first proposed by \cite{pageletal79} and 
is  widely used as an abundance indicator. Several 
calibrations exist for $R_{23}$, revised in the light of new
observations with direct \Te\ determinations or improvements
to photoionization models.Some reformulations of the 
parameter have also been produced in an effort to compensate for the 
effect of the stellar temperature and ionization parameter
\citep{edmundspagel84,zaritskyetal94,kobunickyetal99,
pilyugin01hz,pily01}.

R$_{23}$ has two main drawbacks. Firstly is the fact that the metallicity 
mapping is double valued. For each value of 
R$_{23}$ there is a low metallicity value corresponding to 
the lower branch of the relation and a high metallicity 
estimate corresponding to the upper branch. Besides, the 
transition between branches depends on the calibration and  
is ill defined [see e.g.~Fig.~12 of \cite{kenbregar03}]. Secondly, 
most of the data points tend to lay close to the knee of the 
relation, i.e. in the region in which there is no dependence of the 
R$_{23}$ parameter on metallicity. The R$_{23}$ method is in 
practice a reliable  abundance estimator only for the metal rich 
or very metal poor systems.

Here we used the calibration from \cite{mcgaugh91}, based 
on \HII\ region photoionization models obtained with Cloudy 
\citep{cloudy90} and taking into account the effect of the 
ionization parameter. The analytic expressions for the 
upper and lower branches are taken from \cite{kobuletal99}.
The estimated uncertainty for this calibration ranges from 
0.1  to 0.2 dex when the lines are detected with S/N $>$ 8, 
reaching up to $\pm$0.25 dex in the turnover zone
around 12 +log(O/H) $\sim$ 8.4.

To break the R$_{23}$ degeneracy  the 
[\NII]$\lambda$6584/[\OII]$\lambda$3727 ratio is used as 
an initial guess of metallicity helping in the selection of the 
appropriate R$_{23}$ branch (McCall, Rybski \& Shields 1985).
This ratio 
shows a weak dependence on the ionization parameter and 
a strong correlation with metallicity \citep{kewdop02}. 
Another diagnostic ratio used to distinguish the proper 
R$_{23}$ branch is [\NII]$\lambda$6584/\Ha\ \citep{raimannetal00} 
the N2 index, calibrated in [O/H] by \cite{denicoloetal02}.  
However this ratio has a strong dependence on the ionization 
parameter (see below). 

Fig.~\ref{f33niioii} shows  R$_{23}$  as a function of the 
[\NII]$\lambda$6584/[\OII]$\lambda$3727 ratio. The 
\cite{kewleyellis08} empirical separation between the upper 
and lower branches of R$_{23}$ at 
[\NII]$\lambda$6584/[\OII]$\lambda\lambda$ 3727 =
-1.2 [12 + log (O/H) $\sim$ 8.4] is used. 
For IC~132 the majority of the pixels fall within the lower 
branch regime while for the central zone the opposite 
is observed. The central region seems to follow a 
continuous distribution of R$_{23}$, while in IC~132 most 
of the points are clustered between log(R$_{23}$) $\sim$ -1.2 
and $\sim$ 0.8 and  the distribution shows a particularly 
abrupt cutoff at log(R$_{23}$) $\sim$ 0.8. 

The empirical division by \cite{kewdop02} was obtained 
combining observations and photoionization model
grids. These models show a dispersion below 
[\NII]$\lambda$ 6584/[\OII]$\lambda$ 3727 = 1.2
for different values of ionization parameters and 
metallicities [Fig.~8 from \cite{kewdop02}]. 
However the correlation is remarkably extended up to log(R$_{23}$) 
$>$ 1.6 for the centre. This tail is produced by the faint (diffuse) radiation, 
not included in \cite{kewleyellis08} models.

The resulting maps for the R$_{23}$ estimator are 
shown in the left panels of  Fig.~\ref{f33abundempiricalic} 
for IC 132 and figure \ref{f33abundempiricalker} for the 
centre of M33. 

Another empirical indicator based on the $R_{23}$ method
is proposed by \cite{pilyugin01hz,pily01,pily01a} where the 
excitation parameter P is introduced to compensate for
 $R_{23}$ variations along the region produced by differences in the ionization 
parameter.  The upper branch applies to 12+log(O/H) $>$ 8.2 
and the lower branch to 12+log(O/H) $<$ 8.2 
Results from an estimation spaxel by spaxel are shown in Figs.
~\ref{f33abundempiricalic} and \ref{f33abundempiricalker} for 
IC 132 and the centre respectively. In this case we used the 
[\NII]$\lambda$6584/[\OII]$\lambda$3727 ratio to select the 
branch and break the R$_{23}$ degeneracy.

\begin{figure}
\centering
\begin{tabular}{cc}
\includegraphics[width=7.5cm]{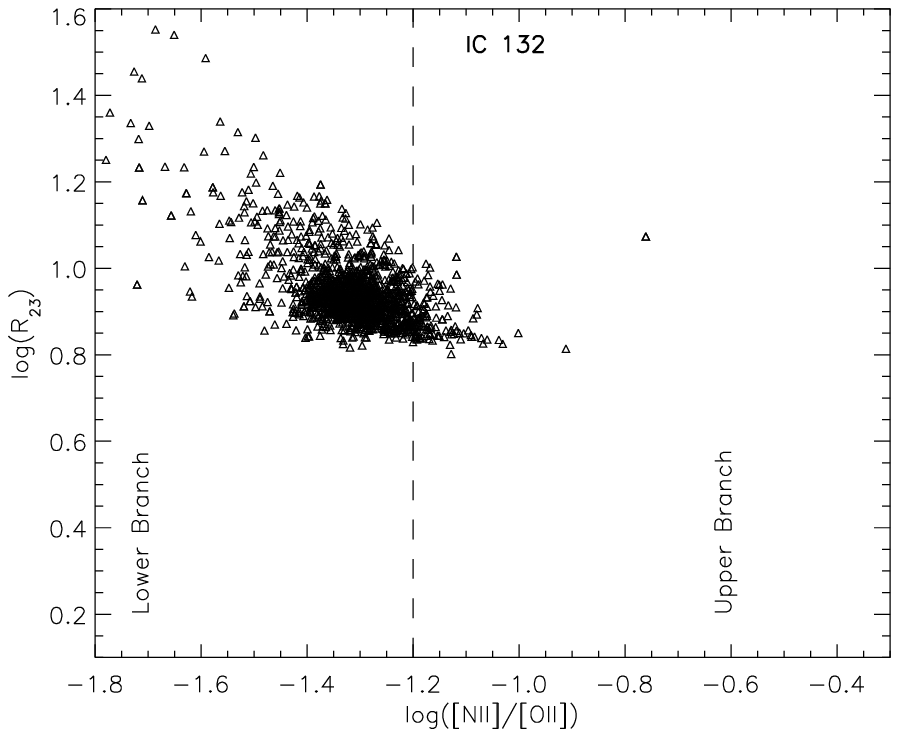} & 
\includegraphics[width=7.5cm]{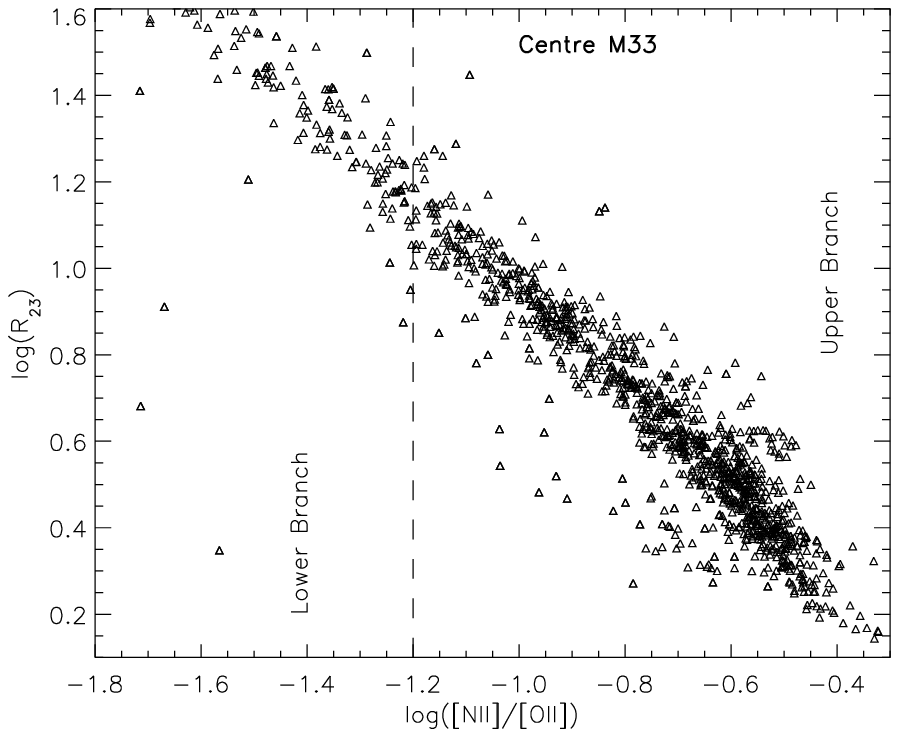}\\  
\end{tabular}
\caption{[NII]/[OII] vs R$_{23}$ to select the upper or lower branch for 
each pixel. Left: IC~132; right: central zone. The dividing dashed line is 
from \protect\cite{kewleyellis08} calibration.
Only those pixels with errors lower than 0.1 dex in each axis are plotted.
} \label{f33niioii}
\end{figure}

\subsubsection{The N2 parameter}

The N2 index, defined by \cite{denicoloetal02} as 
N2=[\NII]6584/\Ha, was previously used by \cite{sbergmannetal94} 
as an empirical abundance estimator for star forming galaxies. 
Its usefulness is based on the fact that it is easy to observe
up to relatively high redshifts and, due to the closeness of 
the lines involved, it is unaffected by reddening and   
can even be measured directly from uncalibrated spectra.
Here we use the N2  fit  from \cite{denicoloetal02}, 
given as 12 + log (O/H)=9.12+0.73N2. The resulting maps 
from the N2 estimator are shown in the lower panels of  
Fig.~\ref{f33abundempiricalic} for IC~132 and in figure 
\ref{f33abundempiricalker} for the centre of M33. 

It is remarkable that the abundance based on N2, shows
practically no variation in the maps. One may conclude from this indicator
that the oxygen abundance has one single value for the whole region.
It seems that all the variation within the region can be traced to the
error which effectively increases with radius both for IC~132 and for the 
bright central condensation hereinafter called
BCLMP~93 as identified in \cite{bclmp74}.

\subsubsection{The O3N2 parameter}

\cite{PP04}  calibrated  the O3N2 ratio \citep{alloinetal79} 
as an abundance estimator, giving the relation 
12 + log (O/H)=8.73 - 0.32 $\times$ O3N2 where O3N2 = 
log([\OIII] $\lambda$5007/\Hb)/([\NII] $\lambda$6584/\Ha).

\begin{figure}
\centering
\begin{tabular}{cc}
\includegraphics[width=6cm]{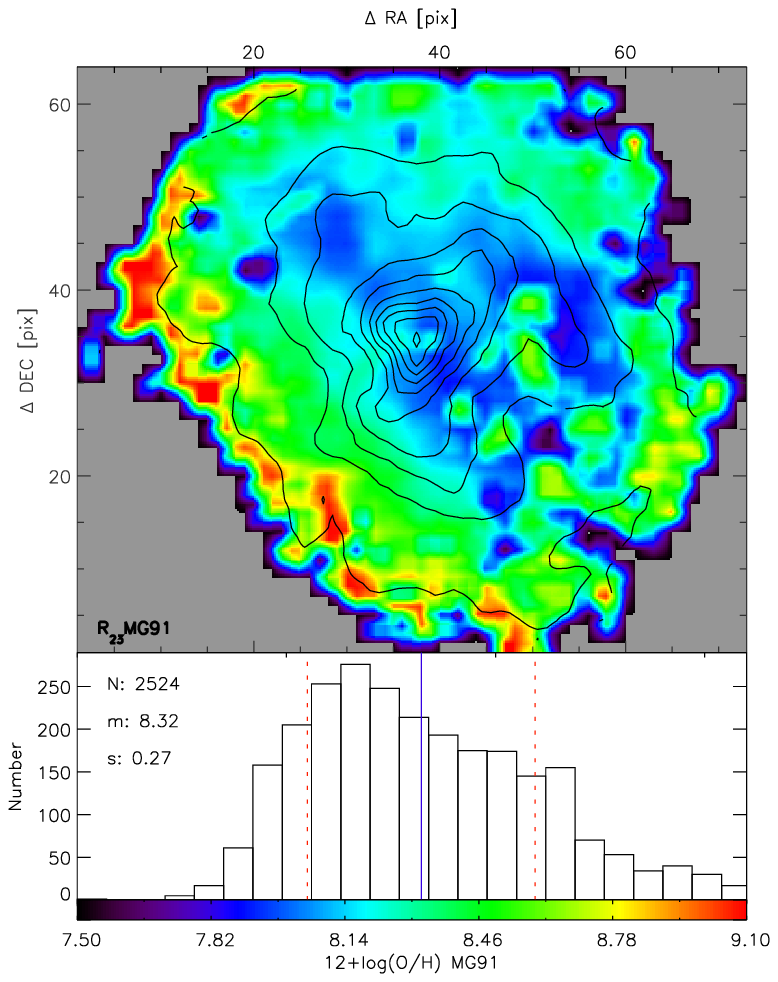} &
\includegraphics[width=6cm]{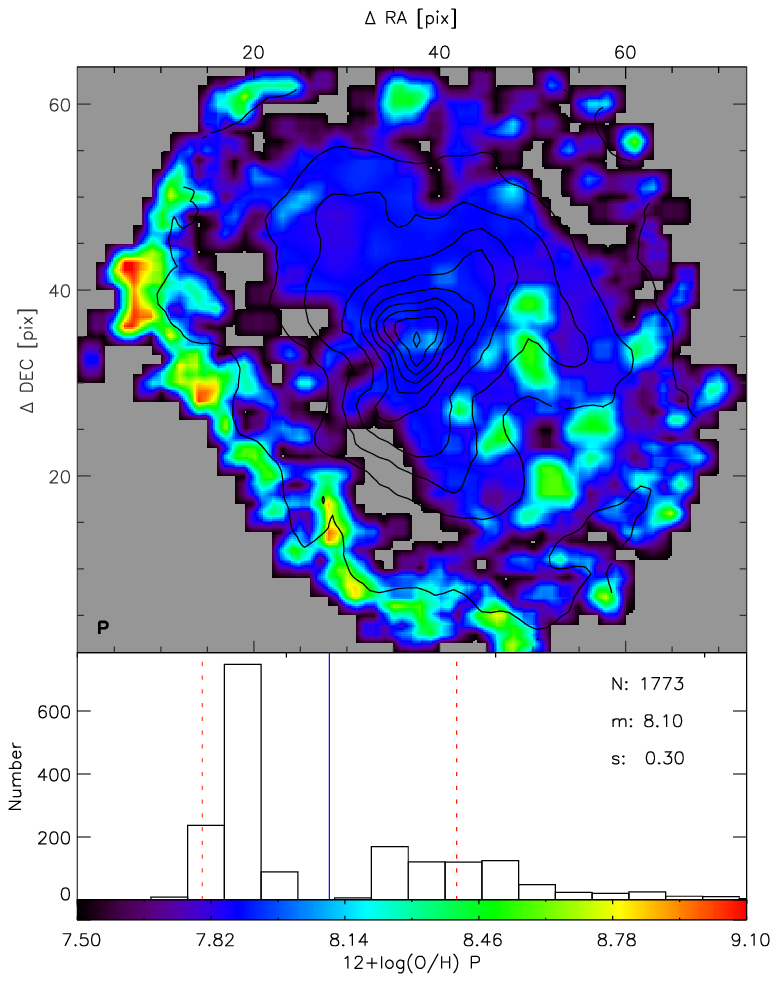} \\
\includegraphics[width=6cm]{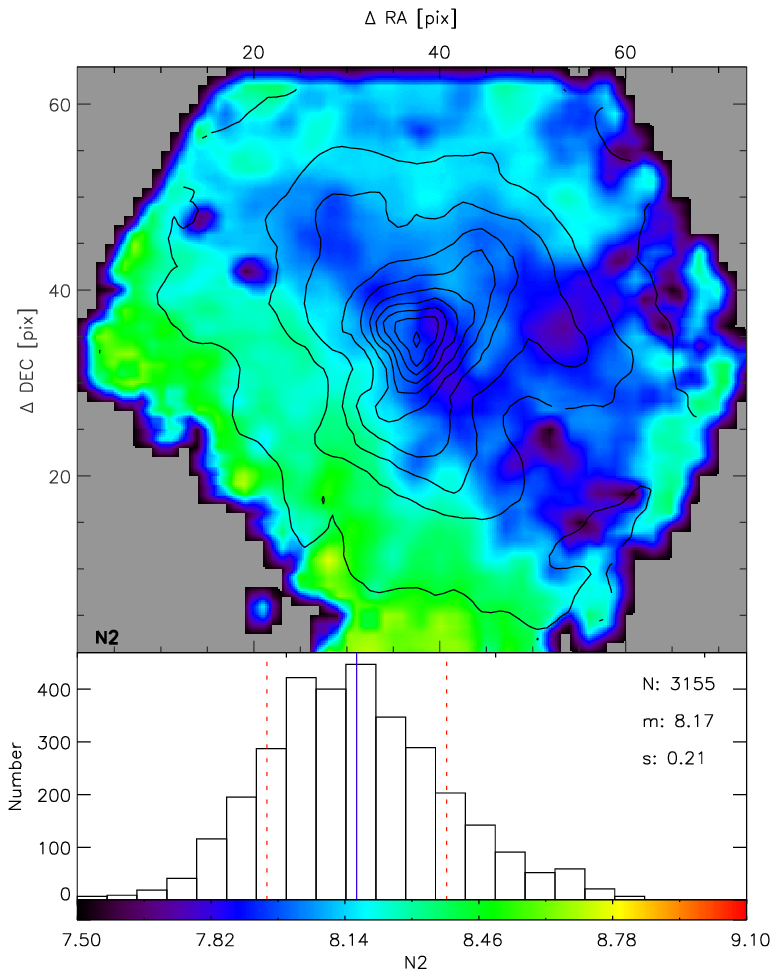}&
\includegraphics[width=6cm]{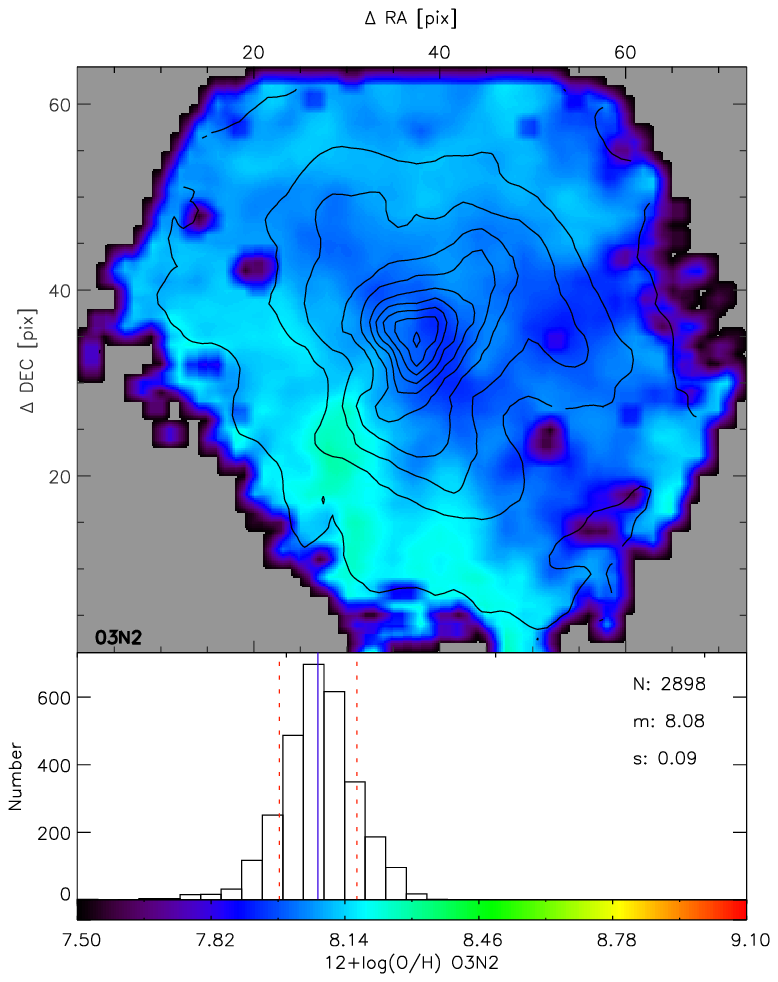} \\
\end{tabular}
\caption{Empirical metallicity estimators for IC~132. Top: R$_{23}$ with the
calibrations from  \protect\cite{mcgaugh91} and \protect\cite{pily01a}. 
Bottom: N2 and O3N2.}
\label{f33abundempiricalic}
\end{figure}

\begin{figure}
\centering
\begin{tabular}{cc}
\includegraphics[width=6cm]{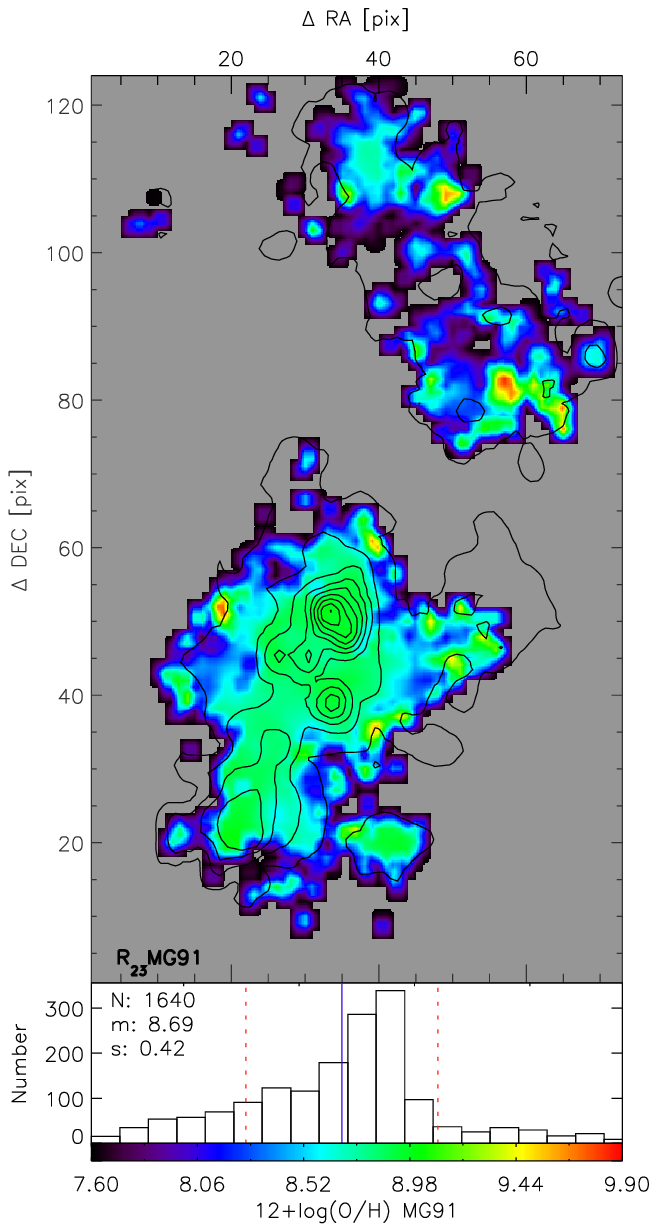} &
\includegraphics[width=6cm]{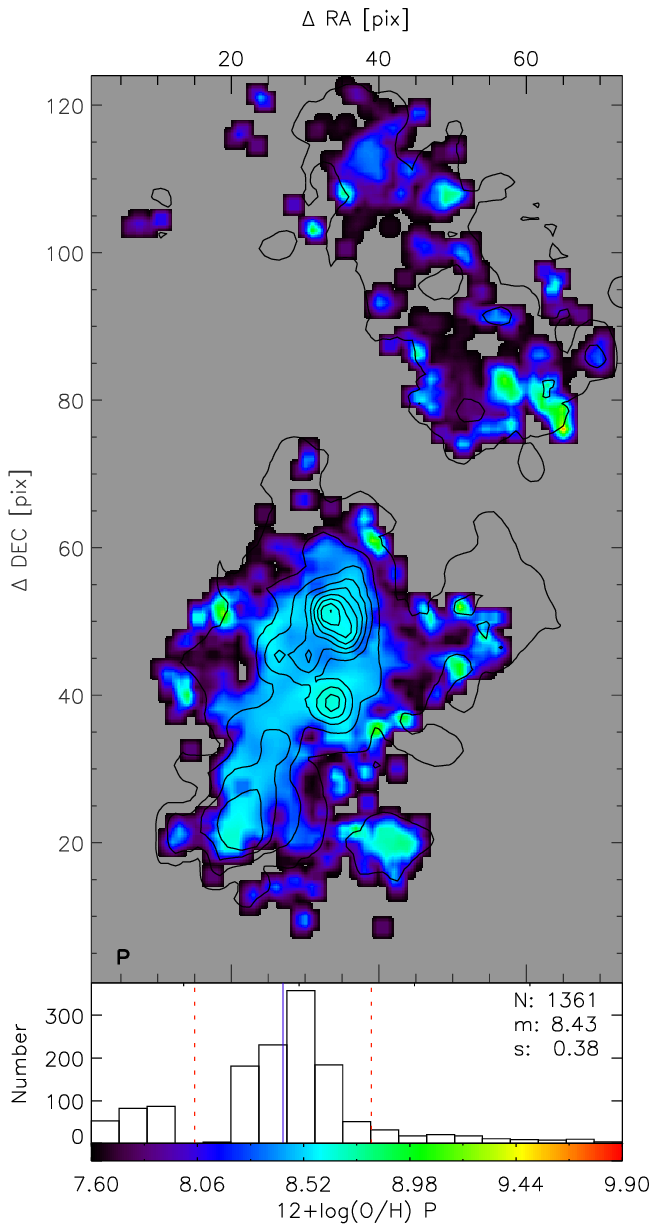} \\

\includegraphics[width=6cm]{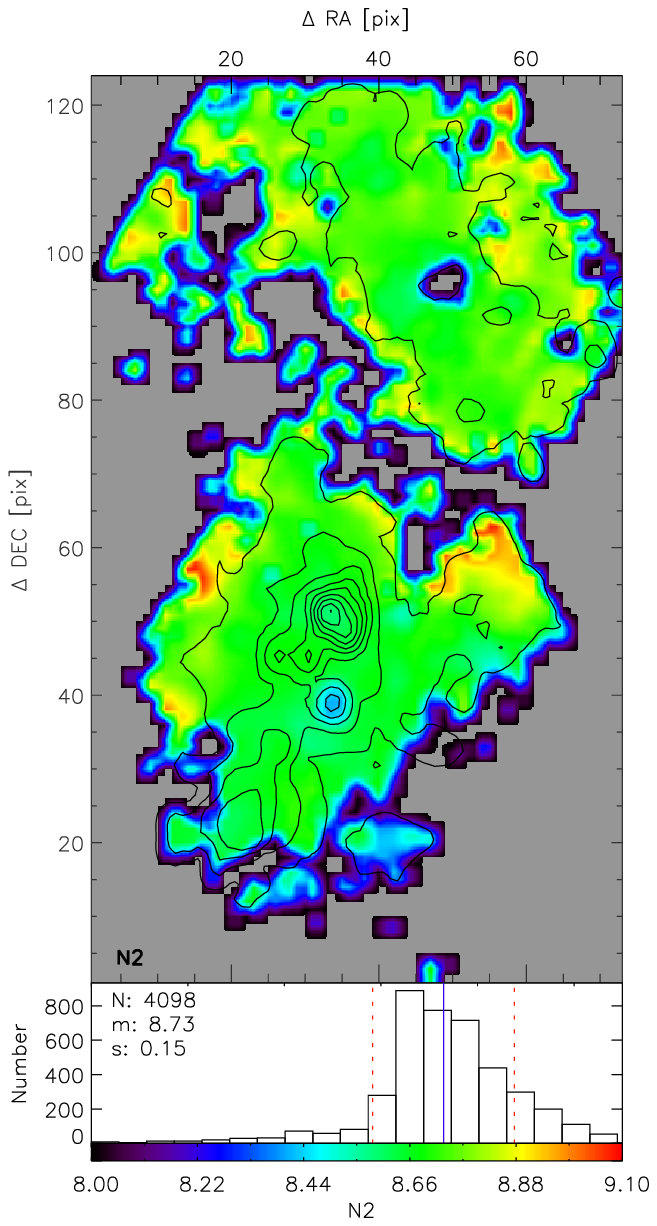} &
\includegraphics[width=6cm]{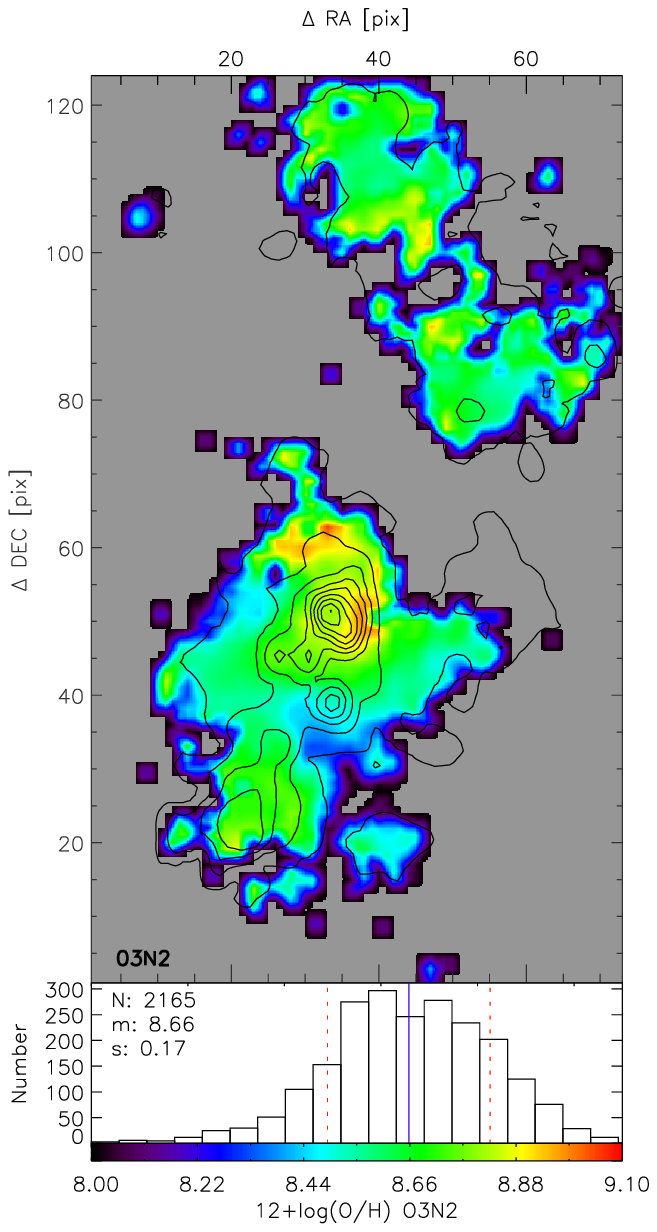} \\
\end{tabular}
\caption{Empirical metallicity estimators for the centre of M33. 
Top: R$_{23}$ with the calibrations from  \protect\cite{mcgaugh91} and 
\protect\cite{pily01a}. Bottom: N2 and O3N2.}
\label{f33abundempiricalker}
\end{figure}

The resulting maps of 12 + log(O/H) from the O3N2 estimator are 
shown in the right panels of  figure~\ref{f33abundempiricalic} for 
IC 132 and figure~\ref{f33abundempiricalker} for the centre of M33.

The distribution of O/H obtained via N2 and O3N2 are very similar;   
the scatter is smaller when the abundance is derived using O3N2  
than when using N2  in IC~132.

\subsubsection{The S$_{23}$ parameter}
The S$_{23}$ parameter, proposed by \cite{vilchezesteban96} that 
uses the strong [\SII] and [\SIII] lines, in analogy to   R$_{23}$ is 
defined as  S$_{23}$=([\SII]6717, 6731 + [\SIII]9069, 9532)/\Hb.

\cite{cristpeter97} calibrated S$_{23}$ as a sulphur abundance
indicator based on observational data compiled from the literature, 
while \cite{diazpm00} proposed S$_{23}$ as an alternative to 
R$_{23}$ as oxygen abundance indicator. S$_{23}$ presents some 
advantages over  R$_{23}$: it is single valued up to solar 
metallicities and has less dependence on the ionization parameter 
and on the effective temperature of the ionizing stars. 
On the other hand, it needs wider spectral coverage to detect the 
near-IR [\SIII] lines and corrections for the unseen S$^{3+}$ ion. 

We estimated the sulphur abundance (S/H) using the relation from 
\cite{pmonteroetal06}, while for (O/H) we used the expression from 
\cite{permodiaz05}. Both are fits obtained from observed
data sets.

\begin{figure}
\centering
\begin{tabular}{cc}
  \includegraphics[width=6cm]{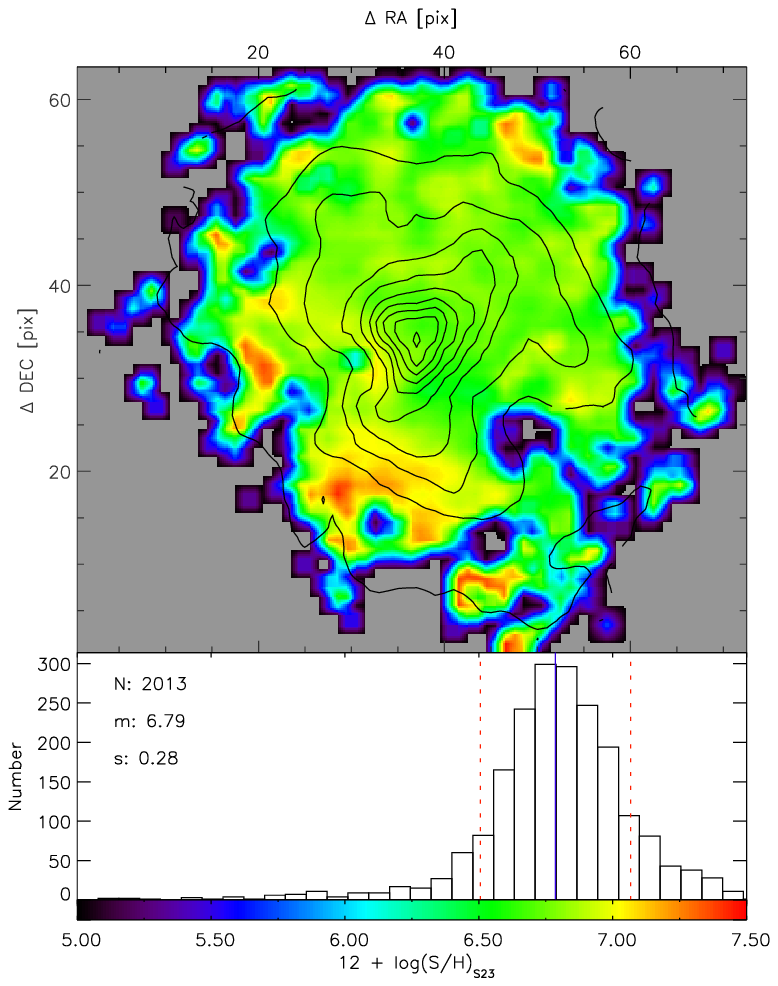} &
  \includegraphics[width=6cm]{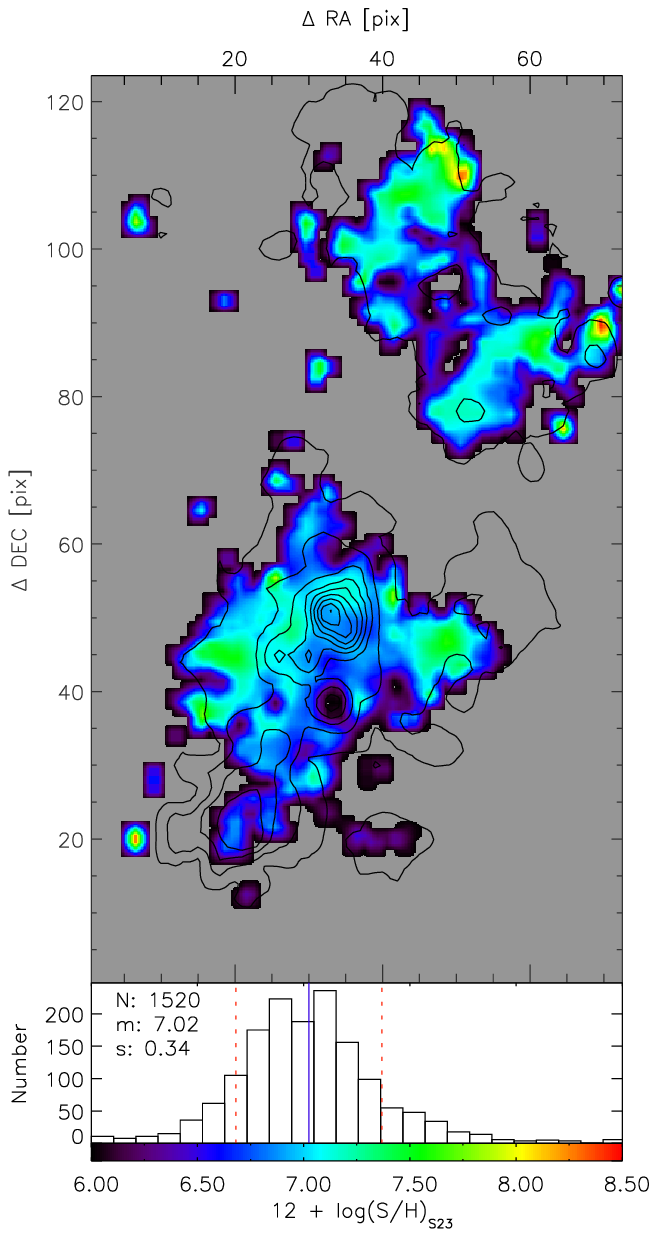} \\
\end{tabular}
\caption{12 + log(S/H) abundance using the S$_{23}$ indicator for 
IC 132 (left) and for the centre of M33 (right).}
\label{fabsulphurs23} 
\end{figure}

\begin{figure}
\centering
\begin{tabular}{cc}
  \includegraphics[width=6cm]{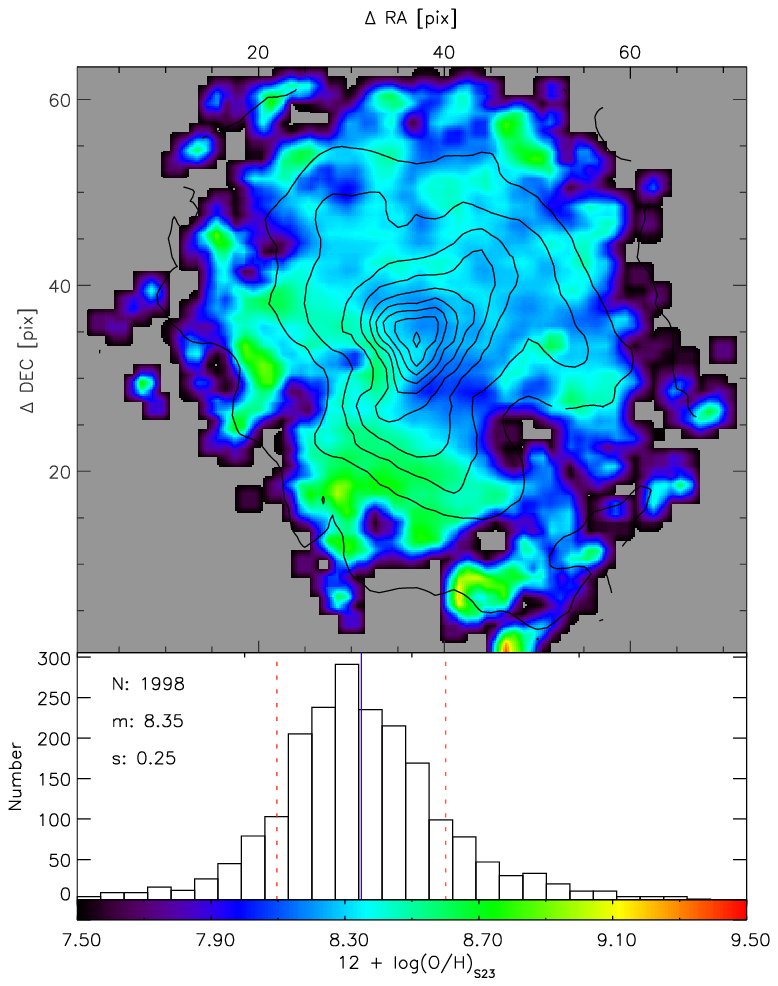} &
  \includegraphics[width=6cm]{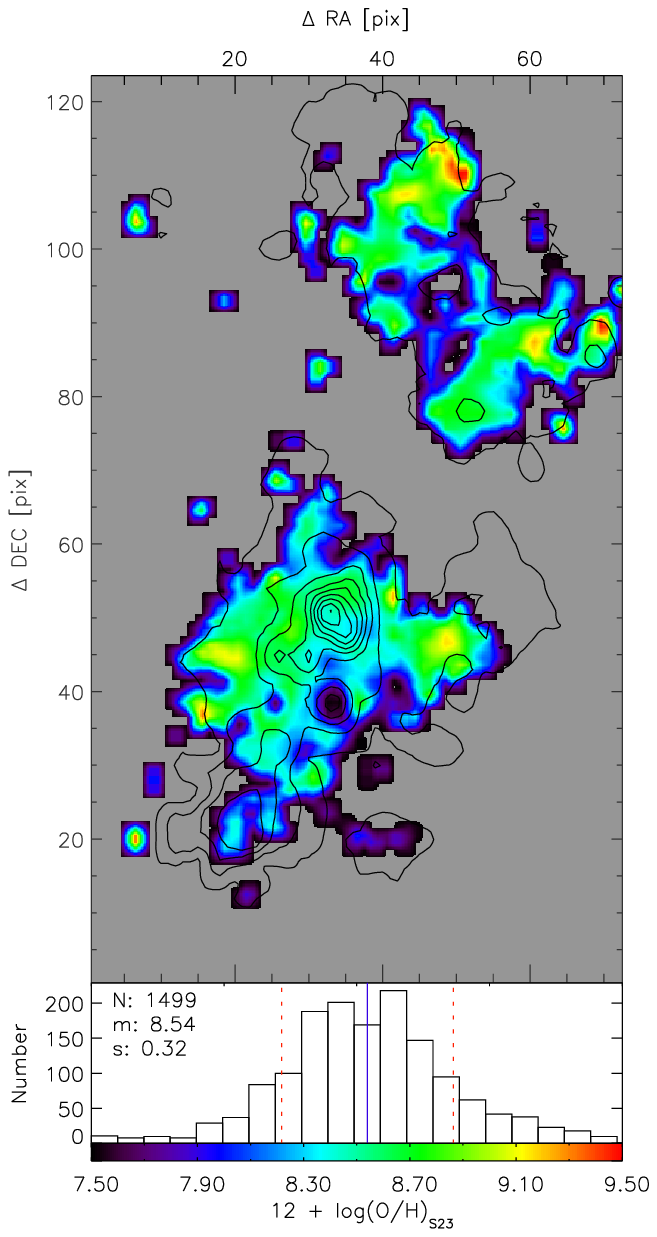} \\
\end{tabular}
\caption{12 + log(O/H) abundance from the S$_{23}$ indicator for  
IC 132 (left) and for the centre of M33 (right).} 
\label{faboxygens23} 
\end{figure}

Sulphur and oxygen abundance maps obtained from the S$_{23}$
indicator are shown in figures \ref{fabsulphurs23} and 
\ref{faboxygens23} respectively.  The mean 12+log(S/H)$_{S23}$ 
value for the central region is 7.02 with a standard deviation of 
0.34 while for IC~132 it is 9.5 with a smaller standard deviation 
of 0.28. The estimates of S/H and O/H for IC132 using S$_{23}$ 
show little scatter, albeit  slightly larger than that for the O/H 
abundance using the N2 or O3N2 estimators.

\subsubsection{Ar$_3$O$_3$ and S$_3$O$_3$ calibrations}
Based on a compilation of \HII\ regions in spiral galaxies and
blue compact galaxies, \cite{sska06ar3s3} proposed the use of
two line ratios 
(Ar$_3$O$_3 \equiv $ [\ArIII]$\lambda$7135/[\OII]$\lambda$5007
and S$_3$O$_3 \equiv $ [\SIII]$\lambda$9069/[\OIII]$\lambda$5007)
as metallicity indicators. The calibration of the relation is based 
on the \cite{pilyugin01hz} method. 
 In  \S \ref{sec_integrated_properties} we will analyze the results 
separating the data in sections of equal \Ha\ intensity in a treatment
 that we call individual shell segmentation. The respective line ratios for
the individual shells segmentation  are given in tables  
\ref{ticphysycalabundshells} and \ref{ticphysycalabundaper} for 
IC~132 and in tables \ref{tcc93physycalabundshell} and 
\ref{tcc93physycalabundaper} for BCLMP 93. The 12 + log(O/H) 
values, also for shell segmentation, are in figures 
\ref{fishellempiricalab} and \ref{fc93empab} for IC~132 and 
BCLMP 93 respectively.

\subsubsection{About the abundance estimations}
The spatially resolved data on the M33 \HII\ regions allows 
the comparison of the different estimators on the assumption 
that the metal content of the ionized gas is well mixed giving 
a uniform total abundance across the \HII\ region. For IC 132 
R$_{23}$ gives an average 12 + log(O/H) = 8.32 with standard deviation (s.d.)=0.27,
the P method gives 12 + log(O/H) = 8.1 with s.d.=0.30,
the N2 method gives an average 12 + log(O/H) = 8.17 with 
s.d.=0.27 and O3N2 gives  
12 + log(O/H) = 8.08 with s.d.=0.09. 

Hence, considering the mean of the abundance distribution we may 
say that the different indicators are consistent with each 
other in the sense that they give the same mean abundance 
value. Caution should arise for applying the P method to IC 132, given 
that the mean of the abundance distribution falls between
the two branches and the turnaround region is not formally
defined by the method.

The results for the central \HII\ regions are equally in
agreement for the mean abundance, but with a larger 
scatter in the three estimators. 
R$_{23}$ gives an average 12 + log(O/H) = 8.69 with s.d.=0.42,
P gives 12 + log(O/H) = 8.43 with s.d.=0.38. The N2 method 
gives an average 12 + log(O/H) = 8.73 with s.d.=0.15 and O3N2 
gives  12 + log(O/H) = 8.66 with s.d.=0.17.

When the whole of the distribution is considered then the 
agreement between indicators seems to hold only for the 
brightest parts of the region and break badly for the spaxels 
at the borders, especially for the R$_{23}$ and P methods.
The uniform abundance assumption is  strongly supported by 
O3N2 and N2, making these methods the preferred ones
for abundance estimation for our data.

The P method for IC 132 is actually
producing two separate distributions because the 
low and high abundance branches are selected for the 
same region. This is also the case for R$_{23}$.  If 
R$_{23}$ is restricted to only one branch the abundance would
be more uniform.  This brings up the question about the validity of 
applying a branch selection for each spaxel, maybe in this case
the selection by the [\NII]/[\OII] criterion only applies to the
integrated region. 
Another point to consider about empirical indicators is their
sensitivity to the ionization parameter. From the maps it 
seems that O3N2 gives low dispersion because it can effectively 
cancel out this effect. On the other hand R$_{23}$ shows high 
values especially in the lower left corner of IC 132 where
we suspect that the diffuse ionized gas is dominant. The 
validity of applying R$_{23}$ in such conditions is questioned.

Similar results are found for the sulphur abundance. Using the 
S$_{23}$ we find for IC 132 a sulphur abundance of 12 + log(S/H) = 6.79 
with s.d.=0.28 and for the central \HII\ regions  12 + log(S/H) = 7.02 
with s.d.=0.34 

Regarding oxygen abundance, S$_{23}$ also shows a relatively small scatter 
albeit higher than that of the oxygen abundance estimates using N2 
and O3N2.

When comparing direct abundance estimated with strong line 
methods, and using global measurements i.e. including 
the whole of the \HII\ regions, \cite{kenbregar03} found that  
(O/H)$_{R_{23}}$ is always higher 
than (O/H)$_{Te}$ while the (O/H)$_{S_{23}}$ method gives results in 
agreement with (O/H)$_{Te}$ except for objects with high abundance 
values (12+log(O/H)$_{Te} > $ 8.5) where (O/H)$_{S_{23}}$ is also larger 
than (O/H)$_{Te}$. \cite{ssska02} noted this systematic overestimates 
by strong line methods, pointing out as possible causes the few number 
of objects with high quality \Te\ determination and offsets between
the temperatures assumed in the photoionization models and the 
temperatures observed. 

For N2, comparing  the \cite{denicoloetal02} calibration with direct 
(O/H) determinations  \cite{permodiaz05} found that (O/H)$_{N2}$ 
overestimates abundances for low metallicity objects and 
underestimates them  at high metallicity with a turnover at around 
12 + log(O/H)$_{Te} \sim$ 8.0.

The selection of the empirical abundance indicator to use is 
primarily dictated by the available lines and their S/N, 
however in our case various indicators are available that 
produce considerable differences between the estimated 
abundance. The question of what empirical indicator can be 
trusted is not a simple one to answer. As the (O/H) calibrations 
are based on a set of objects with oxygen abundance derived 
with the \Te\ method, the reliability of the relation depends 
on the quality, homogeneity  and distribution of the reference 
objects. Ideally to construct an empirical estimator the reference 
sample should cover all the abundance range, in practice the  
reference is generally weighted towards low or high abundance 
objects and with gaps at certain abundances. This situation is 
partially alleviated by the use of photoionization models, 
nevertheless its validity at different metallicities depends
on our knowledge of transitional probabilities and collisional 
strengths for the different species.

A warning may be raised about the use of empirical estimators 
for individual spaxels given that the methodology was developed
for global observations. The comparison of the empirical abundance
maps with the integrated zones shows in general good statistical 
agreement, and this may indicate that the estimators can be used 
for spatially resolved observations, some with more confidence 
(O3N2,N2) than others (R$_{23}$, P). Nevertheless determining 
the range of physical conditions where each empirical method 
is valid requires further investigation.


\section{Diagnostic diagrams in 2D}

When interpreting emission-line spectra, it is important to 
be able to distinguish emission produced by star-forming regions 
from other sources such as planetary nebulae, supernova remnants 
or even an active galactic nucleus. The conventional means for 
quantitatively classifying emission line objects and 
distinguishing between gas ionized by stars or by non-thermal processes 
are diagnostic diagrams \citep[BPT,][]{bpt81,veilloster87} 
involving the ratios of the strongest emission lines in the 
optical spectra.
This traditional approach involves the comparison 
of  the integrated emission line ratios with global photoionization 
models leaving open the question of the validity of using these 
diagnostic diagrams with spatially resolved spectroscopic data.

To search for systematic differences between the inner and 
outer \HII\ regions  as well as detecting regions with line ratios 
indicative of non stellar ionization we computed the 
[\OIII]$\lambda$ 5007/\Hb, [\NII] $\lambda$ 6584/\Ha, 
[\SII]$\lambda$ 6717+6731/\Ha, 
[\OII]$\lambda$ 3727/[\OIII]$\lambda$ 5007 and 
[\OII]$\lambda$ 3727/\Hb\ line ratios both 
spaxel by spaxel and the integrated values, to produce the 
distribution maps and investigate the emission line diagnostic 
diagrams and the excitation structure within the regions. 

\begin{landscape}
\begin{figure}
\centering
\begin{tabular}{ccc}
\includegraphics[width=6.0cm]{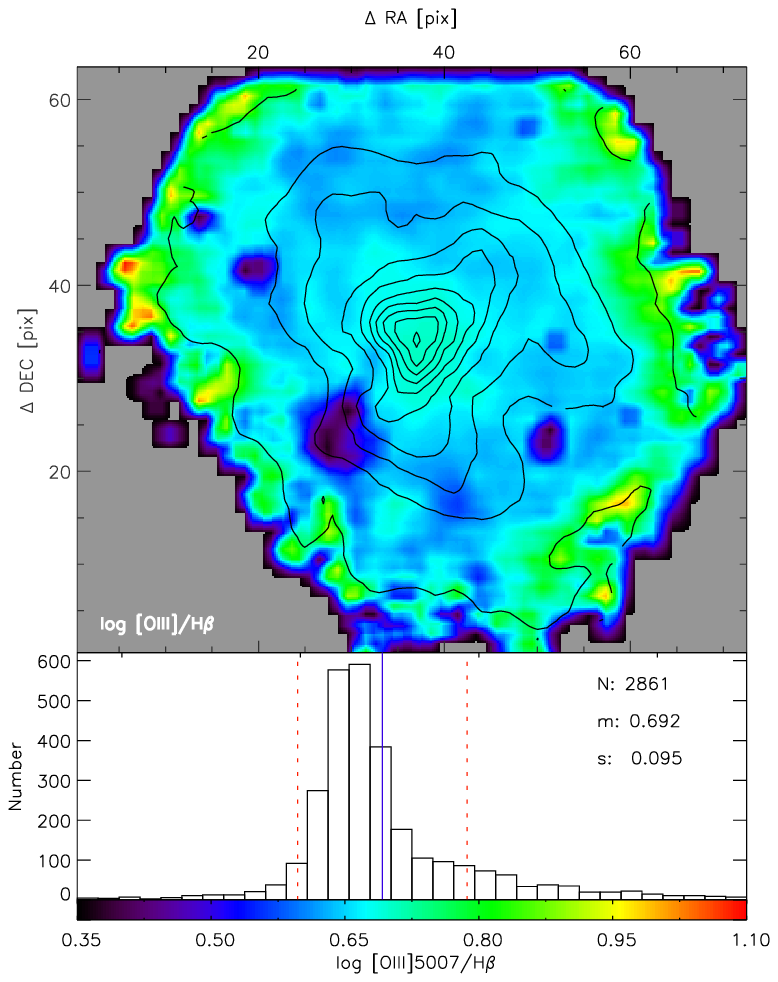} &
\includegraphics[width=6.0cm]{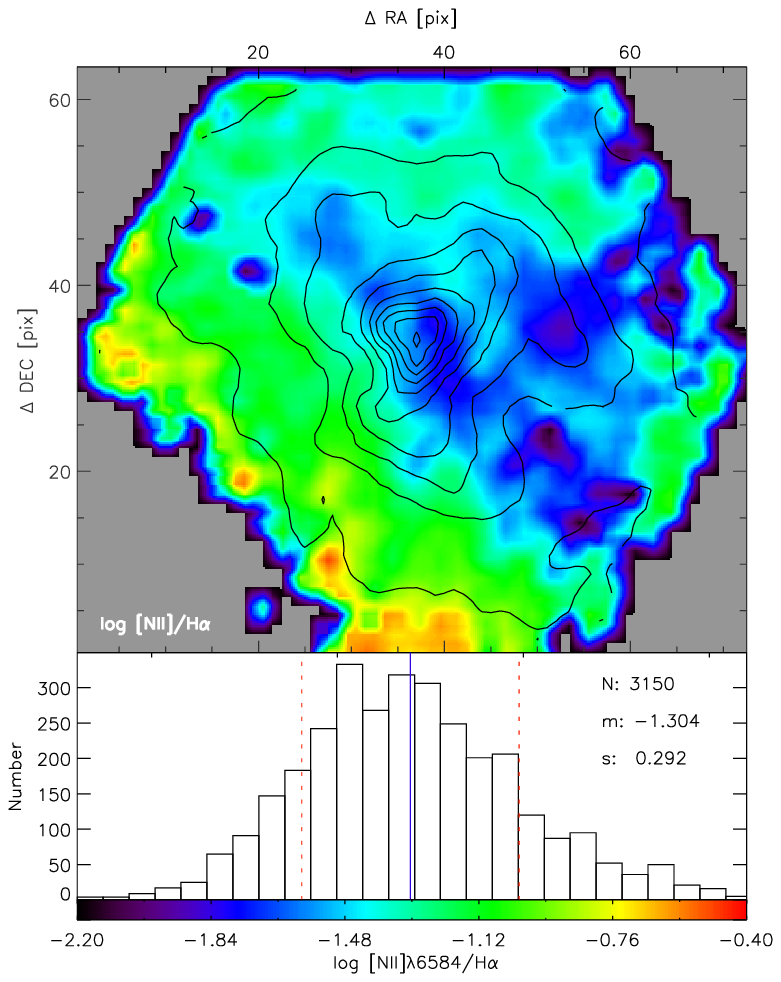} &
\includegraphics[width=6.0cm]{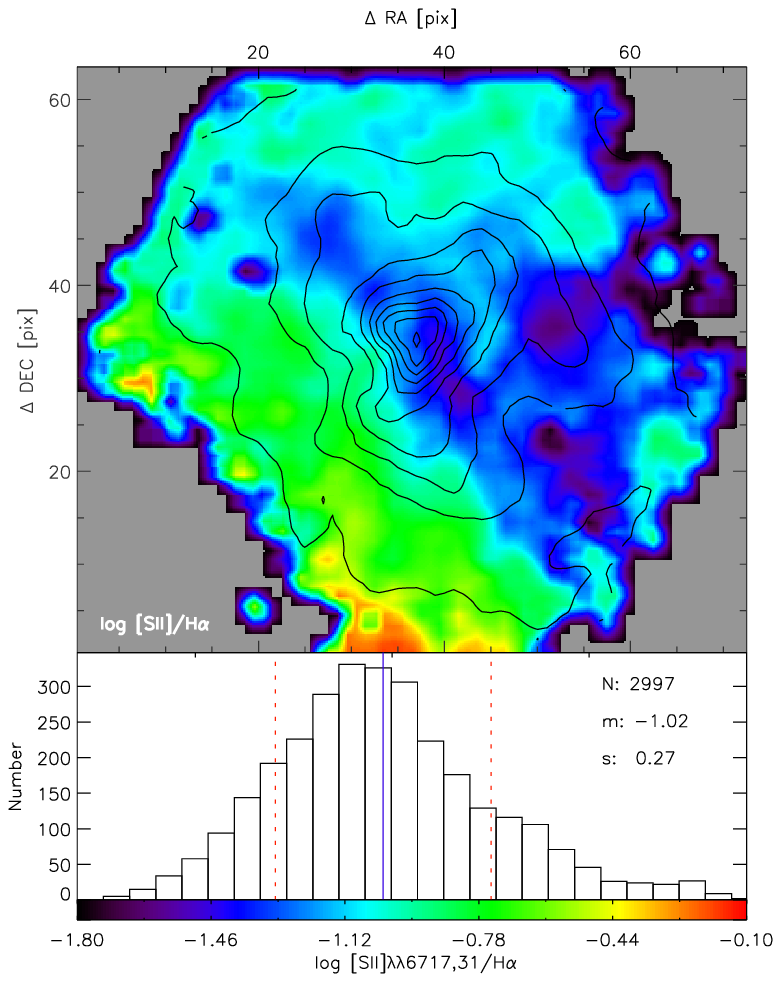} \\
\includegraphics[width=6.0cm]{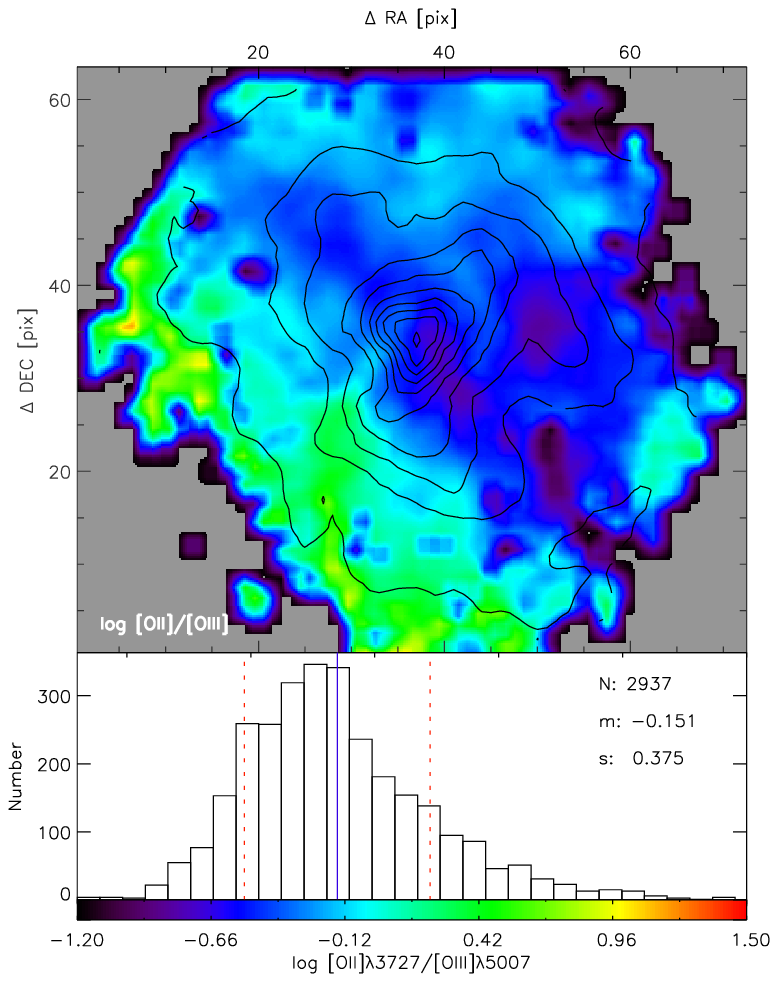} &
\includegraphics[width=6.0cm]{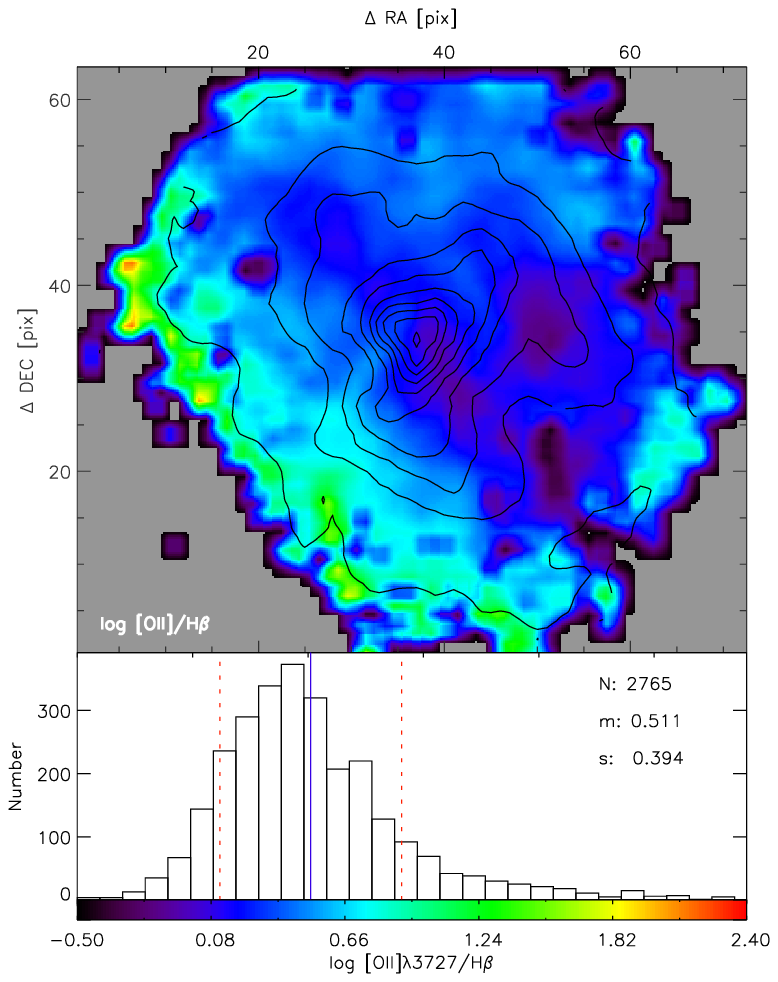} &
\ \\
\end{tabular}
\caption{Top: IC132 maps of the diagnostic line ratios 
[\OIII] $\lambda 5007$/\Hb, [\NII] $\lambda$6584/\Ha
and [\SII] $\lambda\lambda$ 6717,31/\Ha. Bottom:
[\OII]$\lambda$3727/[\OIII]$\lambda$5007 and 
[\OII]$\lambda$3727/\Hb}
\label{f33ionizationratiosic132}
\end{figure}
\end{landscape}

\begin{landscape}
\begin{figure}
\centering
\begin{tabular}{ccc}
\includegraphics[width=6cm]{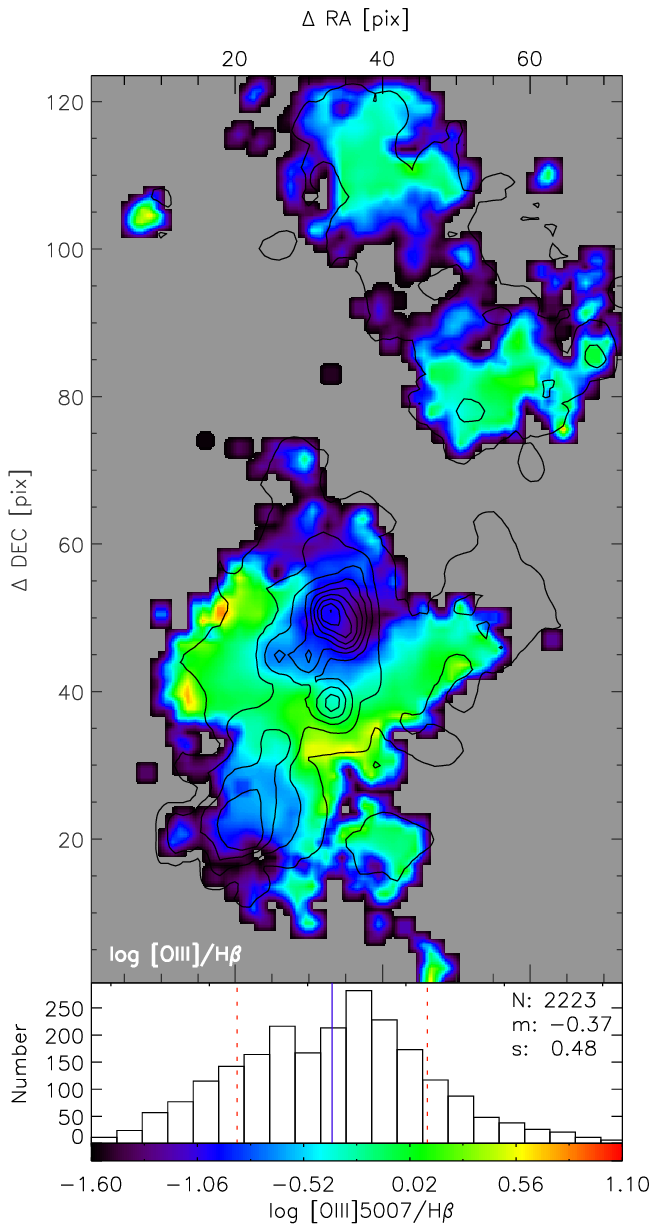} &
\includegraphics[width=6cm]{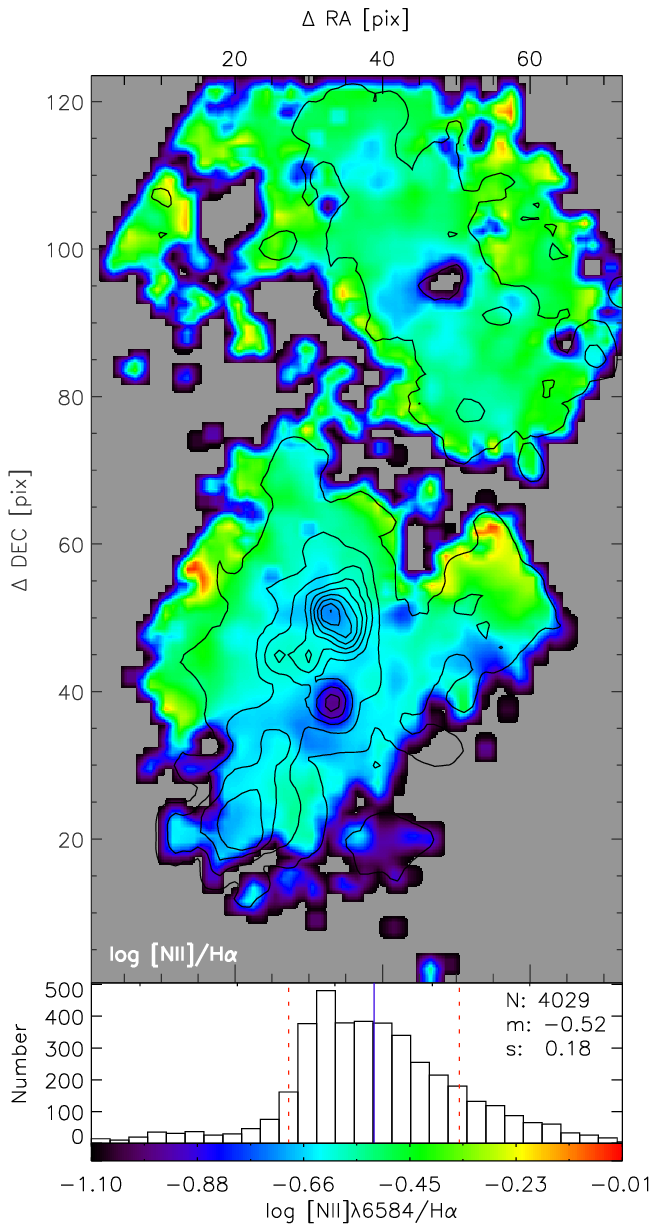} &
\includegraphics[width=6cm]{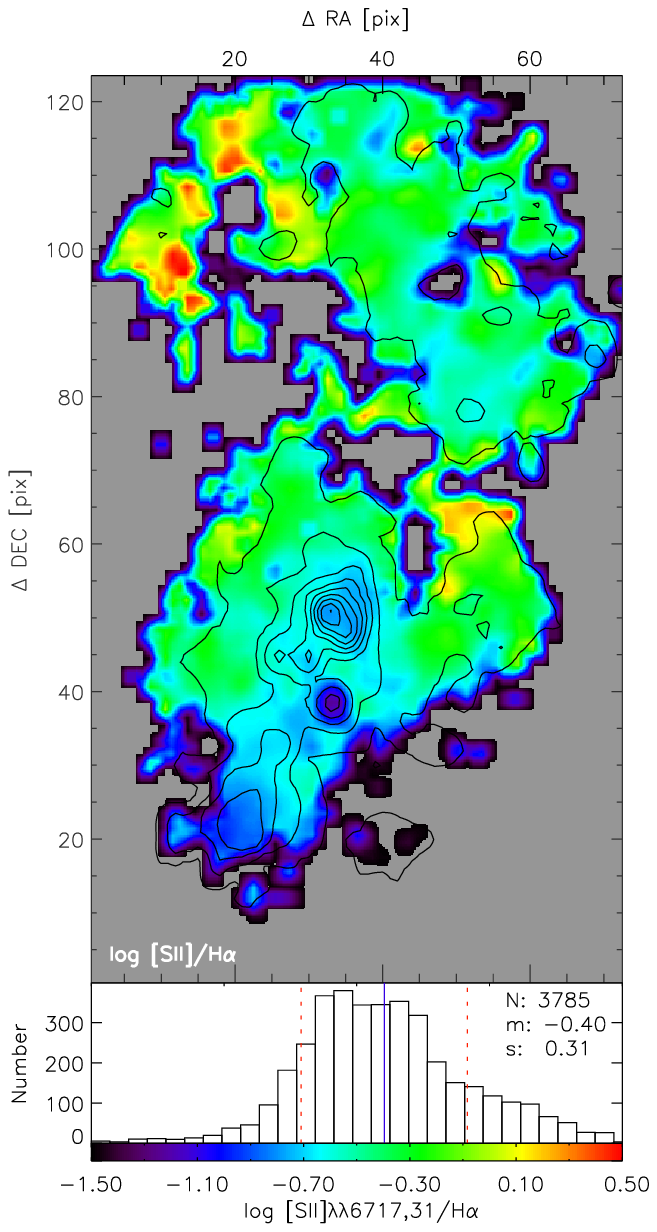} \\
\end{tabular}
\caption{Diagnostic line ratios for the central regions as labelled.}.
\label{f33ionizationratioscore}
\end{figure}
\end{landscape}

\begin{figure}
\centering
\begin{tabular}{cc}
\includegraphics[width=6cm]{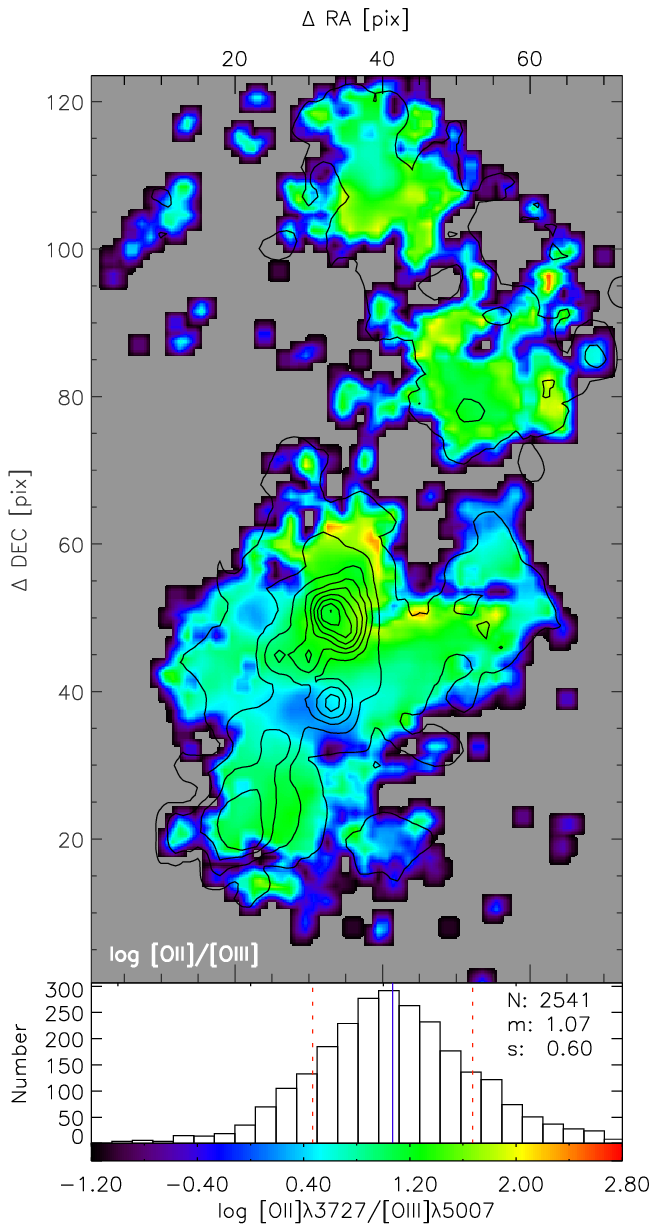} &
\includegraphics[width=6cm]{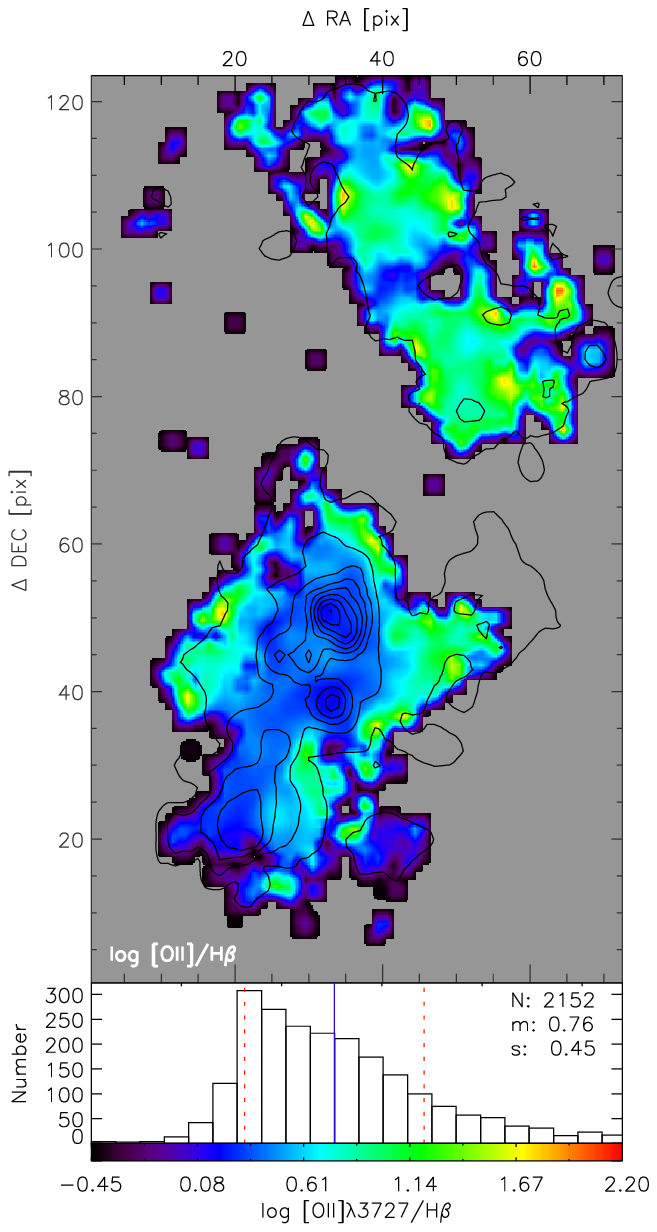} \\
\end{tabular}
\caption{Ionization sensitive ratios [\OII]$\lambda$3727/[\OIII]$\lambda$5007 
and [\OII]$\lambda$3727/\Hb\ for the central region.}
\label{f33ionizationratioscoreoii}
\end{figure}

Given the large ionization potential of [\OIII]  (54.9 eV) the 
[\OIII]$\lambda$ 5007/\Hb\ ratio traces highly ionized gas, 
and grows directly proportional to the degree of ionization. 
On the other hand because of the smaller ionization potential 
of both [\NII] and [\SII] (29.6 and 23.4 eV respectively) their 
line intensities trace the low ionization zones. These line ratio 
maps are shown in figure \ref{f33ionizationratiosic132} for 
IC~132 and in figures \ref{f33ionizationratioscore} and 
\ref{f33ionizationratioscoreoii} for the central regions. 
For IC 132 the ``wall"  in the lower left
part is present in most of the plots and is the strongest zone
in the low ionization indicators maps ([\NII] $\lambda$ 6584/\Ha, 
[\SII]$\lambda$ 6717+6731/\Ha), pointing to the existence
of diffuse ionized gas in this area.

The \Ha\ emission, being a tracer of star formation,  can be expected to be 
spatially related to the position of the ionizing stars (this is 
true also for the highly ionized gas).  Although the gas density 
and kinematics may play a role, in a first order approximation 
the maxima of \Ha\ emission should follow the 
high ionization zones in the ionization ratio maps. From the 
[\OIII]/\Hb\ ratio maps it is possible to see that this is true 
for the brightest part of IC~132  but not for the \HII\ regions in 
the central regions of M33. In particular the behaviour of the 
region BCLMP 93 is just the opposite, the brightest \Ha\ peak 
coincides with a low ionization region and  is partially 
surrounded by higher ionization gas.

\begin{landscape}
\begin{figure}
\begin{tabular}{cc}
\includegraphics[width=11cm]{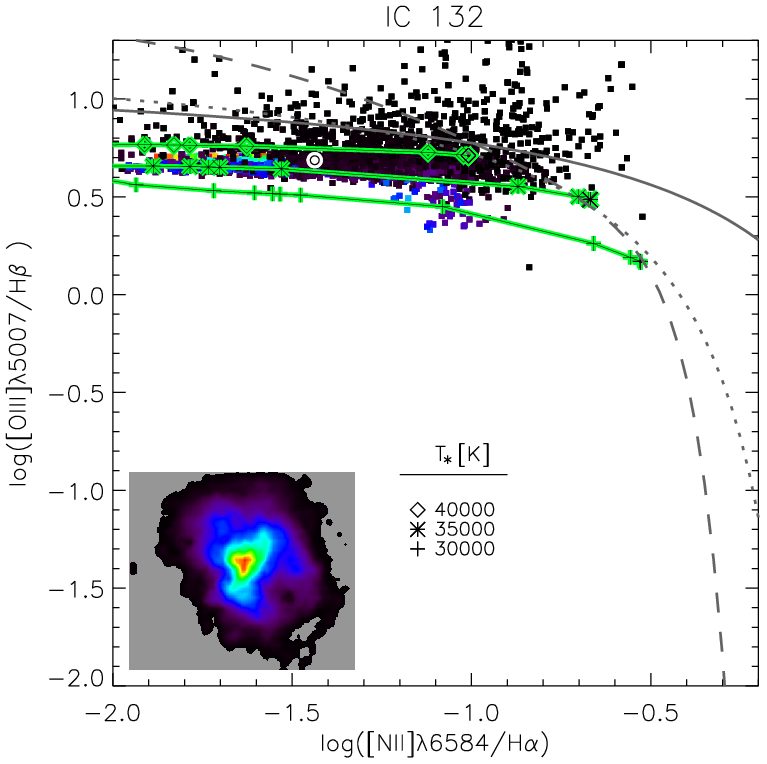} &
\includegraphics[width=11cm]{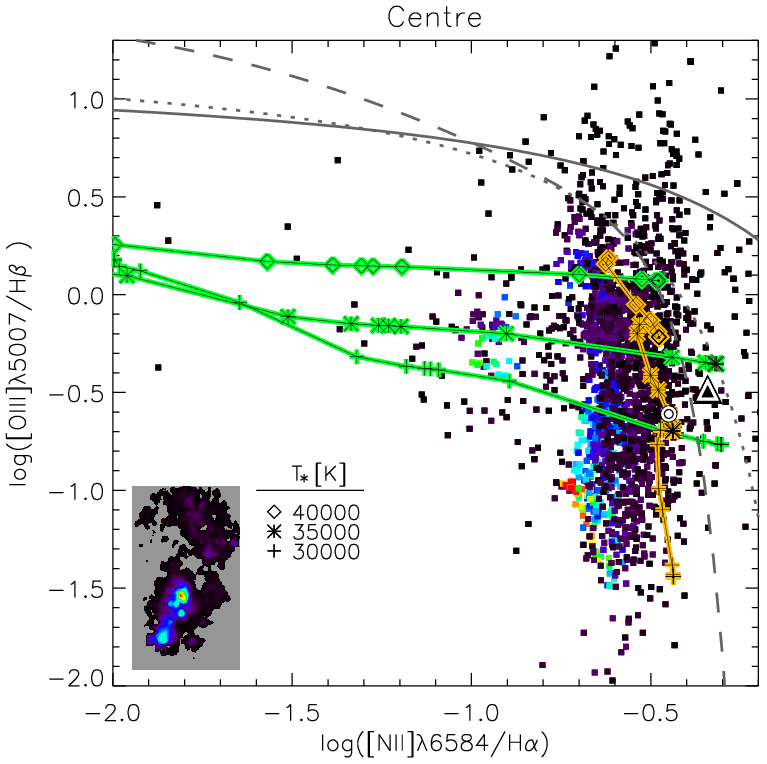} \\
\end{tabular}
\caption{BPT diagnostic diagrams for each spaxel in the external zone
IC~132 (left) and the central zones of M33 (right).Different boundaries 
for photoionization are represented: continuous line from 
\protect\cite{kewdop02}, dotted line  from \protect\cite{kauffea03}  and 
dashed line from \protect\cite{sskaea06}.  The colours of individual 
squares indicate the spatial position according to the inserted map of the 
regions. Black and white concentric circles are the position of the 
integrated field for IC~132 and BCLMP~93. 
Symbols connected by coloured lines are the results from photoionization 
models computed at increasing distance from the ionization source, 
where the symbols closest to the boundary lines are the final result of the 
computation. Three temperatures are given for a combination of 
log U and abundance. For IC 132 the best fit was found with models running
horizontally in the diagram (green lines), while the central zone 
was fitted not only by horizontal but specially by vertical (orange lines) models.
The triangle symbol in the diagram for the centre indicates the 
emission line zone closer to the strong X-ray source (see text).
}
\label{f33bptoiiinii}
\end{figure}
\end{landscape}

Diagnostic  BPT diagrams 
were created from the 
ionization and excitation sensitive line ratio maps shown before. 
In Fig.~\ref{f33bptoiiinii} the
[\NII] $\lambda$6584/\Ha\ vs [\OIII] $\lambda$5007/\Hb\
BPT  diagrams are shown. The individual spaxels are 
plotted in the diagrams, coloured according to the \Ha\ intensity, 
the insert shows the map for reference. Boundary starburst model lines are 
shown as follows: the solid line is that  by \cite{kewdop02}, the dotted
line is from \cite{kauffea03} and the dashed line is from \cite{sskaea06}. 
The location of the integrated spectra is indicated in black and
white concentric circles.

The main and obvious result from an inspection of  Fig.~\ref{f33bptoiiinii} 
is that there is an almost complete dichotomy in the distribution of points 
between the central and outer \HII\ regions. While most of the line ratios of 
the central region are vertically distributed on the right of the diagram, all 
of the points from IC~132 are located horizontally in the upper left part of the 
diagram. The regions occupied by the distributions have hardly any superposition, 
in fact, they are perpendicular to each other, but  in both regions there are 
spaxels that clearly move away from the \HII\ region zone of the diagnostic 
diagram intruding in the AGN zone, mostly the Seyfert zone in the case of 
IC 132 and the LINER and transition zone for  the M33 central region.
Curiously, no anomalous line ratios are detected for the position of individual 
spaxels  near the M33-X8 X-ray source in the central region. The position of the
emission line zone nearest to the X-ray source is indicated with a triangle 
symbol in the left panel of figure ~\ref{f33bptoiiinii}.

\subsection{Matching the diagnostic diagrams to pseudo-3D photoionization models}

Bearing in mind that we are dealing with spatially resolved data and in an 
effort to understand the dichotomy in the distribution of individual spaxel 
line ratios in the BPT diagrams shown in  Fig.~\ref{f33bptoiiinii},  we have 
computed what can be called pseudo-3D photoionization models. Our  grid of 
models  have spherical closed geometry, constant density with a black body
as the ionizing source and a range of  ionization parameters 
(log U) from -2.0 to 0.0, metallicities (Z) from 0.1 to 2.0Z$_\odot$
and stellar temperature (T$_*$) from 30,000 to 50,000 K. As a first approach 
to full 3D models we have used all intermediate shell results from each 
Cloudy \citep{cloudy90} photoionization model to represent the radial behaviour 
of the relevant gas parameters inside the nebula.The models are discussed in 
detail in a forthcoming paper (L\'{o}pez et al. in preparation).

Fig.~\ref{f33bptoiiinii} shows in green-black solid lines the results from the 
photoionization models. Each line corresponds to a model with a 
given combination of log U, Z and  T$_*$,  following the computed
results at increasing distance from the inner face of the cloud. The 
end or outer edge of the computation is marked with the corresponding 
symbol in bold and is generally the one closest to the boundary lines.  

It is possible to see in Fig.~\ref{f33bptoiiinii} that the models are capable 
of reproducing the position and spread of values of both regions even 
though the spaxels in IC~132 show a horizontal dispersion while those of the central zones have
a vertical dispersion.  In particular, the models that best fit the data for 
IC~132 have log(U)=-1, T$_*$= 35,000 and 40,000\degree K and 
Z=0.1Z$_\odot$. T$_*$= 35,000  fits
the high brightness spaxels and T$_*$=40,000 lies above these spaxels
but fits better the low brightness ones. T$_*$=30,000 is below most of the
spaxels but fits a small spur of objects around [\NII]/\Ha =-1.05.

For the central zone, models with the same temperatures (3.0, 3.5 and 4.0
kK), log(U)=-1 and Z=2.0Z$_\odot$ span the location of most of the individual
spaxels. Another set of models with the same T$_*$, but log(U)=-2 and 
Z=0.1Z$_\odot$ also fit the observations but with vertical spread; these
are the orange broken lines in the right panel of figure \ref{f33bptoiiinii}.
These vertical models only fit the rightmost distribution of central spaxels,
corresponding to the faint ones.  The bright spaxels fall on the lower left
tip of the distribution cloud and could not be fitted by our set of models. 
However the integrated position  for BCLMP 93 (concentric circles in the figure)
is well fitted by both sets of models. 

For the same temperatures, the position of the horizontal models
is controlled by the metallicity, the higher it is the lower the 
[\OIII]5007/\Hb\ ratio. The orientation of the models is controlled
by the ionization parameter, being vertical for low values of U.

For the central zone we favour the vertical models as indicators of the
physical conditions in the nebula, mainly because they are consistent with the
high metallicity expected in this region, and the horizontal 
models instead would represent regions of  low metallicity.

As mentioned above while the spaxels with high \Ha\ surface brightness 
are inside the boundary starburst line of the BPT diagram, many of the 
low \Ha\ surface brightness  spaxels that are further away from the ionizing 
source appear outside the boundary; this obviously doesn't mean that an 
AGN exists. Possible explanations are that those pixels have low S/N or 
perhaps diffuse radiation has to be considered. In any case, this raises 
a question about the wisdom of using photoionization models computed 
for integrated systems to define a reference boundary for photoionization 
and apply it to diagnostic diagrams for resolved systems.

\section{Wolf-Rayet stars}
\label{secWR}
Wolf-Rayet (WR) stars are hot and luminous, with broad emission 
lines in the optical range due to a considerable mass loss through 
winds \citep{abbottconti87}.  They represent an evolved stage of O stars 
when the radiation pressure can not be counterbalanced by gravity 
and the outer layers of the stars are blown away.  WR more common 
subtypes are the  WN that show a prominent nitrogen emission  in 
the 4600-4720 \AA~range, known as the blue bump (BB), and the 
WC, with carbon emission in the 5750-5870 \AA~range, known as 
the red bump (RB).  The [\HeII]$\lambda$4686\AA\ emission line 
is present in both types. 

If Wolf-Rayet features are detected then stars more massive than 
M$_{WR} \sim $ 25M${\odot}$ must be present in the cluster if solar metallicity 
is presumed \citep{pindaoetal02}. Assuming an instantaneous 
burst, the WR features are  visible from roughly 3 Myr to almost 
7 Mys of age, their lifetimes getting shorter with lower metallicity 
\citep{sv98}.  The WR feature can serve as a constraint to estimate  the age of the 
burst if its metallicity is known.

\cite{dodorbenv83} studied the integrated UV (IUE) and optical 
(ESO 3.6m) spectrum of IC 132. They found the presence of the blue 
bump ($\lambda$ 4686), indicating the existence of WR stars, 
which they classified as WN4. They estimated the ionizing source 
as a single star with T$_*$=40,000$\pm$5,000\degree K and mass 
larger than 100$M{\odot}$ or a system of 3 components 
(O4V+O9.5I+WN4). They noted the need for spatial resolution to 
determine the extent and nature of the WR stars (few over luminous 
objects vs. many normal WR). 

\begin{landscape}
\begin{figure}
\centering
\includegraphics[width=24cm]{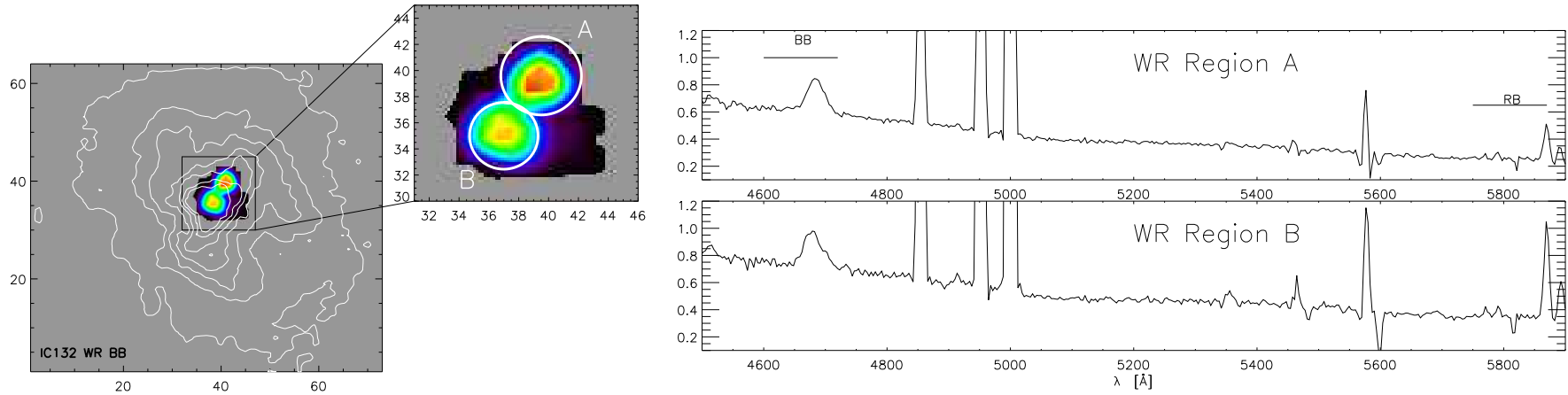}
\caption{Wolf-Rayet detection in IC 132. the left panel shows the 
distribution of the BB with the H$\alpha$ isocontours over plotted. 
The central panel is a zoomed-in version. The right panel shows 
the integrated spectrum for regions A and B. The position of the 
blue and red bumps is indicated. No red bump is detected.}
\label{fg_wric}
\end{figure}
\end{landscape}

We used the IFS observations to map the spatial distribution 
of the WR features by measuring the detected BB in each individual 
spectrum and reconstructing the spatial map in the same 
fashion as the emission line maps were constructed. We 
detected two peaks in the BB distribution in IC~132.  
Fig.~\ref{fg_wric} shows the map of the blue feature flux 
measured by fitting a gaussian in a 70\AA\ interval centred 
at 4686\AA, with the \Ha\ contour overlaid. Also shown is a 
zoom of the BB emission in the region where we detect two 
peaks and the integrated spectra of both regions. The BB 
luminosity, equivalent width and flux normalized to \Hb\ 
are shown in table \ref{tbl_wric}. Assuming  an average 
luminosity for the BB of 10$^{36.5}$  
\cite[for a single WN7 star]{pindaoetal02} the estimated 
number of WN stars in region A is 21 and in region B is 24. 
Taking the total number of Lyman photons estimated by 
\cite{dodorbenv83} N$_{Ly}$ = 8.2 $\times$ 10$^{49}$s$^{-1}$  
and assuming that a single WR produces 1.7 $\times$ 
10$^{49}$s$^{-1}$  \cite[WN]{sv98}, the derived number of 
stars is 5 for the whole region. It is interesting to note the 
discrepancy found in the number of WR computed by different 
methods. This discrepancy may be linked to aperture effects in the observations.

\begin{table}
\centering
\caption{Blue bump luminosity, EW and BB/H$\beta$ for the 
two regions identified in IC 132.}
\label{tbl_wric}
\begin{tabular}{cccc}
region & log(L) & EW(BB) &  BB/H$\beta$ \\
\ & erg s$^{-1}$ & \AA & \ \\
\hline
A & 37.84 & 15 & 0.38 \\
B & 37.89 & 13 & 0.41  \\
\hline
\end{tabular}
\end{table}

Given that the models of WR formation predict that as a result 
of the increased  mass loss rates the number of WR will increase 
with increasing metallicity \citep{sv98,meinetmaeder05WRIX},
it is surprising that no WR features are detected in the higher 
metallicity central region of M 33.
One has to consider that the 
spectrum of the central region BCLMP 93 is of lower S/N than that 
of IC~132 which makes more difficult the detection and measurement of the WR 
bumps. Nevertheless, a limit to the number of WR stars in BCLMP~93 
that may be present but can not be confidently detected can be 
obtained by estimating the blue bump with different continuum 
placements. In this way we find that  BCLMP 93 could contain at most 4
WN7 stars.

\section{Segmentation analysis}
\label{sec_integrated_properties}
Although the intensity maps provide a view of the spatial distribution 
of the emission lines flux, temperatures, abundances and other estimated 
values, an easier quantitative representation of variations with position 
as well as a study of the effect of aperture size is provided by analyzing 
the \HII\ region in concentric layers. To this end we have adopted an approach 
that defines segments or shells of the \HII\ region according to the \Ha\ 
surface brightness level. The segmentation of the observed FOV allows
the study of the faint areas where the spaxel by spaxel analysis can not be
performed  given the low S/N of individual elements. However when the flux 
is integrated in a single spectrum the S/N improves and conditions can 
be estimated, albeit with the loss of spatial  azimuthal resolution. This kind
of analysis is a clear advantage of IFS over long slit observations for the
study of faint areas.

We present the results in two forms: 

I - The individual shell values, i.e.
considering each section as an individual region, and 

II - The cumulative values, i.e. adding the sections in a 
cumulative way from the centre to the outer boundary, going from 
$\overline{{\rm AB}}$=A+B to $\overline{{\rm AJ}}$=A+B+ $\ldots$ +I+J.

Therefore in this notation A is the spectrum of the brightest central
spaxels, while $\overline{{\rm AJ}}$ represents the integrated 
spectrum of the whole region (see figure \ref{fishells} for the segmentation 
of IC~132). The same notation is used for BCLMP~93 but with lowercase
letters going from a to h (see figure \ref{fc93shells}).

\subsection{Integrated properties of IC~132}
\label{sec_integrated_properties_IC}

\begin{figure}
\centering
\includegraphics[width=6.0cm]{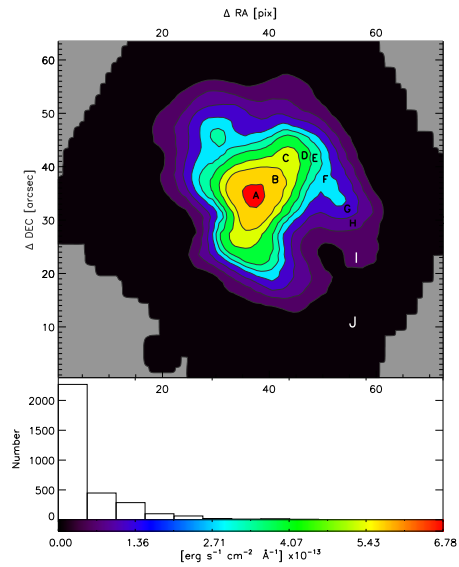}
\caption{IC 132 \Ha\ map with concentric apertures. The single letter 
(A through J) designates the shells area.}
\label{fishells}
\end{figure}

To analyze the radial properties of IC~132 the field was segmented in 
10 sections of equal \Ha\ surface brightness covering from the brightest 
to the faintest spaxels.  The shells were labeled from A to J as shown in 
Fig. ~\ref{fishells}.  The segment sizes were defined to have constant 
integrated flux in \Ha\ after applying the extinction correction.

As can be seen from tables 4 and 8 the integrated flux of \Hb\ is roughly 
the same for all sections in both regions, i.e. about 80$\times$10$^{-13}$ 
erg s$^{-1}$ cm$^{-2}$  for IC 132 and about 20$\times$10$^{-13}$ erg 
s$^{-1}$ cm$^{-2}$  for BCLMP 93; small variations are present because of 
the discrete jumps in flux between spaxels and no interpolation was used 
to define exactly the isocontours.

The emission line measurements, their errors, and the resulting line ratios, 
temperatures and abundances, are given in tables 4 to 7 for the individual 
and cumulative spectra respectively. 
 
\begin{landscape}
\begin{table}
\centering
\caption{Reddening corrected integrated fluxes for IC~132 individual shells. 
Fluxes are normalized to 
\Hb\ = 100}
\label{ticfluxesrings}
\resizebox{24cm}{!}{   
\begin{tabular}{clr@{}lr@{}lr@{}lr@{}lr@{}lr@{}lr@{}lr@{}lr@{}lr@{}l}
\hline
\ & \ &  \multicolumn{20}{c}{Individual} \\
\cline{3-22}
$\lambda$ & Ion & \multicolumn{2}{c}{A}  &  \multicolumn{2}{c}{B}  &
\multicolumn{2}{c}{C}  &  \multicolumn{2}{c}{D}  &  \multicolumn{2}{c}{E}   &
\multicolumn{2}{c}{F}  &  \multicolumn{2}{c}{G}   &  \multicolumn{2}{c}{H}  &
\multicolumn{2}{c}{I}   &  \multicolumn{2}{c}{J} \\
\hline
3727 &	[O II]	& 150.7 	& $\pm$8.0 & 	164.7 	& $\pm$8.9 &   	192.4 	& $\pm$11.7 &   	210.5 	& $\pm$15.7 &   	204.5 	& $\pm$16.5 &   	188.9 	& $\pm$17.6 &   	221.6 	& $\pm$21.0 &   	241.3 	& $\pm$27.7 &   	267.2 	& $\pm$44.8 &   	634.6 	& $\pm$282.6 \\  
4010 &	H$\delta$	& 33.5 	& $\pm$2.4 & 	31.5 	& $\pm$2.6 &   	31.9 	& $\pm$3.1 &   	32.0 	& $\pm$3.9 &   	31.5 	& $\pm$3.9 &   	31.0 	& $\pm$4.0 &   	30.5 	& $\pm$4.8 &   	29.5 	& $\pm$7.0 &   	28.4 	& $\pm$10.8 &   	24.6 	& $\pm$54.3 \\  
4340 &	H$\gamma$	& 50.0 	& $\pm$2.8 & 	48.5 	& $\pm$2.6 &   	49.2 	& $\pm$3.8 &   	49.3 	& $\pm$5.3 &   	48.2 	& $\pm$5.6 &   	47.7 	& $\pm$6.7 &   	47.1 	& $\pm$7.8 &   	45.7 	& $\pm$10.0 &   	43.0 	& $\pm$15.1 &   	31.4 	& $\pm$52.5 \\  
4363 &	[O III]$^1$	& 6.7 	& $\pm$1.2 & 	6.4 	& $\pm$1.3 &   	7.7 	& $\pm$1.7 &   	8.4 	& $\pm$2.3 &   	10.7 	& $\pm$2.7 &   	11.2 	& $\pm$3.3 &   	14.0 	& $\pm$4.3 &   	18.9 	& $\pm$6.3 &   	29.0 	& $\pm$11.6 &   	127.0 	& $\pm$67.6 \\  
4861 &	[H$\beta$]	& 100.0 	& $\pm$3.1 & 	100.0 	& $\pm$4.1 &   	100.0 	& $\pm$5.7 &   	100.0 	& $\pm$7.9 &   	100.0 	& $\pm$7.9 &   	100.0 	& $\pm$8.6 &   	100.0 	& $\pm$9.6 &   	100.0 	& $\pm$12.1 &   	100.0 	& $\pm$18.5 &   	100.0 	& $\pm$48.2 \\  
4959 &	[O III]	& 175.1 	& $\pm$6.4 & 	171.4 	& $\pm$7.3 &   	164.7 	& $\pm$8.0 &   	159.6 	& $\pm$11.8 &   	157.8 	& $\pm$11.6 &   	158.0 	& $\pm$12.3 &   	154.7 	& $\pm$13.4 &   	155.4 	& $\pm$17.4 &   	160.9 	& $\pm$27.5 &   	208.3 	& $\pm$85.4 \\  
5007 &	[O III]	& 500.1 	& $\pm$12.2 & 	485.6 	& $\pm$15.4 &   	463.5 	& $\pm$20.3 &   	449.1 	& $\pm$27.8 &   	443.5 	& $\pm$28.1 &   	445.1 	& $\pm$30.8 &   	435.0 	& $\pm$33.3 &   	433.3 	& $\pm$42.4 &   	444.6 	& $\pm$66.5 &   	576.1 	& $\pm$209.9 \\  
6312 &	[S III]	& 2.7 	& $\pm$1.3 & 	3.4 	& $\pm$2.1 &   	3.9 	& $\pm$2.7 &   	4.6 	& $\pm$3.4 &   	4.9 	& $\pm$4.0 &   	5.5 	& $\pm$4.7 &   	6.1 	& $\pm$5.8 &   	7.4 	& $\pm$7.9 &   	12.0 	& $\pm$13.8 &   	51.9 	& $\pm$60.3 \\  
6548 &	[N II]	& 2.9 	& $\pm$1.1 & 	2.8 	& $\pm$1.2 &   	3.2 	& $\pm$1.3 &   	3.4 	& $\pm$1.8 &   	3.5 	& $\pm$2.2 &   	3.5 	& $\pm$2.1 &   	4.1 	& $\pm$2.7 &   	4.7 	& $\pm$3.4 &   	6.7 	& $\pm$4.4 &   	16.4 	& $\pm$14.8 \\  
6564 &	H$\alpha$	& 284.5 	& $\pm$7.1 & 	284.1 	& $\pm$8.8 &   	284.3 	& $\pm$12.2 &   	284.2 	& $\pm$17.8 &   	284.8 	& $\pm$18.8 &   	284.5 	& $\pm$19.5 &   	284.5 	& $\pm$22.5 &   	284.8 	& $\pm$28.0 &   	285.8 	& $\pm$40.0 &   	317.8 	& $\pm$112.4 \\  
6584 &	[N II]	& 6.7 	& $\pm$0.8 & 	7.7 	& $\pm$0.9 &   	9.1 	& $\pm$1.2 &   	9.8 	& $\pm$2.2 &   	9.6 	& $\pm$2.7 &   	9.1 	& $\pm$2.4 &   	10.1 	& $\pm$3.3 &   	11.3 	& $\pm$4.2 &   	12.7 	& $\pm$5.0 &   	20.0 	& $\pm$15.7 \\  
6717 &	[S II]	& 8.9 	& $\pm$0.8 & 	9.9 	& $\pm$0.9 &   	11.7 	& $\pm$1.2 &   	12.7 	& $\pm$3.2 &   	12.3 	& $\pm$4.2 &   	11.5 	& $\pm$3.0 &   	12.9 	& $\pm$4.6 &   	14.0 	& $\pm$5.4 & 16.1 & $\pm$4.9 & 26.0 & $\pm$16.2  \\  
6731 &	[S II]	& 5.7 	& $\pm$0.4 & 	6.6 	& $\pm$0.7 &   	7.6 	& $\pm$0.9 &   	8.4 	& $\pm$1.5 &   	8.2 	& $\pm$1.7 &   	7.3 	& $\pm$2.0 &   	8.6 	& $\pm$2.5 &   	9.2 	& $\pm$3.4 &  9.5&   $\pm$ 5.5 & 11.6  & $\pm$ 18.8 \\  
7136 &	[Ar III]	& 8.3 	& $\pm$1.0 & 	6.6 	& $\pm$1.2 &   	7.7 	& $\pm$1.4 &   	8.3 	& $\pm$2.9 &   	8.3 	& $\pm$1.8 &   	8.4 	& $\pm$2.2 &   	8.2 	& $\pm$3.0 &   	7.3 	& $\pm$4.8 &   	5.6 	& $\pm$4.3 &   	2.9 	& $\pm$12.6 \\  
9068 &	[\SIII]	& 27.5 	& $\pm$6.3 & 	24.0 	& $\pm$5.4 &   	28.4 	& $\pm$6.5 &   	30.8 	& $\pm$8.2 &   	30.6 	& $\pm$7.6 &   	28.1 	& $\pm$7.5 &   	28.2 	& $\pm$9.3 &   	23.4 	& $\pm$9.9 &   	18.9 	& $\pm$13.2 &   	13.1 	& $\pm$33.5 \\  
\hline
\multicolumn{2}{l}{I(H$\beta$)[{\small erg s$^{-1}$cm$^{-2}$ 10$^{-13}$}]} &  79.29 	& $\pm$0.18 & 	  82.86 	& $\pm$0.24 & 	  79.24 	& $\pm$0.32 & 	  80.21 	& $\pm$0.45 & 	  80.99 	& $\pm$0.45 & 	  80.17 	& $\pm$0.49 & 	  80.52 	& $\pm$0.55 & 	  80.50 	& $\pm$0.69 & 	  80.21 	& $\pm$1.05 & 	  71.20 	& $\pm$2.43 \\ 
\multicolumn{2}{l}{C(H$\beta$)} &  0.50 	& $\pm$0.06 & 	  0.42 	& $\pm$0.04 & 	  0.40 	& $\pm$0.04 & 	  0.40 	& $\pm$0.05 & 	  0.37 	& $\pm$0.05 & 	  0.36 	& $\pm$0.05 & 	  0.36 	& $\pm$0.06 & 	  0.39 	& $\pm$0.06 & 	  0.49 	& $\pm$0.09 & 	  1.30 	& $\pm$0.13 \\ 
\multicolumn{2}{l}{EW(H$\beta$)[\AA]} &  130.34 	& $\pm$11.42 & 	  139.98 	& $\pm$11.83 & 	  194.10 	& $\pm$13.93 & 	  292.21 	& $\pm$17.09 & 	  284.06 	& $\pm$16.85 & 	  266.77 	& $\pm$16.33 & 	  218.40 	& $\pm$14.78 & 	  183.85 	& $\pm$13.56 & 	  132.50 	& $\pm$11.51 & 	  23.46 	& $\pm$4.84 \\ 
\hline
\multicolumn{2}{l}{$^1$Contaminated by mercury street lamps.}
\end{tabular}
}
\end{table}
\end{landscape}

\begin{landscape}
\begin{table}
\caption{Integrated temperatures, abundances and line ratios for IC 132 individual shells.}
\label{ticphysycalabundshells}
\resizebox{22cm}{!}{   
\begin{tabular}{lr@{}lr@{}lr@{}lr@{}lr@{}lr@{}lr@{}lr@{}l}
\hline
\ &  \multicolumn{16}{c}{Individual} \\
\cline{2-17}
\ & \multicolumn{2}{c}{A}   &  \multicolumn{2}{c}{B}  &  \multicolumn{2}{c}{C}  &  
       \multicolumn{2}{c}{D}  &  \multicolumn{2}{c}{E}  &  \multicolumn{2}{c}{F}  &  
       \multicolumn{2}{c}{G}   &  \multicolumn{2}{c}{H} \\
\Te([\OIII])$^a$                     &  12800 	& $\pm$900 & 	  12700 	& $\pm$1000 & 	  13900 	& $\pm$1300 & 	  14600 	& $\pm$1800 & 	  16500 	& $\pm$2100 & 	  16800 	& $\pm$2600 & 	  19200 	& $\pm$3500 & 	  23400 	& $\pm$5700 \\ 
\Te([\OII])$_ {PM03}$   &  11300 	& $\pm$400 & 	  12100 	& $\pm$300 & 	  12800 	& $\pm$400 & 	  13200 	& $\pm$500 & 	  14200 	& $\pm$500 & 	  14400 	& $\pm$600 & 	  15600 	& $\pm$700 & 	  17300 	& $\pm$900 \\ 
\Te([\OII])$_ {G92}$      &  12000 	& $\pm$600 & 	  11900 	& $\pm$700 & 	  12700 	& $\pm$900 & 	  13200 	& $\pm$1300 & 	  14500 	& $\pm$1500 & 	  14800 	& $\pm$1800 & 	  16500 	& $\pm$2400 & 	  19400 	& $\pm$4000 \\ 
\Te([\SIII])$_{6312}$    &  10700 	& $\pm$100 & 	  12800 	& $\pm$200 & 	  12500 	& $\pm$300 & 	  13100 	& $\pm$400 & 	  13500 	& $\pm$500 & 	  15100 	& $\pm$700 & 	  16500 	& $\pm$1000 & 	  22400 	& $\pm$2800 \\ 
\Te([\SIII])$_{SO_{23}}$ &  9300 	    & $\pm$1900 & 	  9600 	& $\pm$1900 & 	  9200 	& $\pm$1700 & 	  9000 	& $\pm$1500 & 	  900 	    & $\pm$1500 & 	  9100 	& $\pm$1600 & 	  9100 	& $\pm$1600 & 	  9600 	& $\pm$1700 \\ 
\Te([\SIII]$_{H06}$       &  12000 	& $\pm$1000 & 	  11900 	& $\pm$1200 & 	  13300 	& $\pm$1600 & 	  14000 	& $\pm$2200 & 	  16400 	& $\pm$2500 & 	  16800 	& $\pm$3100 & 	  19800 	& $\pm$4100 & 	  24700 	& $\pm$6800 \\ 
\hline
12+log(O$^{+}$/H$^{+}$)$^b$ &  7.46 	& $\pm$0.19 & 	  7.50 	& $\pm$0.22 & 	  7.46 	& $\pm$0.26 & 	  7.44 	& $\pm$0.33 & 	  7.29 	& $\pm$0.32 & 	  7.23 	& $\pm$0.39 & 	  7.16 	& $\pm$0.42 & 	  7.01 	& $\pm$0.52 \\ 
12+log(O$^{++}$/H$^{+}$)$^b$ &  7.94 	& $\pm$0.19 & 	  7.93 	& $\pm$0.23 & 	  7.80 	& $\pm$0.26 & 	  7.73 	& $\pm$0.32 & 	  7.60 	& $\pm$0.30 & 	  7.58 	& $\pm$0.36 & 	  7.44 	& $\pm$0.38 & 	  7.27 	& $\pm$0.45 \\ 
12+log(O/H)$^b$ &  8.06 	& $\pm$0.27 & 	  8.07 	& $\pm$0.32 & 	  7.97 	& $\pm$0.37 & 	  7.91 	& $\pm$0.46 & 	  7.77 	& $\pm$0.44 & 	  7.74 	& $\pm$0.53 & 	  7.62 	& $\pm$0.57 & 	  7.46 	& $\pm$0.68 \\ 
12+log(S$^{+}$/H$^{+}$) &  5.49 	& $\pm$0.32 & 	  5.35 	& $\pm$0.36 & 	  5.46 	& $\pm$0.44 & 	  5.46 	& $\pm$0.56 & 	  5.43 	& $\pm$0.66 & 	  5.29 	& $\pm$0.79 & 	  5.29 	& $\pm$1.07 & 	  5.14 	& $\pm$1.73 \\ 
12+log(S$^{++}$/H$^{+}$) &  6.69 	& $\pm$0.18 & 	  6.49 	& $\pm$0.17 & 	  6.58 	& $\pm$0.18 & 	  6.57 	& $\pm$0.22 & 	  6.55 	& $\pm$0.19 & 	  6.45 	& $\pm$0.19 & 	  6.39 	& $\pm$0.26 & 	  6.12 	& $\pm$0.35 \\ 
ICF(S$^+$ + S$^{++}$) &  1.31 	& $\pm$0.08 & 	  1.28 	& $\pm$0.10 & 	  1.22 	& $\pm$0.11 & 	  1.19 	& $\pm$0.13 & 	  1.20 	& $\pm$0.13 & 	  1.22 	& $\pm$0.15 & 	  1.19 	& $\pm$0.16 & 	  1.18 	& $\pm$0.18 \\ 
12+log(S/H) &  6.83 	& $\pm$0.37 & 	  6.63 	& $\pm$0.39 & 	  6.70 	& $\pm$0.47 & 	  6.68 	& $\pm$0.60 & 	  6.66 	& $\pm$0.68 & 	  6.56 	& $\pm$0.81 & 	  6.49 	& $\pm$1.10 & 	  6.24 	& $\pm$1.77 \\ 
log(S/O) &  -1.34 	& $\pm$0.02 & 	  -1.55 	& $\pm$0.02 & 	  -1.35 	& $\pm$0.02 & 	  -1.31 	& $\pm$0.03 & 	  -1.19 	& $\pm$0.03 & 	  -1.27 	& $\pm$0.03 & 	  -1.20 	& $\pm$0.03 & 	  -1.29 	& $\pm$0.04 \\ 
12+log(N$^+$/H$^+$) &  5.92 	& $\pm$0.01 & 	  5.98 	& $\pm$0.01 & 	  5.99 	& $\pm$0.01 & 	  5.98 	& $\pm$0.02 & 	  5.88 	& $\pm$0.03 & 	  5.85 	& $\pm$0.03 & 	  5.80 	& $\pm$0.03 & 	  5.73 	& $\pm$0.04 \\ 
log(N/O) &  -1.54 	& $\pm$0.06 & 	  -1.52 	& $\pm$0.06 & 	  -1.47 	& $\pm$0.06 & 	  -1.46 	& $\pm$0.07 & 	  -1.41 	& $\pm$0.07 & 	  -1.38 	& $\pm$0.08 & 	  -1.36 	& $\pm$0.08 & 	  -1.28 	& $\pm$0.09 \\ 
\hline
12+log(O/H) (N2)  &  8.02 	& $\pm$0.06 & 	  8.05 	& $\pm$0.09 & 	  8.08 	& $\pm$0.20 & 	  8.09 	& $\pm$0.43 & 	  8.09 	& $\pm$0.48 & 	  8.08 	& $\pm$0.50 & 	  8.10 	& $\pm$0.71 & 	  8.12 	& $\pm$1.19 \\ 
12+log(O/H) (R$_{23 lower}$)   &  8.06 	& $\pm$0.08 & 	  8.07 	& $\pm$0.10 & 	  8.10 	& $\pm$0.13 & 	  8.11 	& $\pm$0.18 & 	  8.10 	& $\pm$0.18 & 	  8.07 	& $\pm$0.20 & 	  8.12 	& $\pm$0.22 & 	  8.15 	& $\pm$0.27 \\ 
12+log(O/H)  (Ar$_3$O$_3$) &  8.03 	& $\pm$0.06 & 	  7.91 	& $\pm$0.09 & 	  8.03 	& $\pm$0.08 & 	  8.09 	& $\pm$0.14 & 	  8.10 	& $\pm$0.09 & 	  8.10 	& $\pm$0.11 & 	  8.10 	& $\pm$0.15 & 	  8.04 	& $\pm$0.29 \\ 
12+log(O/H)  (S$_3$O$_3$) &  8.27 	& $\pm$0.05 & 	  8.24 	& $\pm$0.05 & 	  8.29 	& $\pm$0.04 & 	  8.32 	& $\pm$0.05 & 	  8.32 	& $\pm$0.04 & 	  8.31 	& $\pm$0.05 & 	  8.31 	& $\pm$0.06 & 	  8.27 	& $\pm$0.09 \\ 
\hline
RS2 &  1.21 	& $\pm$0.03 & 	  1.45 	& $\pm$0.03 & 	  1.49 	& $\pm$0.03 & 	  1.51 	& $\pm$0.05 & 	  1.71 	& $\pm$0.06 & 
	  1.62 	& $\pm$0.05 & 	  1.86 	& $\pm$0.06 & 	  2.21 	& $\pm$0.07 \\ 
log(SO$_{23}$) &  -0.77 	& $\pm$0.26 & 	  -0.82 	& $\pm$0.26 & 	  -0.74 	& $\pm$0.25 & 	  -0.71 	& $\pm$0.24 & 	  -0.71 	& $\pm$0.24 & 	  -0.73 	& $\pm$0.24 & 	  -0.74 	& $\pm$0.23 & 	  -0.81 	& $\pm$0.23 \\ 
N2 &  -1.63 	& $\pm$0.01 & 	  -1.57 	& $\pm$0.02 & 	  -1.49 	& $\pm$0.05 & 	  -1.46 	& $\pm$0.11 & 	  -1.47 	& $\pm$0.12 & 	  -1.50 	& $\pm$0.12 & 	  -1.45 	& $\pm$0.18 & 	  -1.40 	& $\pm$0.31 \\ 
log(R$_{23}$)  &  0.92 	& $\pm$0.04 & 	  0.91 	& $\pm$0.05 & 	  0.91 	& $\pm$0.06 & 	  0.91 	& $\pm$0.09 & 	  0.91 	& $\pm$0.09 & 	  0.90 	& $\pm$0.10 & 	  0.91 	& $\pm$0.11 & 	  0.92 	& $\pm$0.14 \\ 
log(O$_{32}$)  &  0.65 	& $\pm$0.06 & 	  0.60 	& $\pm$0.06 & 	  0.51 	& $\pm$0.07 & 	  0.46 	& $\pm$0.09 & 	  0.47 	& $\pm$0.10 & 	  0.50 	& $\pm$0.11 & 	  0.43 	& $\pm$0.11 & 	  0.39 	& $\pm$0.14 \\ 
log(Ar$_3$O$_3$) &  -1.78 	& $\pm$0.04 & 	  -1.87 	& $\pm$0.06 & 	  -1.78 	& $\pm$0.06 & 	  -1.73 	& $\pm$0.12 & 	  -1.73 	& $\pm$0.08 & 	  -1.73 	& $\pm$0.10 & 	  -1.72 	& $\pm$0.13 & 	  -1.77 	& $\pm$0.23 \\ 
log(S$_3$O$_3$) &  -1.15 	& $\pm$0.08 & 	  -1.19 	& $\pm$0.07 & 	  -1.10 	& $\pm$0.08 & 	  -1.06 	& $\pm$0.09 & 	  -1.05 	& $\pm$0.08 & 	  -1.08 	& $\pm$0.08 & 	  -1.07 	& $\pm$0.11 & 	  -1.15 	& $\pm$0.15 \\ 
log(u)  ([\SII]/[\SIII]) &  -1.46 	& $\pm$0.16 & 	  -1.58 	& $\pm$0.15 & 	  -1.61 	& $\pm$0.16 & 	  -1.62 	& $\pm$0.21 & 	  -1.59 	& $\pm$0.21 & 	  -1.55 	& $\pm$0.21 & 	  -1.62 	& $\pm$0.26 & 	  -1.81 	& $\pm$0.31 \\ 
log(u)  ([\OII]/[\OIII]) &  -2.50 	& $\pm$0.02 & 	  -2.54 	& $\pm$0.02 & 	  -2.61 	& $\pm$0.02 & 	  -2.65 	& $\pm$0.03 & 	  -2.65 	& $\pm$0.03 & 	  -2.62 	& $\pm$0.04 & 	  -2.68 	& $\pm$0.04 & 	  -2.71 	& $\pm$0.05 \\ 
log($\eta^{\prime}$)  &  0.26 	& $\pm$0.07 & 	  0.24 	& $\pm$0.06 & 	  0.31 	& $\pm$0.07 & 	  0.35 	& $\pm$0.08 & 	  0.36 	& $\pm$0.09 & 	  0.35 	& $\pm$0.10 & 	  0.39 	& $\pm$0.10 & 	  0.32 	& $\pm$0.12 \\ 
\hline
log([\OIII]$\lambda$5007/\Hb) &  0.70 	& $\pm$0.01 & 	  0.69 	& $\pm$0.01 & 	  0.67 	& $\pm$0.02 & 	  0.65 	& $\pm$0.03 & 	  0.65 	& $\pm$0.03 & 	  0.65 	& $\pm$0.03 & 	  0.64 	& $\pm$0.03 & 	  0.64 	& $\pm$0.04 \\ 
log([\NII]$\lambda$6584/\Ha) &  -1.63 	& $\pm$0.05 & 	  -1.57 	& $\pm$0.05 & 	  -1.49 	& $\pm$0.06 & 	  -1.46 	& $\pm$0.09 & 	  -1.47 	& $\pm$0.12 & 	  -1.50 	& $\pm$0.11 & 	  -1.45 	& $\pm$0.14 & 	  -1.40 	& $\pm$0.16 \\ 
log([\SII]$\lambda\lambda$6717,31/\Ha) &  -1.27 	& $\pm$0.06 & 	  -1.25 	& $\pm$0.06 & 	  -1.17 	& $\pm$0.06 & 	  -1.13 	& $\pm$0.08 & 	  -1.15 	& $\pm$0.10 & 	  -1.20 	& $\pm$0.09 & 	  -1.16 	& $\pm$0.10 & 	  -1.13 	& $\pm$0.10 \\ 
\hline
\multicolumn{17}{l}{{\footnotesize References: PM03  \cite{permodiaz03}; G92 \cite{garnett92}; H06 \cite{hageleetal06}}}\\
\multicolumn{17}{l}{{\footnotesize $^a$Upper limit.  $^b$Lower limit.   }}\\
\end{tabular}
} 
\end{table}
\end{landscape}

\begin{landscape}
\begin{table}
\centering
\caption{Reddening corrected fluxes for IC132 accumulated shells. Fluxes are normalized to \Hb\ = 100}
\label{ticfluxesaper}
\resizebox{24cm}{!}{   
\begin{tabular}{clr@{}lr@{}lr@{}lr@{}lr@{}lr@{}lr@{}lr@{}lr@{}lr@{}l}
\hline
\ & \ &  \multicolumn{20}{c}{Cumulative} \\
\cline{3-22} \\
$\lambda$ & Ion & \multicolumn{2}{c}{$\overline{{\rm AA}}$}  &  
\multicolumn{2}{c}{$\overline{{\rm AB}}$}  &
\multicolumn{2}{c}{$\overline{{\rm AC}}$}  & 
\multicolumn{2}{c}{$\overline{{\rm AD}}$}  &  
\multicolumn{2}{c}{$\overline{{\rm AE}}$} &   
\multicolumn{2}{c}{$\overline{{\rm AF}}$}  &  
\multicolumn{2}{c}{$\overline{{\rm AG}}$}  &  
\multicolumn{2}{c}{$\overline{{\rm AH}}$}  & 
\multicolumn{2}{c}{$\overline{{\rm AI}}$}  &   
\multicolumn{2}{c}{$\overline{{\rm AJ}}$}\\
3727 &	[O II]	& 150.7 	& $\pm$8.0 & 	157.8 	& $\pm$8.0 &   	169.2 	& $\pm$8.7 &   	179.5 	& $\pm$9.5 &   	184.5 	& $\pm$10.1 &   	185.2 	& $\pm$10.6 &   	190.4 	& $\pm$11.4 &   	196.7 	& $\pm$13.0 &   	204.5 	& $\pm$15.5 &   	242.8 	& $\pm$33.9 \\  
4010 &	H$\delta$	& 33.5 	& $\pm$2.4 & 	32.5 	& $\pm$2.4 &   	32.3 	& $\pm$2.6 &   	32.2 	& $\pm$2.8 &   	32.1 	& $\pm$2.9 &   	31.9 	& $\pm$3.0 &   	31.7 	& $\pm$3.3 &   	31.4 	& $\pm$3.7 &   	31.1 	& $\pm$4.5 &   	30.8 	& $\pm$9.4 \\  
4340 &	H$\gamma$	& 50.0 	& $\pm$2.8 & 	49.2 	& $\pm$2.2 &   	49.2 	& $\pm$2.5 &   	49.2 	& $\pm$2.9 &   	49.0 	& $\pm$3.3 &   	48.8 	& $\pm$3.8 &   	48.5 	& $\pm$4.3 &   	48.2 	& $\pm$5.0 &   	47.6 	& $\pm$6.2 &   	45.7 	& $\pm$11.4 \\  
4363 &	[O III]$^1$	& 6.7 	& $\pm$1.2 & 	6.6 	& $\pm$1.0 &   	6.9 	& $\pm$1.1 &   	7.3 	& $\pm$1.3 &   	8.0 	& $\pm$1.5 &   	8.5 	& $\pm$1.8 &   	9.2 	& $\pm$2.1 &   	10.4 	& $\pm$2.5 &   	12.4 	& $\pm$3.3 &   	22.6 	& $\pm$7.8 \\  
4861 &	[H$\beta$]	& 100.0 	& $\pm$3.1 & 	100.0 	& $\pm$3.5 &   	100.0 	& $\pm$4.0 &   	100.0 	& $\pm$4.7 &   	100.0 	& $\pm$5.2 &   	100.0 	& $\pm$5.6 &   	100.0 	& $\pm$6.1 &   	100.0 	& $\pm$6.8 &   	100.0 	& $\pm$8.2 &   	100.0 	& $\pm$12.8 \\  
4959 &	[O III]	& 175.1 	& $\pm$6.4 & 	173.2 	& $\pm$6.7 &   	170.4 	& $\pm$6.8 &   	167.7 	& $\pm$7.6 &   	165.7 	& $\pm$8.2 &   	164.4 	& $\pm$8.7 &   	163.0 	& $\pm$9.2 &   	162.0 	& $\pm$10.1 &   	161.9 	& $\pm$12.0 &   	166.1 	& $\pm$19.3 \\  
5007 &	[O III]	& 500.1 	& $\pm$12.2 & 	492.7 	& $\pm$13.4 &   	483.1 	& $\pm$15.1 &   	474.6 	& $\pm$17.3 &   	468.3 	& $\pm$19.0 &   	464.5 	& $\pm$20.6 &   	460.2 	& $\pm$22.2 &   	456.8 	& $\pm$24.7 &   	455.4 	& $\pm$29.8 &   	466.4 	& $\pm$47.9 \\  
6312 &	[S III]	& 2.7 	& $\pm$1.3 & 	3.1 	& $\pm$1.7 &   	3.4 	& $\pm$2.0 &   	3.7 	& $\pm$2.4 &   	3.9 	& $\pm$2.7 &   	4.2 	& $\pm$3.0 &   	4.4 	& $\pm$3.4 &   	4.8 	& $\pm$4.0 &   	5.6 	& $\pm$5.1 &   	9.7 	& $\pm$9.8 \\  
6548 &	[N II]	& 2.9 	& $\pm$1.1 & 	2.8 	& $\pm$1.1 &   	3.0 	& $\pm$1.2 &   	3.1 	& $\pm$1.3 &   	3.2 	& $\pm$1.4 &   	3.2 	& $\pm$1.5 &   	3.3 	& $\pm$1.7 &   	3.5 	& $\pm$1.9 &   	3.8 	& $\pm$2.1 &   	5.0 	& $\pm$3.1 \\  
6564 &	H$\alpha$	& 284.5 	& $\pm$7.1 & 	284.3 	& $\pm$7.7 &   	284.3 	& $\pm$8.8 &   	284.3 	& $\pm$10.4 &   	284.4 	& $\pm$11.7 &   	284.4 	& $\pm$12.7 &   	284.4 	& $\pm$14.0 &   	284.4 	& $\pm$15.7 &   	284.6 	& $\pm$18.4 &   	287.8 	& $\pm$28.2 \\  
6584 &	[N II]	& 6.7 	& $\pm$0.8 & 	7.2 	& $\pm$0.8 &   	7.8 	& $\pm$0.9 &   	8.3 	& $\pm$1.1 &   	8.6 	& $\pm$1.4 &   	8.7 	& $\pm$1.6 &   	8.9 	& $\pm$1.8 &   	9.2 	& $\pm$2.1 &   	9.6 	& $\pm$2.4 &   	10.5 	& $\pm$3.4 \\  
6717 &	[S II]	& 8.6 	& $\pm$1.9 & 	9.2 	& $\pm$1.4 &   	10.0 	& $\pm$0.9 &   	10.7 	& $\pm$1.1 &   	11.0 	& $\pm$1.0 &   	11.3 	& $\pm$2.2 &   	11.6 	& $\pm$1.1 &   	11.7 	& $\pm$1.0 &   	12.2 	& $\pm$2.8 &   	12.8 	& $\pm$1.5 \\  
6731 &	[S II]	& 5.8 	& $\pm$0.3 & 	6.3 	& $\pm$0.3 &   	6.2 	& $\pm$0.5 &   	6.3 	& $\pm$0.6 &   	6.6 	& $\pm$0.8 &   	6.6 	& $\pm$0.9 &   	6.9 	& $\pm$1.0 &   	6.9 	& $\pm$1.2 &   	7.2 	& $\pm$1.4 &   	7.0 	& $\pm$2.0 \\  
7136 &	[Ar III]	& 8.3 	& $\pm$1.0 & 	7.4 	& $\pm$0.9 &   	7.5 	& $\pm$1.0 &   	7.7 	& $\pm$1.3 &   	7.8 	& $\pm$1.3 &   	7.9 	& $\pm$1.3 &   	7.9 	& $\pm$1.3 &   	7.9 	& $\pm$1.5 &   	7.6 	& $\pm$1.7 &   	7.2 	& $\pm$2.1 \\  
9068 &	[\SIII]	& 27.5 	& $\pm$6.3 & 	25.7 	& $\pm$5.0 &   	26.6 	& $\pm$5.3 &   	27.6 	& $\pm$5.8 &   	28.2 	& $\pm$5.9 &   	28.2 	& $\pm$5.9 &   	28.2 	& $\pm$6.3 &   	27.6 	& $\pm$6.5 &   	26.6 	& $\pm$6.9 &   	25.1 	& $\pm$8.9 \\  
\hline
\multicolumn{2}{l}{I(H$\beta$)[{\small erg s$^{-1}$cm$^{-2}$ 10$^{-13}$}]} &  79.29 	& $\pm$0.18 & 	  162.15 	& $\pm$0.40 & 	  241.39 	& $\pm$0.69 & 	  321.61 	& $\pm$1.07 & 	  402.61 	& $\pm$1.47 & 	  482.78 	& $\pm$1.91 & 	  563.35 	& $\pm$2.43 & 	  643.90 	& $\pm$3.10 & 	  724.18 	& $\pm$4.18 & 	  795.15 	& $\pm$7.19 \\ 
\multicolumn{2}{l}{C(H$\beta$)} &  0.50 	& $\pm$0.06 & 	  0.45 	& $\pm$0.03 & 	  0.43 	& $\pm$0.03 & 	  0.42 	& $\pm$0.02 & 	  0.40 	& $\pm$0.02 & 	  0.40 	& $\pm$0.02 & 	  0.39 	& $\pm$0.02 & 	  0.39 	& $\pm$0.02 & 	  0.40 	& $\pm$0.02 & 	  0.49 	& $\pm$0.02 \\ 
\multicolumn{2}{l}{EW(H$\beta$)[\AA]} &  130.34 	& $\pm$11.42 & 	  135.09 	& $\pm$11.62 & 	  150.07 	& $\pm$12.25 & 	  170.79 	& $\pm$13.07 & 	  185.69 	& $\pm$13.63 & 	  195.56 	& $\pm$13.98 & 	  198.54 	& $\pm$14.09 & 	  196.59 	& $\pm$14.02 & 	  186.62 	& $\pm$13.66 & 	  114.99 	& $\pm$10.72 \\ 
\hline
\multicolumn{2}{l}{$^1$Contaminated by mercury street lamps.}
\end{tabular}
}
\end{table}
\end{landscape}

\begin{landscape}
\begin{table}
\caption{Integrated temperatures, abundances and line ratios for IC 132 accumulated shells.}
\label{ticphysycalabundaper}
\resizebox{22cm}{!}{   
\begin{tabular}{lr@{}lr@{}lr@{}lr@{}lr@{}lr@{}lr@{}lr@{}lr@{}lr@{}l}
\hline
\ &  \multicolumn{20}{c}{Cumulative} \\
\cline{2-21} \\
\ & \multicolumn{2}{c}{$\overline{{\rm AA}}$}  &  
\multicolumn{2}{c}{$\overline{{\rm AB}}$}   &
\multicolumn{2}{c}{$\overline{{\rm AC}}$}   &  
\multicolumn{2}{c}{$\overline{{\rm AD}}$}   &  
\multicolumn{2}{c}{$\overline{{\rm AE}}$}    & 
\multicolumn{2}{c}{$\overline{{\rm AF}}$}   &  
\multicolumn{2}{c}{$\overline{{\rm AG}}$}    &  
\multicolumn{2}{c}{$\overline{{\rm AH}}$}  &  
\multicolumn{2}{c}{$\overline{{\rm AI}}$}   &  
\multicolumn{2}{c}{$\overline{{\rm AJ}}$}   \\
\Te([\OIII]) &  12700 	                & $\pm$900 & 	  12700 	& $\pm$700 & 	  13100 	& $\pm$900 & 	  13400 	& $\pm$1000 & 	  14000 	& $\pm$1200 & 	  14500 	& $\pm$1300 & 	  15000 	& $\pm$1600 & 	  16000 	& $\pm$1900 & 	  17600 	& $\pm$2600 & 	  25300 	& $\pm$6960 \\ 
\Te([\OII])$_ {PM03}$ &  12100 	& $\pm$300 & 	  12100 	& $\pm$200 & 	  12300	& $\pm$200 & 	  12600 	& $\pm$300 & 	  12900 	& $\pm$300 & 	  13100 	& $\pm$300 & 	  13500 	& $\pm$400 & 	  14000 	& $\pm$400 & 	  14800 	& $\pm$500 & 	  18000 	& $\pm$1000 \\ 
\Te([\OII])$_ {G92}$ &  11900 	& $\pm$600 & 	  11900 	& $\pm$500 & 	  12200 	& $\pm$600 & 	  12400 	& $\pm$700 & 	  12800 	& $\pm$800 & 	  13100 	& $\pm$900 & 	  13600 	& $\pm$1100 & 	  14200 	& $\pm$1300 & 	  15300 	& $\pm$1800 & 	  20700 	& $\pm$4900 \\ 
\Te([\SIII])$_{6312}$ &  10600 	& $\pm$100 & 	  11600 	& $\pm$200 & 	  12000 	& $\pm$200 & 	  12200 	& $\pm$200 & 	  12500 	& $\pm$200 & 	  13000 	& $\pm$300 & 	  13400 	& $\pm$400 & 	  14300 	& $\pm$500 & 	  16200 	& $\pm$800 & 	  27900 	& $\pm$4500 \\ 
\Te([\SIII])$_{SO_{23}}$ &  9400 	& $\pm$1900 & 	  9500 	& $\pm$1900 & 	  9400 	& $\pm$1800 & 	  9300 	& $\pm$1800 & 	  9200 	& $\pm$1700 & 	  9100 	& $\pm$1700 & 	  9200 	& $\pm$1700 & 	  9200 	& $\pm$1700 & 	  9300 	& $\pm$1700 & 	  9600 	& $\pm$1700 \\ 
\Te([\SIII]$_{H06}$ &  12000 	     & $\pm$1000 & 	  12000 	& $\pm$900 & 	  124000 	& $\pm$1000 & 	  12800 	& $\pm$1200 & 	  13500 	& $\pm$1400 & 	  14000 	& $\pm$1600 & 	  14700 	& $\pm$1900 & 	  15900 	& $\pm$2200 & 	  17700 	& $\pm$3000 & 	  27000 	& $\pm$8300 \\ 
\hline
12+log(O$^{+}$/H$^{+}$) &  7.45 	& $\pm$0.19 & 	  7.47 	& $\pm$0.16 & 	  7.47 	& $\pm$0.18 & 	  7.46 	& $\pm$0.21 & 	  7.43 	& $\pm$0.22 & 	  7.39 	& $\pm$0.25 & 	  7.35 	& $\pm$0.27 & 	  7.30 	& $\pm$0.30 & 	  7.22 	& $\pm$0.36 & 	  6.94 	& $\pm$0.56 \\ 
12+log(O$^{++}$/H$^{+}$) &  7.94 	& $\pm$0.19 & 	  7.94 	& $\pm$0.17 & 	  7.89 	& $\pm$0.19 & 	  7.85 	& $\pm$0.21 & 	  7.80 	& $\pm$0.22 & 	  7.76 	& $\pm$0.24 & 	  7.70 	& $\pm$0.26 & 	  7.64 	& $\pm$0.28 & 	  7.54 	& $\pm$0.33 & 	  7.24 	& $\pm$0.48 \\ 
12+log(O/H) &  8.06 	& $\pm$0.27 & 	  8.06 	& $\pm$0.23 & 	  8.03 	& $\pm$0.26 & 	  8.00 	& $\pm$0.30 & 	  7.95 	& $\pm$0.31 & 	  7.91 	& $\pm$0.34 & 	  7.87 	& $\pm$0.37 & 	  7.80 	& $\pm$0.41 & 	  7.71 	& $\pm$0.48 & 	  7.41 	& $\pm$0.74 \\ 
12+log(S$^{+}$/H$^{+}$) &  5.49 	& $\pm$0.32 & 	  5.43 	& $\pm$0.29 & 	  5.44 	& $\pm$0.32 & 	  5.45 	& $\pm$0.37 & 	  5.44 	& $\pm$0.41 & 	  5.42 	& $\pm$0.45 & 	  5.40 	& $\pm$0.53 & 	  5.35 	& $\pm$0.64 & 	  5.27 	& $\pm$0.88 & 	  4.96 	& $\pm$1.92 \\ 
12+log(S$^{++}$/H$^{+}$) &  6.69 	& $\pm$0.18 & 	  6.59 	& $\pm$0.14 & 	  6.58 	& $\pm$0.15 & 	  6.58 	& $\pm$0.16 & 	  6.57 	& $\pm$0.15 & 	  6.55 	& $\pm$0.15 & 	  6.52 	& $\pm$0.16 & 	  6.47 	& $\pm$0.17 & 	  6.37 	& $\pm$0.20 & 	  6.04 	& $\pm$0.27 \\ 
ICF(S$^+$ + S$^{++}$) &  1.31 	& $\pm$0.08 & 	  1.29 	& $\pm$0.07 & 	  1.27 	& $\pm$0.08 & 	  1.25 	& $\pm$0.09 & 	  1.24 	& $\pm$0.09 & 	  1.23 	& $\pm$0.10 & 	  1.22 	& $\pm$0.11 & 	  1.22 	& $\pm$0.12 & 	  1.21 	& $\pm$0.14 & 	  1.20 	& $\pm$0.20 \\ 
12+log(S/H) &  6.84 	& $\pm$0.37 & 	  6.73 	& $\pm$0.32 & 	  6.72 	& $\pm$0.35 & 	  6.71 	& $\pm$0.40 & 	  6.69 	& $\pm$0.43 & 	  6.67 	& $\pm$0.48 & 	  6.64 	& $\pm$0.55 & 	  6.59 	& $\pm$0.67 & 	  6.48 	& $\pm$0.90 & 	  6.15 	& $\pm$1.93 \\ 
log(S/O) &  -1.34 	& $\pm$0.02 & 	  -1.45 	& $\pm$0.02 & 	  -1.42 	& $\pm$0.02 & 	  -1.39 	& $\pm$0.02 & 	  -1.35 	& $\pm$0.02 & 	  -1.33 	& $\pm$0.02 & 	  -1.31 	& $\pm$0.02 & 	  -1.30 	& $\pm$0.03 & 	  -1.31 	& $\pm$0.03 & 	  -1.34 	& $\pm$0.04 \\ 
12+log(N$^+$/H$^+$) &  5.92 	& $\pm$0.01 & 	  5.95 	& $\pm$0.01 & 	  5.96 	& $\pm$0.01 & 	  5.97 	& $\pm$0.01 & 	  5.95 	& $\pm$0.02 & 	  5.93 	& $\pm$0.02 & 	  5.91 	& $\pm$0.02 & 	  5.89 	& $\pm$0.02 & 	  5.84 	& $\pm$0.02 & 	  5.65 	& $\pm$0.03 \\ 
log(N/O) &  -1.54 	& $\pm$0.06 & 	  -1.53 	& $\pm$0.05 & 	  -1.51 	& $\pm$0.05 & 	  -1.49 	& $\pm$0.05 & 	  -1.47 	& $\pm$0.05 & 	  -1.46 	& $\pm$0.06 & 	  -1.44 	& $\pm$0.06 & 	  -1.42 	& $\pm$0.07 & 	  -1.38 	& $\pm$0.07 & 	  -1.29 	& $\pm$0.10 \\ 
\hline
12+log(O/H) (N2)  &  8.02 	& $\pm$0.06 & 	  8.04 	& $\pm$0.07 & 	  8.05 	& $\pm$0.09 & 	  8.07 	& $\pm$0.13 & 	  8.07 	& $\pm$0.17 & 	  8.07 	& $\pm$0.21 & 	  8.08 	& $\pm$0.25 & 	  8.08 	& $\pm$0.32 & 	  8.09 	& $\pm$0.46 & 	  8.11 	& $\pm$1.14 \\ 
12+log(O/H) (R$_{23 lower}$)   &  8.06 	& $\pm$0.08 & 	  8.07 	& $\pm$0.08 & 	  8.08 	& $\pm$0.10 & 	  8.09 	& $\pm$0.11 & 	  8.09 	& $\pm$0.12 & 	  8.09 	& $\pm$0.13 & 	  8.09 	& $\pm$0.14 & 	  8.10 	& $\pm$0.16 & 	  8.11 	& $\pm$0.19 & 	  8.19 	& $\pm$0.30 \\ 
12+log(O/H)  (Ar$_3$O$_3$) &  8.03 	& $\pm$0.06 & 	  7.98 	& $\pm$0.06 & 	  8.00 	& $\pm$0.06 & 	  8.02 	& $\pm$0.08 & 	  8.03 	& $\pm$0.08 & 	  8.04 	& $\pm$0.07 & 	  8.05 	& $\pm$0.08 & 	  8.05 	& $\pm$0.09 & 	  8.04 	& $\pm$0.10 & 	  7.99 	& $\pm$0.14 \\ 
12+log(O/H)  (S$_3$O$_3$) &  8.27 	& $\pm$0.05 & 	  8.25 	& $\pm$0.04 & 	  8.27 	& $\pm$0.04 & 	  8.28 	& $\pm$0.04 & 	  8.29 	& $\pm$0.04 & 	  8.29 	& $\pm$0.04 & 	  8.30 	& $\pm$0.04 & 	  8.29 	& $\pm$0.04 & 	  8.28 	& $\pm$0.05 & 	  8.26 	& $\pm$0.07 \\ 
\hline
RS2 &  1.50 	& $\pm$0.03 & 	  1.47 	& $\pm$0.02 & 	  1.61 	& $\pm$0.01 & 	  1.69 	& $\pm$0.02 & 	  1.66 	& $\pm$0.01 & 	  1.73 	& $\pm$0.03 & 	  1.68 	& $\pm$0.02 & 	  1.70 	& $\pm$0.02 & 	  1.71 	& $\pm$0.04 & 	  1.83 	& $\pm$0.02 \\ 
log(SO$_{23}$) &  -0.78 	& $\pm$0.26 & 	  -0.80 	& $\pm$0.26 & 	  -0.78 	& $\pm$0.25 & 	  -0.76 	& $\pm$0.25 & 	  -0.75 	& $\pm$0.25 & 	  -0.75 	& $\pm$0.25 & 	  -0.74 	& $\pm$0.25 & 	  -0.75 	& $\pm$0.24 & 	  -0.77 	& $\pm$0.24 & 	  -0.81 	& $\pm$0.23 \\ 
N2 &  -1.63 	& $\pm$0.01 & 	  -1.60 	& $\pm$0.01 & 	  -1.56 	& $\pm$0.02 & 	  -1.53 	& $\pm$0.03 & 	  -1.52 	& $\pm$0.04 & 	  -1.52 	& $\pm$0.05 & 	  -1.51 	& $\pm$0.06 & 	  -1.49 	& $\pm$0.08 & 	  -1.47 	& $\pm$0.11 & 	  -1.44 	& $\pm$0.29 \\ 
log(R$_{23}$)  &  0.92 	& $\pm$0.04 & 	  0.92 	& $\pm$0.04 & 	  0.92 	& $\pm$0.05 & 	  0.91 	& $\pm$0.05 & 	  0.91 	& $\pm$0.06 & 	  0.91 	& $\pm$0.06 & 	  0.91 	& $\pm$0.07 & 	  0.91 	& $\pm$0.08 & 	  0.91 	& $\pm$0.09 & 	  0.94 	& $\pm$0.15 \\ 
log(O$_{32}$)  &  0.65 	& $\pm$0.06 & 	  0.63 	& $\pm$0.06 & 	  0.59 	& $\pm$0.06 & 	  0.55 	& $\pm$0.06 & 	  0.54 	& $\pm$0.06 & 	  0.53 	& $\pm$0.07 & 	  0.52 	& $\pm$0.07 & 	  0.50 	& $\pm$0.08 & 	  0.48 	& $\pm$0.09 & 	  0.42 	& $\pm$0.16 \\ 
log(Ar$_3$O$_3$) &  -1.78 	& $\pm$0.04 & 	  -1.82 	& $\pm$0.04 & 	  -1.81 	& $\pm$0.05 & 	  -1.79 	& $\pm$0.06 & 	  -1.78 	& $\pm$0.06 & 	  -1.77 	& $\pm$0.06 & 	  -1.76 	& $\pm$0.06 & 	  -1.76 	& $\pm$0.07 & 	  -1.78 	& $\pm$0.08 & 	  -1.81 	& $\pm$0.11 \\ 
log(S$_3$O$_3$) &  -1.15 	& $\pm$0.08 & 	  -1.17 	& $\pm$0.06 & 	  -1.15 	& $\pm$0.06 & 	  -1.12 	& $\pm$0.07 & 	  -1.11 	& $\pm$0.07 & 	  -1.10 	& $\pm$0.07 & 	  -1.10 	& $\pm$0.07 & 	  -1.11 	& $\pm$0.08 & 	  -1.12 	& $\pm$0.09 & 	  -1.15 	& $\pm$0.12 \\ 
log(u)  ([\SII]/[\SIII]) &  -1.42 	& $\pm$0.17 & 	  -1.52 	& $\pm$0.12 & 	  -1.53 	& $\pm$0.12 & 	  -1.54 	& $\pm$0.13 & 	  -1.55 	& $\pm$0.12 & 	  -1.56 	& $\pm$0.15 & 	  -1.58 	& $\pm$0.13 & 	  -1.60 	& $\pm$0.14 & 	  -1.66 	& $\pm$0.19 & 	  -1.70 	& $\pm$0.22 \\ 
log(u)  ([\OII]/[\OIII]) &  -2.50 	& $\pm$0.02 & 	  -2.52 	& $\pm$0.02 & 	  -2.55 	& $\pm$0.02 & 	  -2.58 	& $\pm$0.02 & 	  -2.59 	& $\pm$0.02 & 	  -2.60 	& $\pm$0.02 & 	  -2.61 	& $\pm$0.02 & 	  -2.62 	& $\pm$0.03 & 	  -2.64 	& $\pm$0.03 & 	  -2.69 	& $\pm$0.06 \\ 
log($\eta^{\prime}$)  &  0.28 	& $\pm$0.07 & 	  0.25 	& $\pm$0.06 & 	  0.28 	& $\pm$0.06 & 	  0.31 	& $\pm$0.06 & 	  0.32 	& $\pm$0.06 & 	  0.32 	& $\pm$0.06 & 	  0.32 	& $\pm$0.07 & 	  0.33 	& $\pm$0.07 & 	  0.31 	& $\pm$0.08 & 	  0.35 	& $\pm$0.14 \\ 
\hline
log([\OIII]$\lambda$5007/\Hb) &  0.70 	& $\pm$0.01 & 	  0.69 	& $\pm$0.01 & 	  0.68 	& $\pm$0.01 & 	  0.68 	& $\pm$0.02 & 	  0.67 	& $\pm$0.02 & 	  0.67 	& $\pm$0.02 & 	  0.66 	& $\pm$0.02 & 	  0.66 	& $\pm$0.02 & 	  0.66 	& $\pm$0.03 & 	  0.67 	& $\pm$0.04 \\ 
log([\NII]$\lambda$6584/\Ha) &  -1.63 	& $\pm$0.05 & 	  -1.60 	& $\pm$0.05 & 	  -1.56 	& $\pm$0.05 & 	  -1.53 	& $\pm$0.06 & 	  -1.52 	& $\pm$0.07 & 	  -1.52 	& $\pm$0.08 & 	  -1.51 	& $\pm$0.09 & 	  -1.49 	& $\pm$0.10 & 	  -1.47 	& $\pm$0.10 & 	  -1.44 	& $\pm$0.14 \\ 
log([\SII]$\lambda\lambda$6717,31/\Ha) &  -1.30 	& $\pm$0.06 & 	  -1.27 	& $\pm$0.04 & 	  -1.24 	& $\pm$0.03 & 	  -1.22 	& $\pm$0.03 & 	  -1.21 	& $\pm$0.03 & 	  -1.20 	& $\pm$0.06 & 	  -1.19 	& $\pm$0.03 & 	  -1.18 	& $\pm$0.03 & 	  -1.17 	& $\pm$0.07 & 	  -1.16 	& $\pm$0.05 \\ 
\hline
\multicolumn{10}{l}{{\footnotesize References: PM03  \cite{permodiaz03}; G92 \cite{garnett92}; H06 \cite{hageleetal06}}}
\end{tabular}
} 
\end{table}
\end{landscape}

\begin{figure}
\centering
\includegraphics[width=10.0cm]{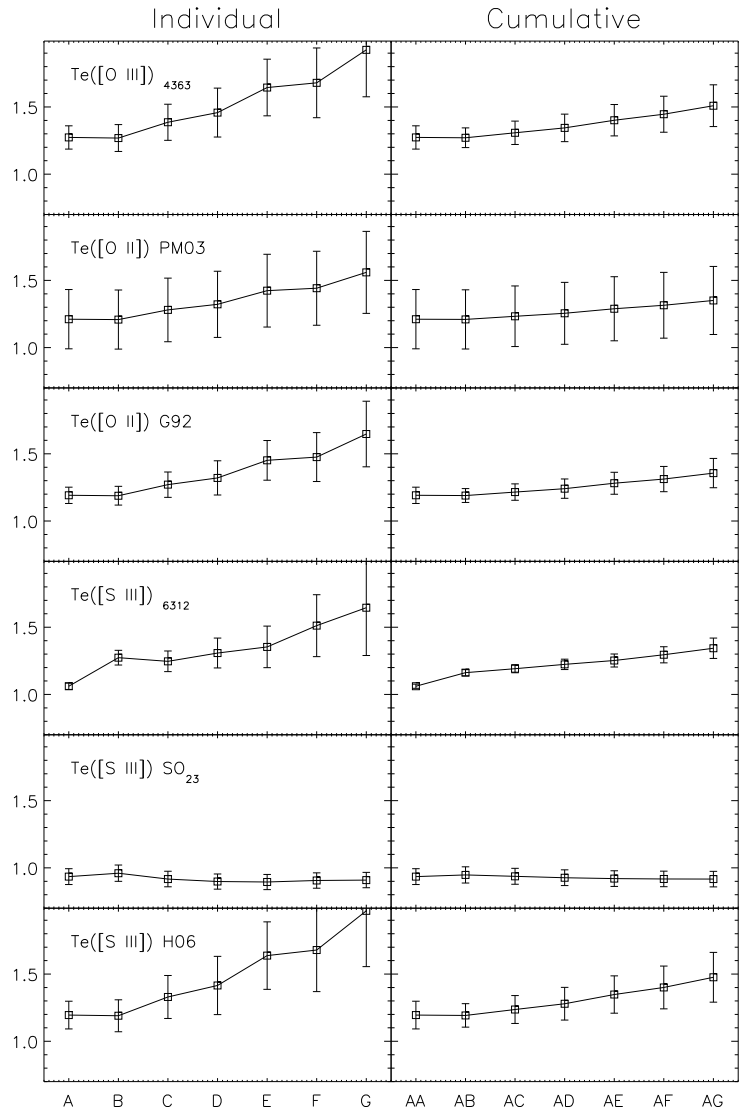}
\caption{IC~132: temperatures estimated for individual 
and accumulated shells as labeled in each panel.
Vertical lines are the associated error.}
\label{fishelltemps}
\end{figure}

\begin{figure}
\centering
\includegraphics[width=10.0cm]{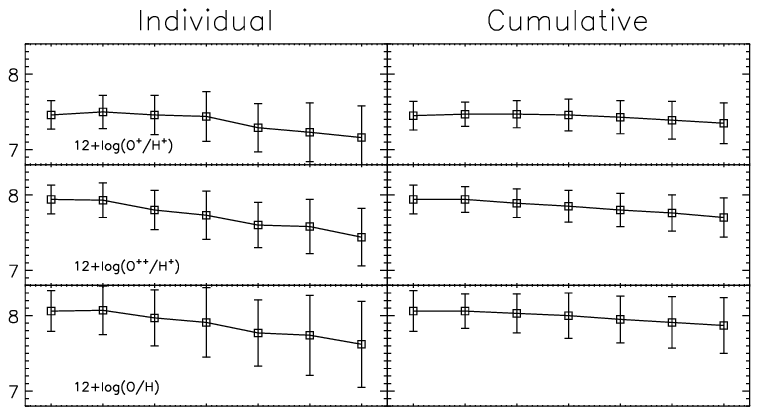}
\caption{IC 132: ionic and total oxygen abundances for individual 
and accumulated shells.}
\label{fishellabunds_OXY}
\end{figure}

\begin{figure}
\centering
\includegraphics[width=10.0cm]{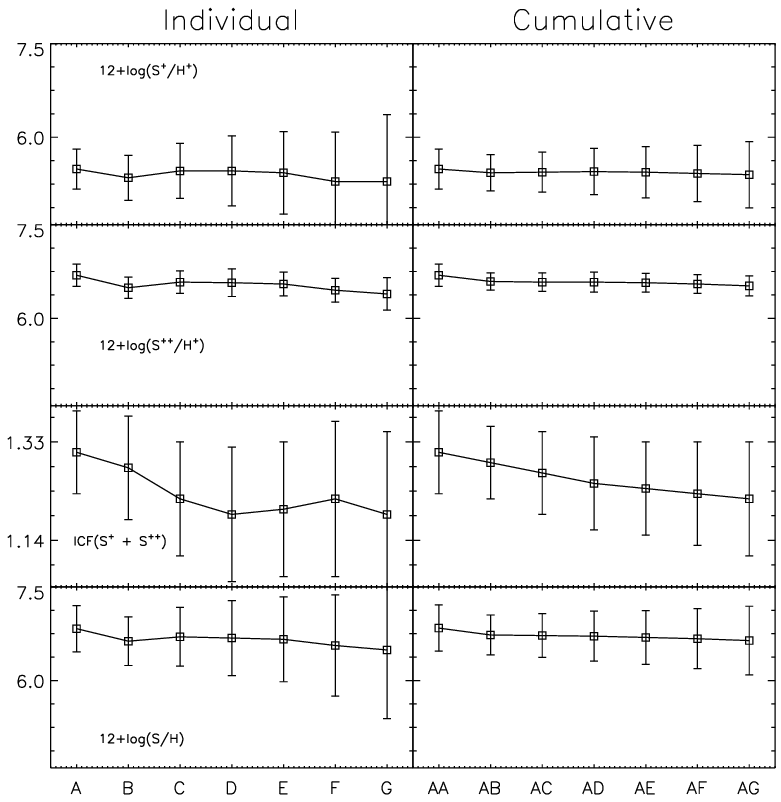}
\caption{IC 132: ionic and total oxygen abundances for individual 
and accumulated shells.}
\label{fishellabunds_SUL}
\end{figure}

\begin{figure}
\centering
\includegraphics[width=10.0cm]{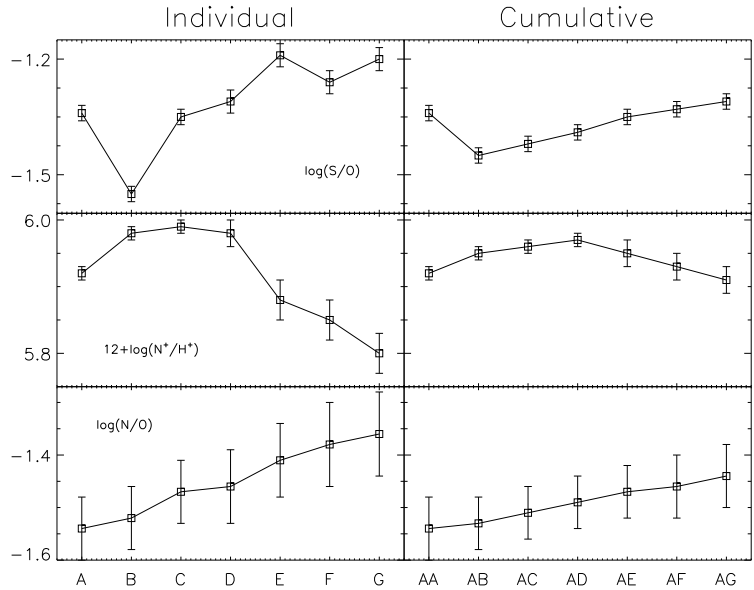}
\caption{IC 132: ionic ratios and N$^+$ abundances for individual 
and accumulated shells.}
\label{fishellabunds_SON}
\end{figure}

\begin{figure}
\centering
\includegraphics[width=10.0cm]{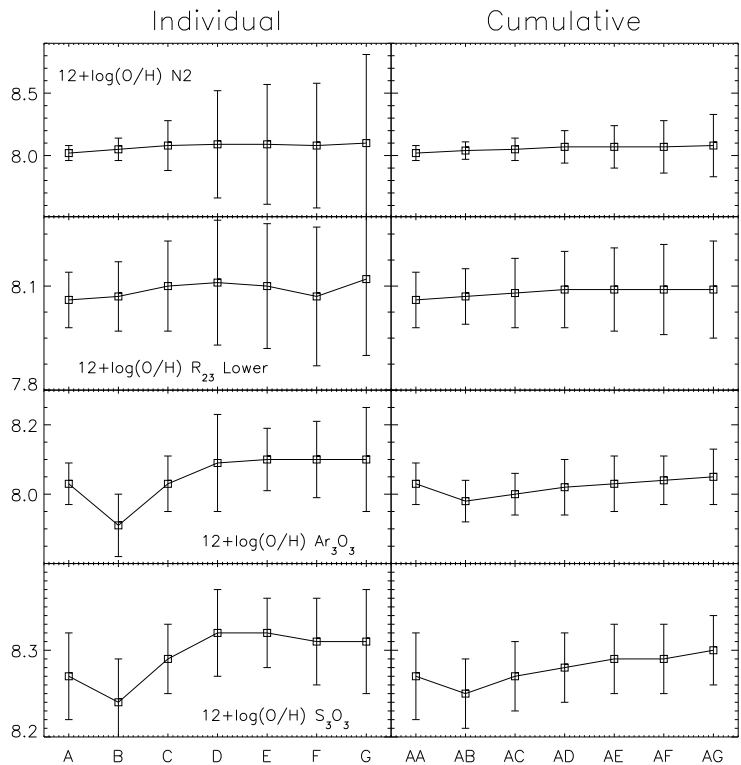}
\caption{IC~132 empirical abundance indicators  for 
individual and accumulated shells.}
\label{fishellempiricalab}
\end{figure}

\begin{figure}
\centering
\includegraphics[width=10.0cm]{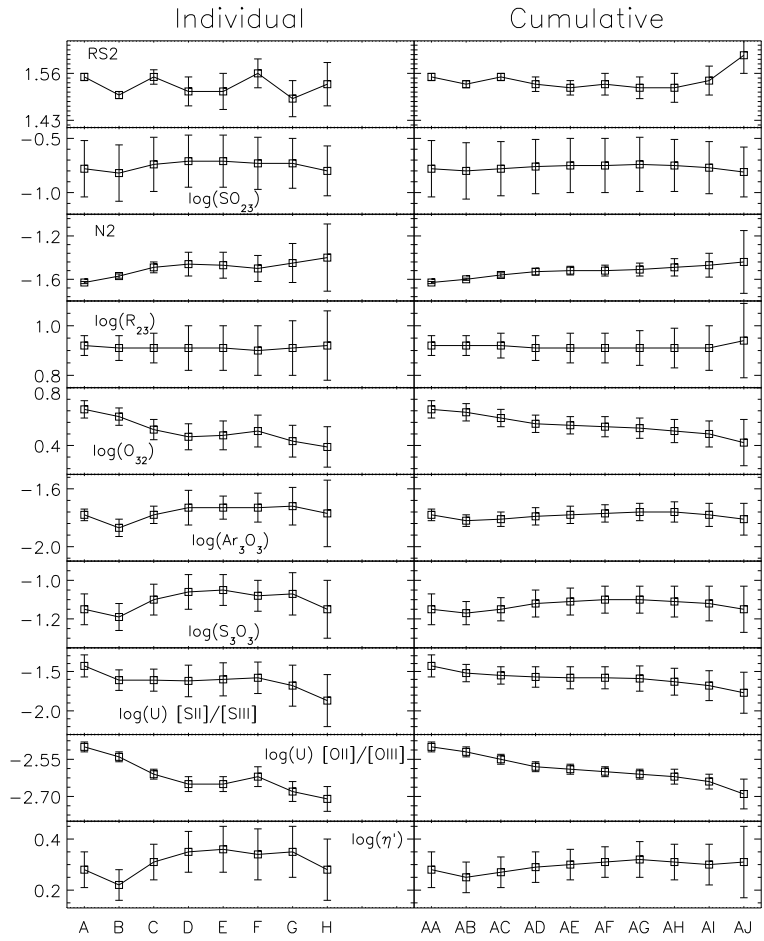}
\caption{IC~132 various line ratios  for individual and 
accumulated shells.}
\label{fishellratios}
\end{figure}

\begin{figure}
\centering
\includegraphics[width=10.0cm]{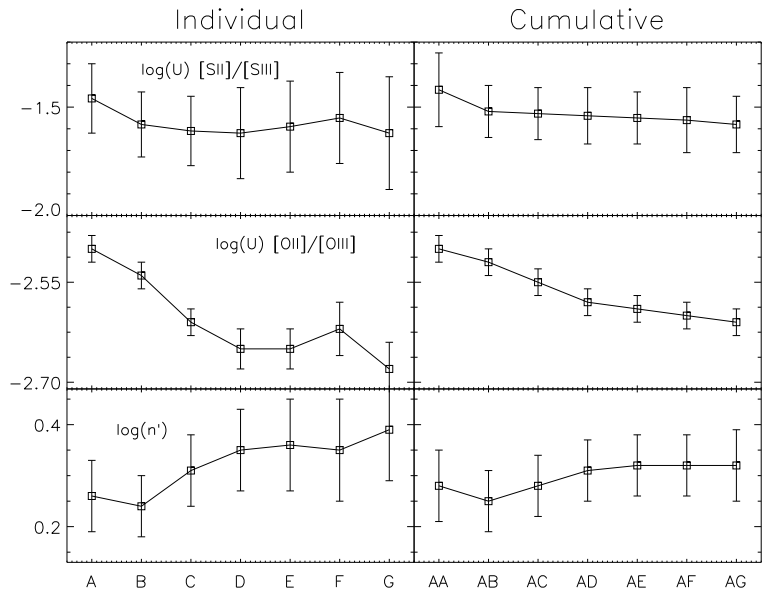}
\caption{IC~132 ionization parameter  from sulphur and oxygen and the 
parameter $\eta^{\prime}$ (see text) for 
individual and accumulated shells.}
\label{fishellionparam}
\end{figure}

\begin{figure}
\centering
\includegraphics[width=10.0cm]{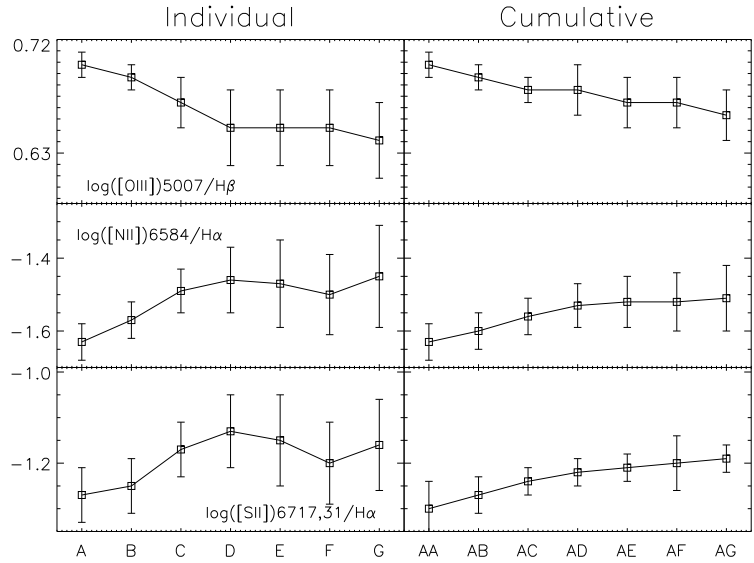}
\caption{IC~132 ionization sensitive line ratios  for 
individual and accumulated shells.}
\label{fishellionrat}
\end{figure}

\begin{figure}
\centering
\includegraphics[width=10.0cm]{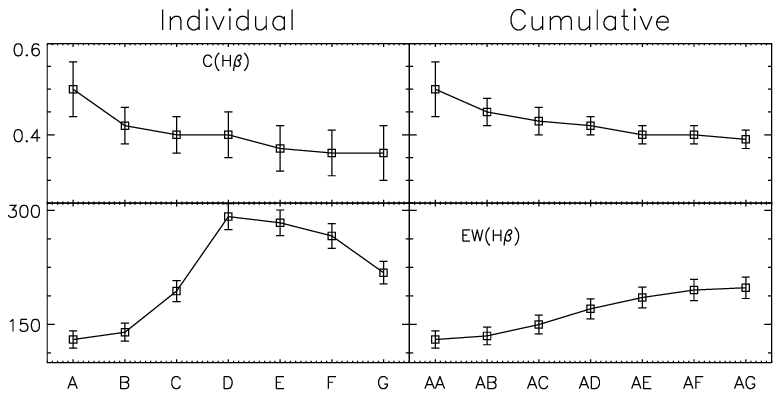}
\caption{IC~132 extinction parameter C(\Hb) 
and equivalent width EW(\Hb)  for individual 
and accumulated shells.}
\label{fishellexeq}
\end{figure}

The resulting integrated properties for IC 132  are shown in plots 
\ref{fishelltemps} to   \ref{fishellexeq},  for               
electron temperatures, ionic and total abundances, empirical 
abundances, line ratios, ionization sensitive ratios, extinction 
and EW(\Hb). Each figure shows the values for individual 
shells on the left and for cumulative segmentation on the right panels.  
Individual shells I an J are not shown given the large errors obtained. 
Errors were calculated by formal propagation of each line flux 
associated error. For \Te~([\OII])$_{PM03}$, also the error in 
\Ne\ determination at the low density limit was added in 
quadrature, to take into account the spread of the models in the original paper.

Figure \ref{fishelltemps} shows the radial trend of the electron 
temperatures. Inside the errors there is a clear tendency for the 
temperature to increase with radius, except for \Te([\SIII]) computed
with the SO$_{23}$ method, which shows no temperature gradient. 
The \Te\ increasing tendency is reflected in the radial trend of 
abundance shown in Fig.~\ref{fishellabunds_OXY} where it is 
possible to see  that the estimated O/H goes to smaller abundances 
in the outer regions. However, given the large errors obtained, 
the radial behaviour is also compatible with a constant value
for the whole region. In such case the total (O/H) may be set 
between 7.8 and 8.2. 

For figure \ref{fishellabunds_SUL}, although a slight
decrease with radius is noted, the ionic and 
total abundances are compatible with a constant value within errors.
Total (S/H)
is between 6.4 and 7.1. The mean value from the 2D map of 6.62
may as well be adopted as the representative value of the region.

Interestingly, in figure \ref{fishellabunds_SON} 
S/O has a rather flat trend while N/O displays   
an apparent increase with radial distance in the individual 
shells analysis. This is contrary to the inverse trend naively expected
given that the Wolf Rayet stars are located at the centre of
IC~132 (see \S \ref{secWR}).  On the other hand, it might simply 
mean that the WR phenomenon didn't have enough time to pollute 
the ionized gas or that, even if some contamination occurs, within 
the errors it cannot be observed. The radial increase
may be explained by a) SF was propagated from the border
to the central part of the region b) Enrichment of an old 
population (by shocks or SN) `escaped' the central zone:
and is still propagating outside.
 However it may be noted that (N/O) is smoothed in the cumulative 
case and also becomes compatible with a constant ratio.

The empirical abundances in figure \ref{fishellempiricalab} 
all  hint at radial increase, but again when the errors are
considered all are compatible with a constant value through the region. All the 
indicators, except the one based on S$_3$O$_3$ may be set to 
a value of 8.0, in agreement with the mean value from the 
2D maps. S$_3$O$_3$ reaches higher values by almost 0.25 dex even
within the errors.

Figure \ref{fishellratios} shows the various line ratios used.
RS2 is in almost all cases above the theoretical limit for the [\SII] 
ratio, although in the cumulative case the dispersion is lower than in
the individual one and the values are closer to the upper 
theoretical limit,  a systematic effect that produces ``unrealistic"
results is evident. When comparing with the 2D maps also
the trend is different as the spaxels at the borders of the map
have low RS2 ($\sim$ 0.7). This reflects the problems with 
the line deblending,  its dependence with the S/N or 
limitations in the atomic values to estimate \Ne\ in the low density regime, as discussed 
previously in \S \ref{electrodensity}.

Log(SO$_{23}$) is constant through the region; taking into
account the errors the range lies within the fit produced
by \cite{diazetal07} for \Te\ determination. 
It is worth  noticing that  \cite{diazetal07} only used 
high metallicity \HII\ regions in their fit. This sets 
a limit log(SO$_{23}$)=-1.0 for the validity of the relation
although many objects exist below this limit even in
the compilation they present. If the fit is extended to low
metallicity objects  that may shift the relation to produce higher temperatures.
The values of log(R$_{23}$), log(Ar$_3$O$_3$) and log(S$_3$O$_3$)
are compatible, within the errors, with no gradient. The ionization sensitive ratio 
log(O$_{32}$) (McGaugh 1991)
indicates that the ionization decreases with radius.

Figure \ref{fishellionparam} shows the softness parameter 
$\eta^{\prime}$={\small([\OII]3727/[\OIII]4959,5007)/([\SII]6717,6731/[\SIII]9069,9532)} , 
defined by \cite{vilchezetal88} as an indicator of the effective 
temperature. To first order insensitive to the ionization parameter in the nebula,
it grows inversely proportional to the temperature. 
$\eta^{\prime}$ is also compatible with a constant value when errors are 
considered, and an increase with radius is not expected considering \Te\
trends in figure \ref{fishelltemps}. 

Ionization sensitive ratios are plotted in figure \ref{fishellionrat}, a
decrease of ionization with radius is concluded from the three 
indicators.

In general when comparing the shell analysis with the 2D maps 
statistical agreement is found between  the mean of the
distribution and the standard deviation (s). However
some evident features in the maps disappear in the shell
analysis. A clear example is the ``wall" that runs diagonally 
in the south-east section of many 2D maps for IC~132. Hence 
the shell analysis must always be complemented with the maps for 
a full analysis of the region and exploitation of the IFS 
capabilities.


\subsection{Integrated properties of the Individual central regions}
\label{sec_central_regions}

\begin{figure}
\centering
\includegraphics[width=6.0cm]{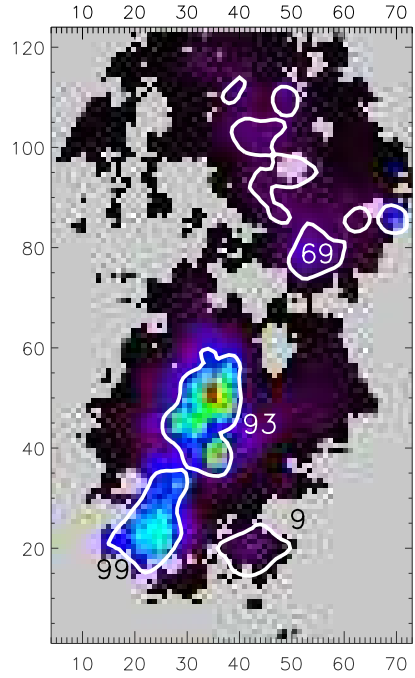}
\caption{Identification of regions from the \protect\cite{HoSkAs02} 
catalog. Labels are the BCLMP numbers (see text).}
\label{fkBCLMPregs}
\end{figure}

In contrast to the IC~132 frame that shows a single region, our observed frames in the centre of M33  
contain several localized \HII\ regions.  We have used the charts from 
\cite{HoSkAs02} to match separate regions in the \Ha\ map. 
The  identification of the regions together with the number from 
\cite{bclmp74} (which we called BCLMP) is shown in Fig.~\ref{fkBCLMPregs}. 

\begin{figure}
\centering
\includegraphics[width=6.0cm]{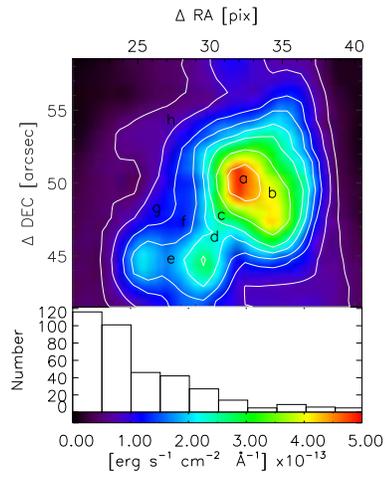}
\caption{BCLMP 93 shell segmentation.}
\label{fc93shells}
\end{figure}

\begin{figure}
\centering
\includegraphics[width=10.0cm]{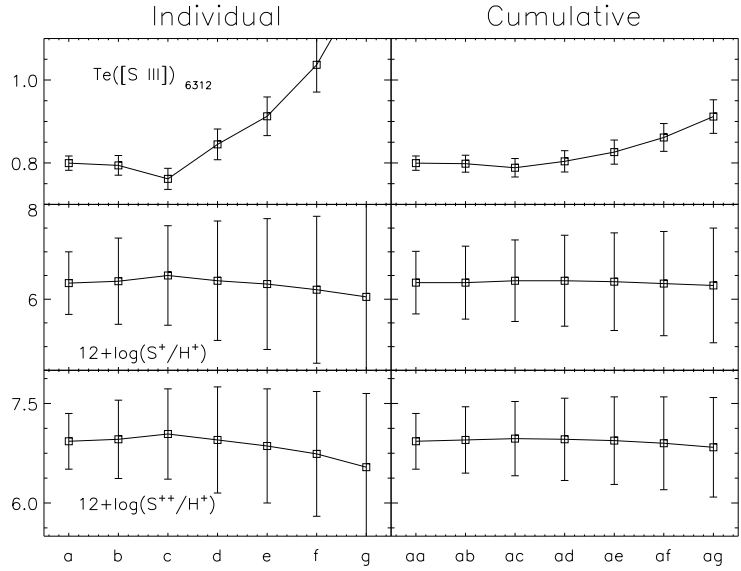}
\caption{BCLMP 93 temperature and ionic abundances estimated from the 
sulphur lines for the individual and cumulative shells.}
\label{fc93tempion}
\end{figure}

\begin{figure}
\centering
\includegraphics[width=10.0cm]{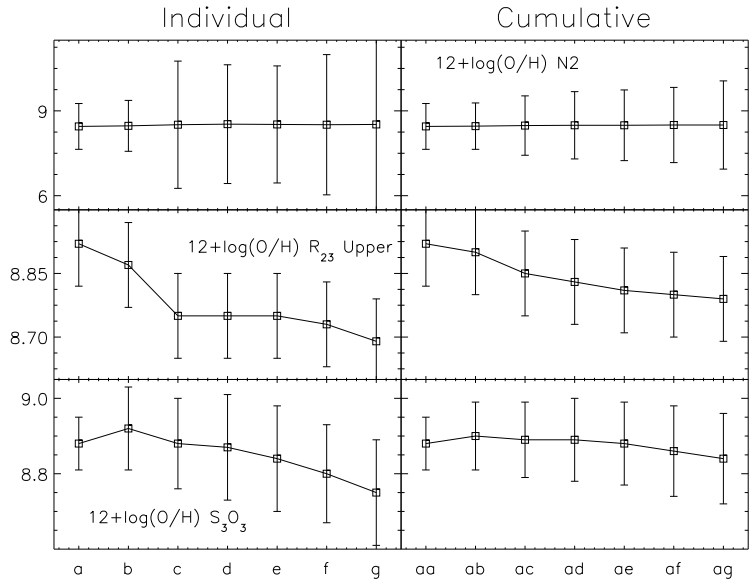}
\caption{BCLMP 93 empirical abundance estimates  for the individual 
and cumulative shells.}
\label{fc93empab}
\end{figure}

\begin{figure}
\centering
\includegraphics[width=10.0cm]{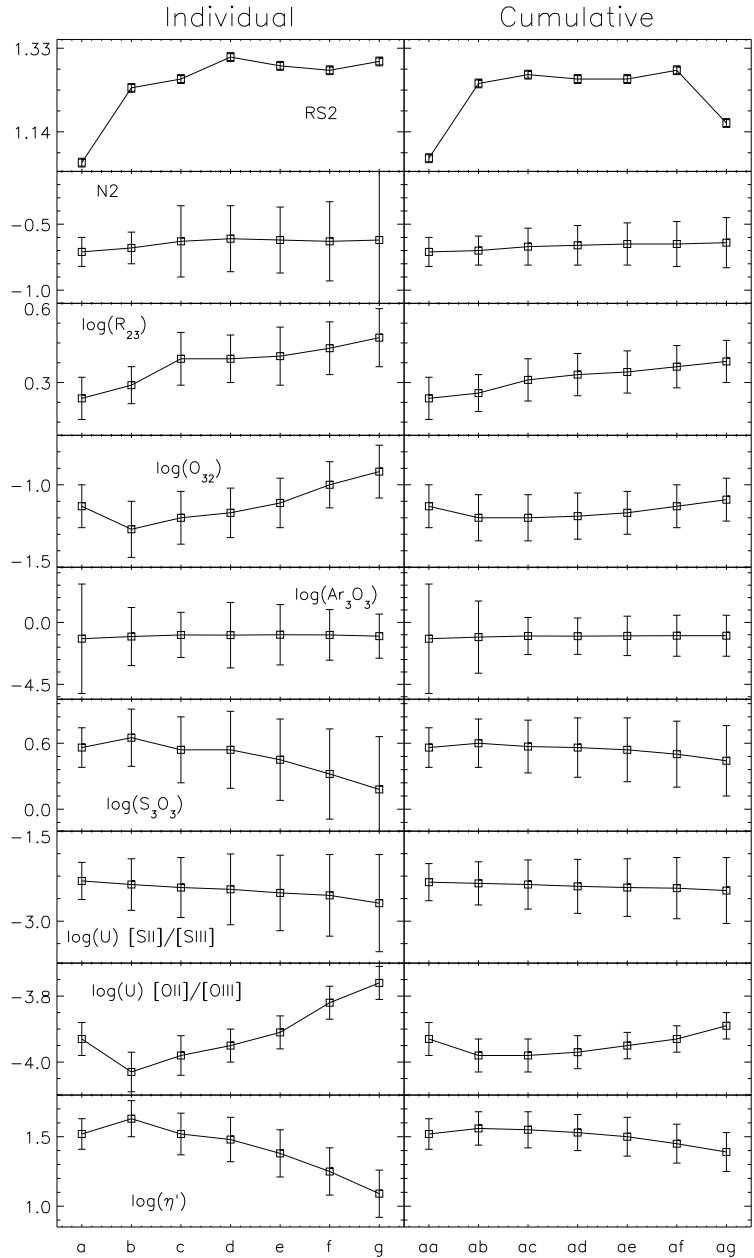}
\caption{BCLMP 93 line ratios  for the individual and  cumulative shells.}
\label{fc93linrat}
\end{figure}

\begin{figure}
\centering
\includegraphics[width=10.0cm]{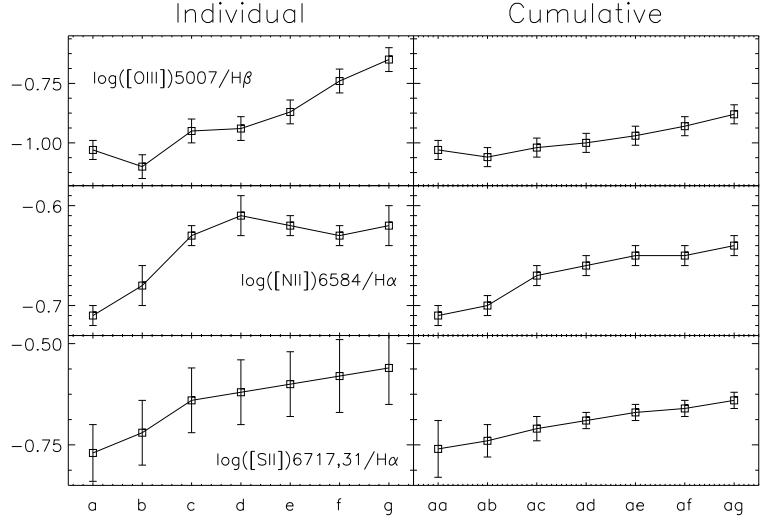}
\caption{BCLMP 93 ionization ratios  for the individual and cumulative shells.}
\label{fc93ddiag}
\end{figure}

A segmentation analysis  was performed 
for BCLMP~93; the rest of the central regions are too faint to obtain 
spectra with S/N high enough if segmented. 
The \Ha\ levels used to measure BCLMP~93 integrated shells are shown in 
Fig.~\ref{fc93shells}. The measured line fluxes and derived quantities 
are given in tables \ref{tcc93fluxesshell}, \ref{tcc93physycalabundshell},
\ref{tcc93fluxesaper} and \ref{tcc93physycalabundaper}, while the 
corresponding plots are in figures \ref{fc93tempion}, \ref{fc93empab},
\ref{fc93linrat} and \ref{fc93ddiag}.

Figure \ref{fc93tempion} shows a steep increase of \Te([\SIII]) for the
individual shells but a moderate one for the cumulative analysis. The sulphur ionic 
abundances show a slight decrease with radius although they are consistent with being uniform when errors are considered. 

Empirical abundance indicators are represented in figure \ref{fc93empab}. 
N2, albeit with large errors, is compatible with  constant abundance.
The R$_{23}$ based abundance decreases for the individual shells 
but becomes compatible with constant in the cumulative case. The same can be concluded for 
S$_3$O$_3$. Within the errors, they are all compatible with being constant.

For the line ratios in figure \ref{fc93linrat}, RS2 is always below the 
low density limit indicating  higher densities in the central region than in the outskirts. They show that \Ne\ values decrease with radius in the individual 
case while for the cumulative shells the last point shows an increase 
in \Ne\ for the outermost spaxels. Even when the rest of the ratios indicate some variation 
trend, in general they are compatible with a constant value, except for the 
last two ratios. The log(U)$_{[\OII]/[\OIII]}$ increases while $\eta^{\prime}$
decreases. The decrease in $\eta^{\prime}$ is compatible with the trend displayed by \Te([\SIII])
in figure \ref{fc93tempion}.

All the ionization sensitive ratios in figure \ref{fc93ddiag} show a
clear increase with radius. This is puzzling, as one would expect that when [\OIII]/\Hb\ increases, both 
[\NII]/\Ha\ and  [\SII]/\Ha\ should decrease.
Possible causes may be a projection effect of two zones 
physically separated but on the same line of sight or in general, multiplicity of the region or even the 
meddling effect of diffuse radiation. 

The large error bars in some line ratio plots are a consequence of the 
relatively small S/N of the data. We can nevertheless distinguish 
some differences between IC~132 and BCLMP~93. Firstly the higher 
Balmer decrement C(\Hb ) and smaller EW(\Hb )  in BCLMP 93  
indicate higher extinction and either an older \HII\ region or, more 
likely, a larger contamination due to the underlying  stellar population unrelated to the ionizing cluster
in the central regions of M33, i.e. the bulge population. Regarding the 
electron temperatures, only Te([\SIII])$_{6312}$ is computed for both
regions and provides a result that suggests an outwards increasing temperature
for each region.

\begin{landscape}
\begin{table}
\centering
\caption{Reddening corrected fluxes for BCLMP 93 with integration over individual \Ha\ shells. Lines are normalized to \Hb\ = 100}
\label{tcc93fluxesshell}
\resizebox{22cm}{!}{   
\begin{tabular}{clr@{}lr@{}lr@{}lr@{}lr@{}lr@{}lr@{}lr@{}l}
\hline
\ & \ &  \multicolumn{16}{c}{Individual} \\
\cline{3-18}
$\lambda$ & Ion & \multicolumn{2}{c}{A}  &  \multicolumn{2}{c}{B}  &
\multicolumn{2}{c}{C}  &  \multicolumn{2}{c}{D}  &  \multicolumn{2}{c}{E}  &
\multicolumn{2}{c}{F}  &  \multicolumn{2}{c}{G}  &  \multicolumn{2}{c}{H}  \\
3727 &	[O II]	& 160.7 	& $\pm$11.4 & 	183.9 	& $\pm$12.6 &   	230.8 	& $\pm$21.4 &   	232.2 	& $\pm$18.9 &   	235.0 	& $\pm$24.4 &   	244.8 	& $\pm$21.9 &   	264.8 	& $\pm$25.1 &   	294.7 	& $\pm$52.4 \\  
4010 &	H$\delta$	& 23.4 	& $\pm$3.2 & 	24.3 	& $\pm$3.2 &   	21.4 	& $\pm$4.5 &   	21.2 	& $\pm$5.5 &   	21.0 	& $\pm$5.1 &   	21.0 	& $\pm$5.9 &   	17.4 	& $\pm$7.1 &   	13.8 	& $\pm$11.7 \\  
4340 &	H$\gamma$	& 43.0 	& $\pm$2.3 & 	43.3 	& $\pm$2.0 &   	42.3 	& $\pm$3.0 &   	43.0 	& $\pm$2.9 &   	41.6 	& $\pm$3.5 &   	43.0 	& $\pm$4.4 &   	41.7 	& $\pm$5.1 &   	41.3 	& $\pm$9.5 \\  
4363 &	[O III]$^1$	& -4.6 	& $\pm$1.3 & 	-4.0 	& $\pm$1.4 &   	-5.8 	& $\pm$2.0 &   	-4.9 	& $\pm$1.9 &   	-5.9 	& $\pm$2.1 &   	-7.8 	& $\pm$2.5 &   	-12.6 	& $\pm$2.9 &   	-21.3 	& $\pm$4.6 \\  
4861 &	[H$\beta$]	& 100.0 	& $\pm$3.5 & 	100.0 	& $\pm$3.4 &   	100.0 	& $\pm$5.2 &   	100.0 	& $\pm$4.7 &   	100.0 	& $\pm$4.8 &   	100.0 	& $\pm$5.5 &   	100.0 	& $\pm$7.0 &   	100.0 	& $\pm$10.8 \\  
4959 &	[O III]	& 2.4 	& $\pm$1.0 & 	2.0 	& $\pm$1.2 &   	3.6 	& $\pm$1.6 &   	4.4 	& $\pm$1.5 &   	4.9 	& $\pm$1.3 &   	6.6 	& $\pm$1.9 &   	9.5 	& $\pm$2.9 &   	15.2 	& $\pm$3.8 \\  
5007 &	[O III]	& 9.4 	& $\pm$0.8 & 	7.9 	& $\pm$0.9 &   	11.1 	& $\pm$1.2 &   	11.4 	& $\pm$1.3 &   	13.5 	& $\pm$1.4 &   	18.0 	& $\pm$2.0 &   	22.1 	& $\pm$2.6 &   	35.3 	& $\pm$4.0 \\  
6312 &	[S III]	& 1.3 	& $\pm$0.9 & 	1.3 	& $\pm$1.1 &   	1.3 	& $\pm$1.3 &   	1.8 	& $\pm$1.7 &   	2.1 	& $\pm$2.1 &   	2.8 	& $\pm$2.7 &   	3.5 	& $\pm$3.7 &   	5.0 	& $\pm$5.7 \\  
6548 &	[N II]	& 10.1 	& $\pm$1.2 & 	11.0 	& $\pm$1.5 &   	11.8 	& $\pm$1.5 &   	13.3 	& $\pm$1.5 &   	12.0 	& $\pm$1.4 &   	12.2 	& $\pm$1.2 &   	13.2 	& $\pm$2.0 &   	12.6 	& $\pm$3.5 \\  
6564 &	H$\alpha$	& 283.8 	& $\pm$7.6 & 	283.8 	& $\pm$7.5 &   	281.2 	& $\pm$10.8 &   	280.1 	& $\pm$10.1 &   	282.3 	& $\pm$10.1 &   	282.2 	& $\pm$11.2 &   	284.8 	& $\pm$14.8 &   	282.3 	& $\pm$22.6 \\  
6584 &	[N II]	& 55.5 	& $\pm$2.1 & 	59.0 	& $\pm$2.4 &   	66.0 	& $\pm$3.2 &   	68.6 	& $\pm$3.3 &   	68.0 	& $\pm$3.1 &   	66.8 	& $\pm$3.1 &   	69.0 	& $\pm$4.4 &   	68.9 	& $\pm$6.9 \\  
6717 &	[S II]	& 25.0 	& $\pm$1.3 & 	30.2 	& $\pm$1.4 &   	35.5 	& $\pm$2.0 &   	38.2 	& $\pm$2.2 &   	40.0 	& $\pm$2.4 &   	41.4 	& $\pm$2.6 &   	44.5 	& $\pm$3.2 &   	52.1 	& $\pm$5.3 \\  
6731 &	[S II]	& 23.3 	& $\pm$1.1 & 	24.3 	& $\pm$1.2 &   	28.2 	& $\pm$1.6 &   	29.2 	& $\pm$1.8 &   	31.0 	& $\pm$2.0 &   	32.4 	& $\pm$2.1 &   	34.2 	& $\pm$2.6 &   	38.6 	& $\pm$4.1 \\  
7136 &	[Ar III]	& 0.6 	& $\pm$7.1 & 	0.7 	& $\pm$4.6 &   	1.4 	& $\pm$6.5 &   	1.3 	& $\pm$9.3 &   	1.7 	& $\pm$10.9 &   	2.2 	& $\pm$11.9 &   	2.2 	& $\pm$10.3 &   	8.3 	& $\pm$13.5 \\  
9068 &	[\SIII]	& 27.1 	& $\pm$14.4 & 	29.0 	& $\pm$20.7 &   	31.7 	& $\pm$25.3 &   	32.6 	& $\pm$30.6 &   	32.9 	& $\pm$32.9 &   	30.3 	& $\pm$32.6 &   	26.8 	& $\pm$35.1 &   	15.7 	& $\pm$44.6 \\  
\hline
\multicolumn{2}{l}{I(H$\beta$)[{\small erg s$^{-1}$cm$^{-2}$ 10$^{-13}$}]} &  21.28 	& $\pm$0.05 & 	  18.95 	& $\pm$0.05 & 	  18.61 	& $\pm$0.07 & 	  18.66 	& $\pm$0.06 & 	  18.48 	& $\pm$0.06 & 	  18.08 	& $\pm$0.07 & 	  18.25 	& $\pm$0.09 & 	  18.63 	& $\pm$0.14 \\ 
\multicolumn{2}{l}{C(H$\beta$)} &  0.66 	& $\pm$0.06 & 	  0.52 	& $\pm$0.06 & 	  0.68 	& $\pm$0.08 & 	  0.84 	& $\pm$0.07 & 	  1.02 	& $\pm$0.07 & 	  1.24 	& $\pm$0.08 & 	  1.25 	& $\pm$0.09 & 	  1.08 	& $\pm$0.11 \\ 
\multicolumn{2}{l}{EW(H$\beta$)[\AA]} &  38.08 	& $\pm$6.17 & 	  46.96 	& $\pm$6.85 & 	  47.47 	& $\pm$6.89 & 	  55.80 	& $\pm$7.47 & 	  50.28 	& $\pm$7.09 & 	  48.44 	& $\pm$6.96 & 	  42.13 	& $\pm$6.49 & 	  33.43 	& $\pm$5.78 \\ 
\hline
\multicolumn{2}{l}{$^1$Contaminated by mercury street lamps.}
\end{tabular}
}
\end{table}
\end{landscape}

\begin{landscape}
\begin{table}
\caption{Integrated temperatures, abundances and line ratios for BCLMP 93 individual \Ha\ shells.}
\label{tcc93physycalabundshell}

\resizebox{22cm}{!}{   
\begin{tabular}{lr@{}lr@{}lr@{}lr@{}lr@{}lr@{}lr@{}lr@{}l}
\hline
\ &  \multicolumn{16}{c}{Individual} \\
\cline{2-17}
\ & \multicolumn{2}{c}{A}  &  \multicolumn{2}{c}{B}  &
\multicolumn{2}{c}{C}  &  \multicolumn{2}{c}{D}  &  \multicolumn{2}{c}{E}   &
\multicolumn{2}{c}{F}  &  \multicolumn{2}{c}{G}   &  \multicolumn{2}{c}{H} \\
\Te([\SIII])$_{6312}$ &  8000 	& $\pm$200 & 	  8000 	& $\pm$200 & 	  7600 	& $\pm$300 & 	  8400 	& $\pm$400 & 	  9100 	& $\pm$500 & 	  10300 	& $\pm$700 & 	  12400 	& $\pm$1100 & 	  16000 	& $\pm$2700 \\ 
\Te([\SIII])$_{SO_{23}}$ &  6000 	& $\pm$300 & 	  6100 	& $\pm$300 & 	  6200 	& $\pm$300 & 	  6200 	& $\pm$300 & 	  6200 	& $\pm$300 & 	  6300 	& $\pm$300 & 	  6500 	& $\pm$400 & 	  6700 	& $\pm$400 \\ 
\hline
12+log(S$^{+}$/H$^{+}$) &  6.34 	& $\pm$0.66 & 	  6.38 	& $\pm$0.91 & 	  6.50 	& $\pm$1.05 & 	  6.39 	& $\pm$1.26 & 	  6.32 	& $\pm$1.38 & 	  6.20 	& $\pm$1.55 & 	  6.05 	& $\pm$1.95 & 	  5.90 	& $\pm$3.04 \\ 
12+log(S$^{++}$/H$^{+}$) &  6.93 	& $\pm$0.42 & 	  6.96 	& $\pm$0.59 & 	  7.04 	& $\pm$0.68 & 	  6.95 	& $\pm$0.80 & 	  6.86 	& $\pm$0.86 & 	  6.74 	& $\pm$0.94 & 	  6.54 	& $\pm$1.11 & 	  6.33 	& $\pm$1.57 \\ 
\hline
12+log(O/H) (N2)  &  8.45 	& $\pm$0.81 & 	  8.47 	& $\pm$0.90 & 	  8.51 	& $\pm$2.25 & 	  8.53 	& $\pm$2.10 & 	  8.52 	& $\pm$2.07 & 	  8.51 	& $\pm$2.48 & 	  8.52 	& $\pm$4.50 & 	  8.53 	& $\pm$10.60 \\ 
12+log(O/H) (R$_{23 upper}$)   &  8.92 	& $\pm$0.10 & 	  8.87 	& $\pm$0.10 & 	  8.75 	& $\pm$0.10 & 	  8.75 	& $\pm$0.10 & 	  8.75 	& $\pm$0.10 & 	  8.73 	& $\pm$0.10 & 	  8.69 	& $\pm$0.10 & 	  8.63 	& $\pm$0.10 \\ 
12+log(O/H)  (Ar$_3$O$_3$) &  8.55 	& $\pm$2.16 & 	  8.63 	& $\pm$0.89 & 	  8.67 	& $\pm$0.57 & 	  8.67 	& $\pm$0.85 & 	  8.68 	& $\pm$0.74 & 	  8.67 	& $\pm$0.64 & 	  8.64 	& $\pm$0.64 & 	  8.75 	& $\pm$0.13 \\ 
12+log(O/H)  (S$_3$O$_3$) &  8.88 	& $\pm$0.07 & 	  8.92 	& $\pm$0.11 & 	  8.88 	& $\pm$0.12 & 	  8.87 	& $\pm$0.14 & 	  8.84 	& $\pm$0.14 & 	  8.80 	& $\pm$0.13 & 	  8.75 	& $\pm$0.14 & 	  8.69 	& $\pm$0.19 \\ 
\hline
RS2 &  1.07 	& $\pm$0.01 & 	  1.24 	& $\pm$0.01 & 	  1.26 	& $\pm$0.01 & 	  1.31 	& $\pm$0.01 & 	  1.29 	& $\pm$0.01 & 	  1.28 	& $\pm$0.01 & 	  1.30 	& $\pm$0.01 & 	  1.35 	& $\pm$0.02 \\ 
\Ne &  328 	& \ \ \ --- & 	 137 	& \ \ \ --- & 	 127 	& \ \ \ --- & 	 85 	& \ \ \ --- & 	 101 	& \ \ \ --- & 	 109 	& \ \ \ --- & 	 91 	& \ \ \ --- & 	 55 	& \ \ \ --- \\ 
log(SO$_{23}$) &  -0.01 	& $\pm$0.03 & 	  -0.03 	& $\pm$0.03 & 	  -0.09 	& $\pm$0.03 & 	  -0.08 	& $\pm$0.03 & 	  -0.09 	& $\pm$0.03 & 	  -0.12 	& $\pm$0.04 & 	  -0.18 	& $\pm$0.04 & 	  -0.24 	& $\pm$0.06 \\ 
N2 &  -0.71 	& $\pm$0.11 & 	  -0.68 	& $\pm$0.12 & 	  -0.63 	& $\pm$0.27 & 	  -0.61 	& $\pm$0.25 & 	  -0.62 	& $\pm$0.25 & 	  -0.63 	& $\pm$0.30 & 	  -0.62 	& $\pm$0.53 & 	  -0.61 	& $\pm$1.25 \\ 
log(R$_{23}$)  &  0.24 	& $\pm$0.08 & 	  0.29 	& $\pm$0.07 & 	  0.39 	& $\pm$0.10 & 	  0.39 	& $\pm$0.09 & 	  0.40 	& $\pm$0.11 & 	  0.43 	& $\pm$0.10 & 	  0.47 	& $\pm$0.11 & 	  0.54 	& $\pm$0.19 \\ 
log(O$_{32}$)  &  -1.13 	& $\pm$0.13 & 	  -1.27 	& $\pm$0.17 & 	  -1.20 	& $\pm$0.16 & 	  -1.17 	& $\pm$0.15 & 	  -1.11 	& $\pm$0.15 & 	  -1.00 	& $\pm$0.14 & 	  -0.92 	& $\pm$0.16 & 	  -0.77 	& $\pm$0.21 \\ 
log(Ar$_3$O$_3$) &  -1.19 	& $\pm$3.97 & 	  -1.03 	& $\pm$2.11 & 	  -0.91 	& $\pm$1.64 & 	  -0.93 	& $\pm$2.38 & 	  -0.90 	& $\pm$2.19 & 	  -0.91 	& $\pm$1.84 & 	  -1.00 	& $\pm$1.60 & 	  -0.63 	& $\pm$0.56 \\ 
log(S$_3$O$_3$) &  0.56 	& $\pm$0.18 & 	  0.65 	& $\pm$0.26 & 	  0.54 	& $\pm$0.30 & 	  0.54 	& $\pm$0.35 & 	  0.45 	& $\pm$0.37 & 	  0.32 	& $\pm$0.41 & 	  0.18 	& $\pm$0.48 & 	  -0.05 	& $\pm$0.68 \\ 
log(u)  ([\SII]/[\SIII]) &  -2.33 	& $\pm$0.31 & 	  -2.39 	& $\pm$0.43 & 	  -2.44 	& $\pm$0.50 & 	  -2.47 	& $\pm$0.59 & 	  -2.53 	& $\pm$0.63 & 	  -2.57 	& $\pm$0.68 & 	  -2.70 	& $\pm$0.81 & 	  -2.85 	& $\pm$1.15 \\ 
log(u)  ([\OII]/[\OIII]) &  -3.93 	& $\pm$0.05 & 	  -4.03 	& $\pm$0.06 & 	  -3.98 	& $\pm$0.06 & 	  -3.95 	& $\pm$0.05 & 	  -3.91 	& $\pm$0.05 & 	  -3.82 	& $\pm$0.05 & 	  -3.76 	& $\pm$0.05 & 	  -3.63 	& $\pm$0.07 \\ 
log($\eta^{\prime}$)  &  1.52 	& $\pm$0.11 & 	  1.63 	& $\pm$0.13 & 	  1.52 	& $\pm$0.15 & 	  1.48 	& $\pm$0.16 & 	  1.38 	& $\pm$0.17 & 	  1.25 	& $\pm$0.17 & 	  1.09 	& $\pm$0.17 & 	  0.85 	& $\pm$0.24 \\ 
\hline
log([\OIII]$\lambda$5007/\Hb) &  -1.03 	& $\pm$0.04 & 	  -1.10 	& $\pm$0.05 & 	  -0.95 	& $\pm$0.05 & 	  -0.94 	& $\pm$0.05 & 	  -0.87 	& $\pm$0.05 & 	  -0.74 	& $\pm$0.05 & 	  -0.65 	& $\pm$0.05 & 	  -0.45 	& $\pm$0.05 \\ 
log([\NII]$\lambda$6584/\Ha) &  -0.71 	& $\pm$0.01 & 	  -0.68 	& $\pm$0.02 & 	  -0.63 	& $\pm$0.01 & 	  -0.61 	& $\pm$0.02 & 	  -0.62 	& $\pm$0.01 & 	  -0.63 	& $\pm$0.01 & 	  -0.62 	& $\pm$0.02 & 	  -0.61 	& $\pm$0.03 \\ 
log([\SII]$\lambda\lambda$6717,31/\Ha) &  -0.77 	& $\pm$0.07 & 	  -0.72 	& $\pm$0.08 & 	  -0.64 	& $\pm$0.08 & 	  -0.62 	& $\pm$0.08 & 	  -0.60 	& $\pm$0.08 & 	  -0.58 	& $\pm$0.09 & 	  -0.56 	& $\pm$0.09 & 	  -0.49 	& $\pm$0.09 \\ 

\hline
\end{tabular}
}
\end{table}
\end{landscape}

\begin{landscape}
\begin{table}
\centering
\caption{Reddening corrected fluxes for BCLMP 93 with integration over cumulative \Ha\ shells. Lines are normalized to \Hb\ =100}
\label{tcc93fluxesaper}
\resizebox{22cm}{!}{   
\begin{tabular}{clr@{}lr@{}lr@{}lr@{}lr@{}lr@{}lr@{}lr@{}l}
\hline
\ &  \multicolumn{17}{c}{Cumulative} \\
\cline{3-18} \\
\multicolumn{2}{c}{} & \multicolumn{2}{c}{$\overline{{\rm AA}}$}  &  
\multicolumn{2}{c}{$\overline{{\rm AB}}$}   &
\multicolumn{2}{c}{$\overline{{\rm AC}}$}   &  
\multicolumn{2}{c}{$\overline{{\rm AD}}$}   &  
\multicolumn{2}{c}{$\overline{{\rm AE}}$}    & 
\multicolumn{2}{c}{$\overline{{\rm AF}}$}   &  
\multicolumn{2}{c}{$\overline{{\rm AG}}$}    &  
\multicolumn{2}{c}{$\overline{{\rm AH}}$}  \\

3727 &	[O II]	& 160.7 	& $\pm$11.4 & 	171.6 	& $\pm$11.4 &   	190.3 	& $\pm$13.8 &   	200.4 	& $\pm$14.1 &   	207.1 	& $\pm$14.7 &   	213.1 	& $\pm$15.2 &   	220.2 	& $\pm$16.1 &   	229.4 	& $\pm$17.7 \\  
4010 &	H$\delta$	& 23.4 	& $\pm$3.2 & 	23.8 	& $\pm$3.1 &   	23.0 	& $\pm$3.1 &   	22.6 	& $\pm$3.5 &   	22.3 	& $\pm$3.7 &   	22.0 	& $\pm$3.9 &   	21.4 	& $\pm$4.2 &   	20.4 	& $\pm$4.9 \\  
4340 &	H$\gamma$	& 43.0 	& $\pm$2.3 & 	43.1 	& $\pm$2.1 &   	42.9 	& $\pm$2.1 &   	42.9 	& $\pm$2.2 &   	42.6 	& $\pm$2.3 &   	42.7 	& $\pm$2.5 &   	42.6 	& $\pm$2.6 &   	42.4 	& $\pm$3.2 \\  
4363 &	[O III]$^1$	& -4.6 	& $\pm$1.3 & 	-4.3 	& $\pm$1.3 &   	-4.8 	& $\pm$1.4 &   	-4.8 	& $\pm$1.4 &   	-5.1 	& $\pm$1.5 &   	-5.5 	& $\pm$1.6 &   	-6.4 	& $\pm$1.7 &   	-8.1 	& $\pm$2.0 \\  
4861 &	[H$\beta$]	& 100.0 	& $\pm$3.5 & 	100.0 	& $\pm$3.4 &   	100.0 	& $\pm$3.7 &   	100.0 	& $\pm$3.9 &   	100.0 	& $\pm$3.9 &   	100.0 	& $\pm$4.1 &   	100.0 	& $\pm$4.4 &   	100.0 	& $\pm$5.0 \\  
4959 &	[O III]	& 2.4 	& $\pm$1.0 & 	2.2 	& $\pm$1.1 &   	2.7 	& $\pm$1.1 &   	3.1 	& $\pm$1.2 &   	3.4 	& $\pm$1.2 &   	3.9 	& $\pm$1.2 &   	4.7 	& $\pm$1.4 &   	6.0 	& $\pm$1.6 \\  
5007 &	[O III]	& 9.4 	& $\pm$0.8 & 	8.7 	& $\pm$0.8 &   	9.5 	& $\pm$0.9 &   	9.9 	& $\pm$0.9 &   	10.6 	& $\pm$1.0 &   	11.8 	& $\pm$1.1 &   	13.2 	& $\pm$1.2 &   	15.9 	& $\pm$1.3 \\  
6312 &	[S III]	& 1.3 	& $\pm$0.9 & 	1.3 	& $\pm$1.0 &   	1.3 	& $\pm$1.1 &   	1.4 	& $\pm$1.2 &   	1.5 	& $\pm$1.4 &   	1.7 	& $\pm$1.6 &   	2.0 	& $\pm$1.9 &   	2.4 	& $\pm$2.4 \\  
6548 &	[N II]	& 10.1 	& $\pm$1.2 & 	10.5 	& $\pm$1.3 &   	11.1 	& $\pm$1.3 &   	11.4 	& $\pm$1.3 &   	11.5 	& $\pm$1.3 &   	11.5 	& $\pm$1.2 &   	11.8 	& $\pm$1.3 &   	11.9 	& $\pm$1.5 \\  
6564 &	H$\alpha$	& 283.8 	& $\pm$7.6 & 	283.8 	& $\pm$7.4 &   	283.0 	& $\pm$8.0 &   	282.3 	& $\pm$8.3 &   	282.3 	& $\pm$8.4 &   	282.3 	& $\pm$8.6 &   	282.6 	& $\pm$9.2 &   	282.6 	& $\pm$10.5 \\  
6584 &	[N II]	& 55.5 	& $\pm$2.1 & 	57.1 	& $\pm$2.2 &   	60.0 	& $\pm$2.4 &   	62.1 	& $\pm$2.5 &   	63.2 	& $\pm$2.5 &   	63.8 	& $\pm$2.5 &   	64.5 	& $\pm$2.7 &   	65.0 	& $\pm$3.1 \\  
6717 &	[S II]	& 25.7 	& $\pm$1.3 & 	28.8 	& $\pm$1.3 &   	30.6 	& $\pm$1.4 &   	32.3 	& $\pm$1.5 &   	33.6 	& $\pm$1.6 &   	34.7 	& $\pm$1.7 &   	34.6 	& $\pm$1.8 &   	38.2 	& $\pm$2.1 \\  
6731 &	[S II]	& 23.7 	& $\pm$1.1 & 	23.1 	& $\pm$1.1 &   	24.2 	& $\pm$1.1 &   	25.7 	& $\pm$1.2 &   	26.7 	& $\pm$1.3 &   	27.0 	& $\pm$1.4 &   	29.9 	& $\pm$1.5 &   	29.5 	& $\pm$1.7 \\  
7136 &	[Ar III]	& 0.6 	& $\pm$7.1 & 	0.7 	& $\pm$5.6 &   	1.0 	& $\pm$3.8 &   	1.0 	& $\pm$3.8 &   	1.1 	& $\pm$4.7 &   	1.3 	& $\pm$5.6 &   	1.4 	& $\pm$6.1 &   	1.7 	& $\pm$6.9 \\  
9068 &	[\SIII]	& 27.1 	& $\pm$14.4 & 	27.7 	& $\pm$17.2 &   	28.8 	& $\pm$19.5 &   	29.9 	& $\pm$22.0 &   	30.7 	& $\pm$24.0 &   	30.7 	& $\pm$25.1 &   	30.1 	& $\pm$26.5 &   	29.8 	& $\pm$29.1 \\  
\hline
\multicolumn{2}{l}{I(H$\beta$)[{\small erg s$^{-1}$cm$^{-2}$ 10$^{-13}$}]} &  21.28 	& $\pm$0.05 & 	  40.23 	& $\pm$0.10 & 	  58.83 	& $\pm$0.16 & 	  77.49 	& $\pm$0.21 & 	  95.97 	& $\pm$0.27 & 	  114.05 	& $\pm$0.33 & 	  132.30 	& $\pm$0.41 & 	  150.92 	& $\pm$0.53 \\ 
\multicolumn{2}{l}{C(H$\beta$)} &  0.66 	& $\pm$0.06 & 	  0.60 	& $\pm$0.04 & 	  0.62 	& $\pm$0.04 & 	  0.68 	& $\pm$0.03 & 	  0.75 	& $\pm$0.03 & 	  0.83 	& $\pm$0.03 & 	  0.89 	& $\pm$0.03 & 	  0.92 	& $\pm$0.03 \\ 
\multicolumn{2}{l}{EW(H$\beta$)[\AA]} &  38.08 	& $\pm$6.17 & 	  41.80 	& $\pm$6.47 & 	  43.44 	& $\pm$6.59 & 	  45.89 	& $\pm$6.77 & 	  46.67 	& $\pm$6.83 & 	  46.94 	& $\pm$6.85 & 	  46.21 	& $\pm$6.80 & 	  44.13 	& $\pm$6.64 \\ 
\hline
\multicolumn{2}{l}{$^1$Contaminated by mercury street lamps.}
\end{tabular}
}
\end{table}
\end{landscape}

\begin{landscape}
\begin{table}
\caption{Integrated temperatures, abundances and line ratios for BCLMP 93 accumulated \Ha\  shells.}
\label{tcc93physycalabundaper}
\resizebox{22cm}{!}{   
\begin{tabular}{lr@{}lr@{}lr@{}lr@{}lr@{}lr@{}lr@{}lr@{}l}
\hline
\ &  \multicolumn{15}{c}{Cumulative} \\
\cline{2-17}
\ & \multicolumn{2}{c}{$\overline{{\rm AA}}$}  &  
\multicolumn{2}{c}{$\overline{{\rm AB}}$}   &
\multicolumn{2}{c}{$\overline{{\rm AC}}$}   &  
\multicolumn{2}{c}{$\overline{{\rm AD}}$}   &  
\multicolumn{2}{c}{$\overline{{\rm AE}}$}    & 
\multicolumn{2}{c}{$\overline{{\rm AF}}$}   &  
\multicolumn{2}{c}{$\overline{{\rm AG}}$}    &  
\multicolumn{2}{c}{$\overline{{\rm AH}}$}  \\
\hline

\Te([\SIII])$_{6312}$ &  8000 	& $\pm$200 & 	  8000 	& $\pm$200 & 	  7900 	& $\pm$200 & 	  8000 	& $\pm$300 & 	  8200 	& $\pm$300 & 	  8600 	& $\pm$300 & 	  9100 	& $\pm$400 & 	  9900 	& $\pm$500 \\ 
\Te([\SIII])$_{SO_{23}}$ &  6000 	& $\pm$300 & 	  6000 	& $\pm$300 & 	  6100 	& $\pm$300 & 	  6100 	& $\pm$300 & 	  6200 	& $\pm$300 & 	  6200 	& $\pm$300 & 	  6200 	& $\pm$300 & 	  6300 	& $\pm$300 \\ 
\hline
12+log(S$^{+}$/H$^{+}$) &  6.35 	& $\pm$0.66 & 	  6.35 	& $\pm$0.77 & 	  6.39 	& $\pm$0.86 & 	  6.39 	& $\pm$0.96 & 	  6.37 	& $\pm$1.03 & 	  6.33 	& $\pm$1.10 & 	  6.29 	& $\pm$1.21 & 	  6.21 	& $\pm$1.36 \\ 
12+log(S$^{++}$/H$^{+}$) &  6.93 	& $\pm$0.42 & 	  6.95 	& $\pm$0.50 & 	  6.97 	& $\pm$0.56 & 	  6.96 	& $\pm$0.62 & 	  6.94 	& $\pm$0.66 & 	  6.90 	& $\pm$0.70 & 	  6.84 	& $\pm$0.75 & 	  6.76 	& $\pm$0.83 \\ 
\hline
12+log(O/H) (N2)  &  8.45 	& $\pm$0.81 & 	  8.46 	& $\pm$0.82 & 	  8.48 	& $\pm$1.05 & 	  8.49 	& $\pm$1.19 & 	  8.49 	& $\pm$1.25 & 	  8.50 	& $\pm$1.33 & 	  8.50 	& $\pm$1.56 & 	  8.50 	& $\pm$2.06 \\ 
12+log(O/H) (R$_{23 upper}$)   &  8.92 	& $\pm$0.10 & 	  8.90 	& $\pm$0.10 & 	  8.85 	& $\pm$0.10 & 	  8.83 	& $\pm$0.10 & 	  8.81 	& $\pm$0.10 & 	  8.80 	& $\pm$0.10 & 	  8.79 	& $\pm$0.10 & 	  8.77 	& $\pm$0.10 \\ 
12+log(O/H)  (Ar$_3$O$_3$) &  8.55 	& $\pm$2.16 & 	  8.61 	& $\pm$1.18 & 	  8.64 	& $\pm$0.53 & 	  8.64 	& $\pm$0.53 & 	  8.65 	& $\pm$0.55 & 	  8.65 	& $\pm$0.56 & 	  8.65 	& $\pm$0.57 & 	  8.65 	& $\pm$0.52 \\ 
12+log(O/H)  (S$_3$O$_3$) &  8.88 	& $\pm$0.07 & 	  8.90 	& $\pm$0.09 & 	  8.89 	& $\pm$0.10 & 	  8.89 	& $\pm$0.11 & 	  8.88 	& $\pm$0.11 & 	  8.86 	& $\pm$0.12 & 	  8.84 	& $\pm$0.12 & 	  8.81 	& $\pm$0.12 \\ 
\hline
RS2 &  1.08 	& $\pm$0.01 & 	  1.25 	& $\pm$0.01 & 	  1.27 	& $\pm$0.01 & 	  1.26 	& $\pm$0.01 & 	  1.26 	& $\pm$0.01 & 	  1.28 	& $\pm$0.01 & 	  1.16 	& $\pm$0.01 & 	  1.30 	& $\pm$0.01 \\ 
\Ne &  312 	& \ \ \ --- & 	 133 	& \ \ \ --- & 	 119 	& \ \ \ --- & 	 127 	& \ \ \ --- & 	 122 	& \ \ \ --- & 	 103 	& \ \ \ --- & 	 224 	& \ \ \ --- & 	 95 	& \ \ \ --- \\ 
log(SO$_{23}$) &  -0.01 	& $\pm$0.03 & 	  -0.02 	& $\pm$0.03 & 	  -0.05 	& $\pm$0.03 & 	  -0.06 	& $\pm$0.03 & 	  -0.07 	& $\pm$0.03 & 	  -0.08 	& $\pm$0.03 & 	  -0.09 	& $\pm$0.03 & 	  -0.11 	& $\pm$0.04 \\ 
N2 &  -0.71 	& $\pm$0.11 & 	  -0.70 	& $\pm$0.11 & 	  -0.67 	& $\pm$0.14 & 	  -0.66 	& $\pm$0.15 & 	  -0.65 	& $\pm$0.16 & 	  -0.65 	& $\pm$0.17 & 	  -0.64 	& $\pm$0.19 & 	  -0.64 	& $\pm$0.25 \\ 
log(R$_{23}$)  &  0.24 	& $\pm$0.08 & 	  0.26 	& $\pm$0.07 & 	  0.31 	& $\pm$0.08 & 	  0.33 	& $\pm$0.08 & 	  0.34 	& $\pm$0.08 & 	  0.36 	& $\pm$0.08 & 	  0.38 	& $\pm$0.08 & 	  0.40 	& $\pm$0.09 \\ 
log(O$_{32}$)  &  -1.13 	& $\pm$0.13 & 	  -1.20 	& $\pm$0.14 & 	  -1.20 	& $\pm$0.14 & 	  -1.19 	& $\pm$0.14 & 	  -1.17 	& $\pm$0.13 & 	  -1.13 	& $\pm$0.13 & 	  -1.09 	& $\pm$0.13 & 	  -1.02 	& $\pm$0.12 \\ 
log(Ar$_3$O$_3$) &  -1.19 	& $\pm$3.97 & 	  -1.07 	& $\pm$2.62 & 	  -0.99 	& $\pm$1.34 & 	  -1.00 	& $\pm$1.32 & 	  -0.98 	& $\pm$1.43 & 	  -0.97 	& $\pm$1.49 & 	  -0.97 	& $\pm$1.49 & 	  -0.97 	& $\pm$1.38 \\ 
log(S$_3$O$_3$) &  0.56 	& $\pm$0.18 & 	  0.60 	& $\pm$0.22 & 	  0.57 	& $\pm$0.24 & 	  0.56 	& $\pm$0.27 & 	  0.54 	& $\pm$0.29 & 	  0.50 	& $\pm$0.30 & 	  0.44 	& $\pm$0.32 & 	  0.35 	& $\pm$0.36 \\ 
log(u)  ([\SII]/[\SIII]) &  -2.35 	& $\pm$0.31 & 	  -2.37 	& $\pm$0.36 & 	  -2.39 	& $\pm$0.41 & 	  -2.42 	& $\pm$0.45 & 	  -2.44 	& $\pm$0.48 & 	  -2.45 	& $\pm$0.51 & 	  -2.49 	& $\pm$0.55 & 	  -2.54 	& $\pm$0.61 \\ 
log(u)  ([\OII]/[\OIII]) &  -3.93 	& $\pm$0.05 & 	  -3.98 	& $\pm$0.05 & 	  -3.98 	& $\pm$0.05 & 	  -3.97 	& $\pm$0.05 & 	  -3.95 	& $\pm$0.04 & 	  -3.93 	& $\pm$0.04 & 	  -3.89 	& $\pm$0.04 & 	  -3.84 	& $\pm$0.04 \\ 
log($\eta^{\prime}$)  &  1.52 	& $\pm$0.11 & 	  1.56 	& $\pm$0.12 & 	  1.55 	& $\pm$0.13 & 	  1.53 	& $\pm$0.13 & 	  1.50 	& $\pm$0.14 & 	  1.45 	& $\pm$0.14 & 	  1.39 	& $\pm$0.14 & 	  1.29 	& $\pm$0.15 \\ 
\hline
log([\OIII]$\lambda$5007/\Hb) &  -1.03 	& $\pm$0.04 & 	  -1.06 	& $\pm$0.04 & 	  -1.02 	& $\pm$0.04 & 	  -1.00 	& $\pm$0.04 & 	  -0.97 	& $\pm$0.04 & 	  -0.93 	& $\pm$0.04 & 	  -0.88 	& $\pm$0.04 & 	  -0.80 	& $\pm$0.04 \\ 
log([\NII]$\lambda$6584/\Ha) &  -0.71 	& $\pm$0.01 & 	  -0.70 	& $\pm$0.01 & 	  -0.67 	& $\pm$0.01 & 	  -0.66 	& $\pm$0.01 & 	  -0.65 	& $\pm$0.01 & 	  -0.65 	& $\pm$0.01 & 	  -0.64 	& $\pm$0.01 & 	  -0.64 	& $\pm$0.01 \\ 
log([\SII]$\lambda\lambda$6717,31/\Ha) &  -0.76 	& $\pm$0.07 & 	  -0.74 	& $\pm$0.04 & 	  -0.71 	& $\pm$0.03 & 	  -0.69 	& $\pm$0.02 & 	  -0.67 	& $\pm$0.02 & 	  -0.66 	& $\pm$0.02 & 	  -0.64 	& $\pm$0.02 & 	  -0.62 	& $\pm$0.02 \\ 

\hline
\end{tabular}
}
\end{table}
\end{landscape}

\begin{figure}
\centering
\includegraphics[width=10.0cm]{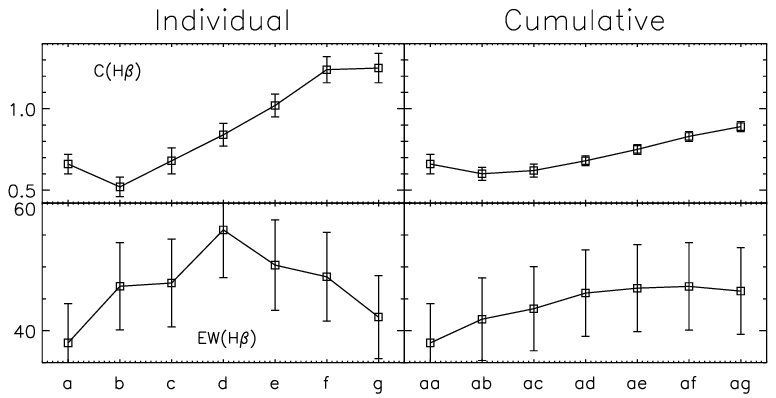}
\caption{BCLMP 93 extinction parameter and  EW(\Hb)  for the 
individual and cumulative shells.}
\label{fc93exew}
\end{figure}


Due to the radial decrease in the ionization parameter, we expect to observe 
a radial decrease in ionization;   IC~132 (fig.~\ref{fishellionrat}) 
shows a decreasing [\OIII]/\Hb\ and increasing [\NII]/\Ha\ and [SII]/\Ha, 
but in  BCLMP~93 (fig.~\ref{fc93ddiag}) these line ratios seem to 
increase outwards. We have already discussed a possible explanation due to confusion in the central region.

\section{Conclusions}

The main objective of this work was to study with spatially resolved spectroscopy the 
differences in the characteristics between the innermost and outermost regions of 
star formation in nearby spiral galaxies with the aim of exploring the differences 
in the properties of intense star formation activity  in widely contrasting  
environments.

We have presented  the results of our IFS data obtained with PMAS in PPak mode 
at the CAHA 3.5m telescope, of \HII\ regions in M33, selected to 
cover a wide range in environment and properties of the star forming regions.
Our sample includes the luminous \HII\ region IC 132 in the outer disc and the 
innermost \HII\ regions in a field of 500pc $\times$ 300pc  that surrounds the 
galaxy centre. The results are presented in the form of intensity maps of the
\HII\ regions intrinsic parameters: line ratios, excitation, extinction, electron 
density, electron temperatures and elemental abundances with a sampling of 
1$^{\prime\prime}\times$1$^{\prime\prime}$.   We also studied  internal 
gradients of the intrinsic parameters of the \HII\ regions for apertures or shells 
defined according to their average surface brightness in \Ha .

Our main findings are: 

\ \\
\noindent{\bf Extinction}\\
We find a clear difference in the extinction and its radial 
gradient between the innermost and outer regions in M 33.
While IC 132 shows a global value of C(\Hb) of about 0.3 and indications of  
lower extinction at larger radius,  C(\Hb) for BCLMP 93 is 
globally  around 1.3, decreasing outwards in the region.

\ \\
\noindent{\bf Equivalent widths and ages}\\

In both systems the EW(\Hb ) 
is smaller in the centre of the region and grows to a maximum in the 
intermediate shells before dropping in the outer shells. The radial 
behaviour of the EW(\Hb ) can be understood by a combination of 
factors, firstly the continuum in the inner regions is strongly affected 
by the massive ionizing stars, secondly the outer regions are affected 
by the disc (IC 132) or bulge (BCLMP 93) component  and globally the 
brightness of the emission lines decreases radially outwards. Thus the 
EW(\Hb)  is probably peaking in the transition zone between the edge 
of the ionizing cluster and the regions where the surface brightness of 
\Hb\ drops below that of the older population.

We find a  smaller equivalent width of \Hb\  in BCLMP 93 than in IC 132 indicating 
higher extinction and either an older \HII\ region or more likely a larger 
contamination due to the presence of an old stellar population in the central 
regions of M 33, i.e. the bulge population. 

It is interesting to note that the maximum EW(\Hb)  $\sim$ 300 \AA\ in 
IC 132 is close to the  maximum of $\sim$ 400 \AA\, for
SB99 models \citep{sb99}, assuming instantaneous star formation with 
upper mass limit = 100M$_\odot$, zero age main sequence, an initial 
mass function with $\alpha$ = 2.35 and Z=0.020Z$_\odot$.
The asymptotic maximum of the cumulative EW(\Hb ) reaches $\sim$ 200 \AA\  
indicating an ionizing cluster with an age of less than 3 Myrs  that is 
in agreement with the detection of WN stars that are expected to appear after 
2.5~Myrs of evolution of a young massive cluster.

For BCLMP 93 the same procedure suggests an upper limit for the age of the cluster of 
about 5 Myrs, bearing in mind that the large contamination due to the 
underlying bulge population precludes a more precise age estimate. This 
age makes the central region older than IC 132 and could explain the lack of present day WR 
features in the region.

\ \\
\noindent{\bf Electron density}\\
The electron density (\Ne) was estimated from the 
RS2=[\SII]$\lambda$6717/[\SII]$\lambda$6731 line ratio after deblending the doublet. This diagnostic 
situates the \HII\ regions in the low density limit. Even if an exact value 
for \Ne\ is not possible, a variation in the ratio with position is observed,
supporting the fact that to assume a single \Ne\ value to characterize a whole
\HII\ region is too simplistic. Better spectral resolution, access to 
alternative density diagnostics and revision of the theoretical estimation based on new atomic data for collisional strengths
are necessary to shed light on the subject.

\ \\
\noindent{\bf Electron  temperature}\\
For the central zone of IC~132 the electron temperature (\Te) 
was estimated from the detected  [\OIII]$\lambda$4363 and 
[\SIII]$\lambda$6312 lines allowing the use of the 
[\OIII]$\lambda\lambda$4959,5007/[\OIII]$\lambda$4363 
and [\SIII]$\lambda\lambda$9069,9532/[\SIII]$\lambda$6013 
ratios to compute \Te([\OIII]) and Te([\SIII])
respectively. For BCLMP~93 the lack of detection of 
[\OIII]$\lambda$4363 implies that only Te([\SIII]) was determined
in its central zone. The blueshift of the galaxy means that the 
[\OIII]$\lambda$ 4363 line is only an upper limit, as it is affected 
by the street light  mercury line (Hg I $\lambda$ 4358.3).
 
 The 2D distribution of  \Te([\OIII]) and \Te([\OII]) in IC 132 appears to be 
rather uniform in the high surface brightness region 
surrounded by slightly hotter gas. The mean values are 16100 K for 
\Te([\OIII]) and between 15300 K and 13700 K for \Te([\OII]) with an 
rms  scatter per spaxel of about 3200 K in all cases. For \Te([\SIII]) 
we find a mean value of 16200 K with an rms of 3800 K over 279 spaxels.
 Given that the maps include about 300 spaxels, the error of the mean 
temperatures is less than 600 K.  These  results are confirmed by the 
segmentation analysis that shows a positive gradient in all temperature 
estimates and a global cumulative value of 13500 K  for \Te([\OII]) and 
15100 for \Te([\OIII]) while \Te([\SIII]) reaches an asymptotic value of 13600 K.
 
 Higher \Te([\SIII]) is derived for IC 132 than for BCLMP 93. This is in
agreement with the centre being more metallic than the disc. Consequently  
in central regions cooling is more efficient through fine structure 
lines in the infrared.

The  analysis in shells indicates  that the direct method is reliable for the 
brightest  zones. However as the radial distance within the regions 
increases the lines are still detected in the  integrated spectra and 
can be included in the usual temperature computation but  
the inclusion of fainter zones produces physically unrealistic 
temperature values. 

This can be seen in tables \ref{ticphysycalabundaper} and 
\ref{tcc93physycalabundaper} and in figures 
\ref{fishelltemps} and \ref{fc93tempion}
where the outer ring drives the 
temperatures up. This is true for temperature determinations 
using [\OIII]$\lambda$4363 and [\OII]$\lambda$6312 and 
hence can not be attributed just to the contamination of 
[\OIII]$\lambda$4363 by street lights due to the particular blueshift of M33. 

 \ \\
\noindent{\bf Abundances}\\
The smallest dispersion in O/H determined from  empirical indicators is obtained for  O3N2 followed by N2. Although the related line ratio may have 
a large dispersion along the region, whether or not it is transferred  to
(O/H) depends on the calibration of the relation. If the conversion
is lineal, the steepness of the slope dictates the dispersion in 
(O/H). This is why the oxygen abundances from O3N2 and N2 are 
almost constant {\it despite the fact that the line ratios have been 
discarded as good indicators because their larger dependence on 
excitation \citep{relanoetl10}.}

R$_{23}$ displays a dispersion of $\sim$ 0.3 dex for the central
zone, however when individual spaxels are converted to (O/H)
an extra difficulty is found because the upper and lower branches
are included for the same object, thus the spaxels in the transition 
zone have larger uncertainty depending on the calibration chosen.
This is worse for the \cite{pilyugin01hz} method because
a first estimation of (O/H) is required and the area of transition between 
branches is not defined.
{\it The fact that for the same region both branches are involved is
suspicious and raises the question about the validity of the method
when applied to individual spaxels.}

Sulphur is the only direct indicator in common for IC 132 
and BCLMP 93. It shows higher abundance for the central region coinciding with the results for the rest of the
empirical indicators.

\ \\
\noindent{\bf Diagnostic diagrams}\\
Our search for systematic differences between the inner and outer \HII\ 
regions in M33  was complemented with the use of diagnostic diagrams.  
In particular we focused in the [\OIII]$\lambda$ 5007/\Hb,
 vs.~[\NII] $\lambda$ 6584/H$\alpha$ diagram and found that the individual
spaxels display a wide range of ionization conditions, extending
by more than 2.5 dex from the location of the brightest spot. 
The distribution of points for the two \HII\ regions considered displays 
a striking clear dichotomy. While the  points for BCLMP 93 show a vertical scatter, 
the points for IC 132  have a horizontal spread orthogonal to that of BCLMP 93.
The regions occupied by the distributions have hardly any superposition while 
some points  clearly move away from the \HII\ region zone of the diagnostic 
diagram intruding in the AGN and transition zone.
This intrusion raises a question about the wisdom of using the results of global 
models in the diagnostic diagrams for individual spaxels of resolved 
systems, including the definition of the HII boundary line.

Thus it is important to be cautious when applying or comparing model 
predictions  for global observations to spatially resolved data. In this case 
the BPT diagnostic diagrams may give misleading results for spaxels with 
ionization conditions that may not be reached in global spectra  
\citep{ercolanoetal12}. 

Bearing this in mind we  computed pseudo 3D photoionization models and  
compared the distributions of observed line ratios in the diagnostic diagram 
with the results of our photoionization models and found that
 it was possible to reproduce the spread observed in the [\OIII]5007/\Hb\ 
versus [\NII]6584/\Ha\ diagram for both zones. 
 We conclude that  the position and horizontal spread of points in IC 132  
is reproduced by models with high U,  high $T_*$ and low Z
while the position and the vertical scatter of the observed points in the 
central \HII\ regions is better reproduced by models with  lower U and  $T_*$ 
and higher Z. 
The contribution of the central X-ray source to the ionization budget is not
clear, since not very  anomalous line ratios were detected in its proximity.

\ \\
\noindent{\bf Wolf Rayet stars}\\
Blue WR features were isolated and characterized for IC~132, the spatial
resolution is enough to distinguish two concentrations. 
No WR features were detected in the central zone in apparent 
discordance with theoretical  predictions that higher metallicity 
regions should have  larger WR content, but perhaps consistent with 
the estimated larger age of the central region.

\section*{Acknowledgments}

We are grateful to Sebasti\'an 
S\'{a}nchez, resident astronomer at  CAHA, for his help 
during the observations and in the process of data 
reduction. We thank an extremely  thorough referee 
whose comments greatly contributed to improving this paper.
JL thanks Fabio Bresolin for illuminating 
discussions during his study stay at the Institute for Astronomy
of the University of Hawaii and for extra financial support.

RT , ET and DRG are grateful to the Mexican research council 
(CONACYT) for supporting this research under grants 
CB-2005-01-49847, CB-2007-01-84746, CB-2008-103365-F and
CB-2011-01-167281-F3.
JLH is grateful to CONACYT  and  to the Research Council of the State of Puebla (CONCYTEP)
for  financial support both through national scholarships and
an International Research Visiting program.

We acknowledge  financial support  from the Estallidos 
collaboration of the Spanish Ministerio de Educaci\'{o}n 
y Ciencia (AYA2007- 67965-C03-03 and 02).
JLH, ET and RT enjoyed the hospitality of the Astrophysics 
Group of the Theoretical Physics Department of the 
Universidad Aut\'onoma de Madrid during the process 
that led to the reduction of the data analyzed in this paper.

Based on observations collected at the 
Centro Astron\'omico Hispano Alem\'an (CAHA) at Calar 
Alto, operated jointly by the Max-Planck Institut f\"ur 
Astronomie and the Instituto de Astrof\'\i sica de Andaluc\'\i a (CSIC).
This research draws upon data provided by {\it The Resolved 
Stellar Content of Local Group Galaxies Currently Forming Stars} 
PI: Dr. Philip Massey,  as distributed by the NOAO Science Archive. 
NOAO is operated by the Association of Universities for Research in 
Astronomy (AURA), Inc. under a cooperative agreement with the 
National Science Foundation.


\include{M33biblio}

\end{document}
